\documentclass[superscriptaddress,nofootinbib,12pt]{revtex4-2}
\usepackage[utf8]{inputenc}
\usepackage{graphicx}
\usepackage{epsf}
\usepackage{bm}
\usepackage{amsmath}
\usepackage{amsfonts}
\usepackage{amssymb}
\usepackage{subfigure}
\usepackage{scalerel}
\newcommand{\abs}[1]{\left\lvert#1\right\rvert}
\usepackage[font=scriptsize,labelfont=bf]{caption}
\RequirePackage{latexsym}
\usepackage{hyperref}
\hypersetup{
    colorlinks=true,
    linkcolor=red,
    citecolor=blue,
    filecolor=magenta,      
    urlcolor=cyan,
    pdftitle={Global dynamics of two models for Quintom Friedman-Lemaître-Robertson-Walker Universes},
    pdfpagemode=FullScreen}

\begin{document}

\title{Global dynamics of two models for Quintom Friedman-Lemaître-Robertson-Walker Universes}

\author{\textsc{Genly Leon}}
\email{genly.leon@ucn.cl}
\affiliation{Departamento de Matem\'aticas, Universidad Cat\'olica del Norte, \\ Avda. Angamos 0610, Casilla 1280 Antofagasta, Chile}
\affiliation{Institute of Systems Science, Durban University of Technology, PO Box 1334, Durban 4000, South Africa}

\author{\textsc{Alan Coley}}
\email{alancoley@dal.ca}
\affiliation{Department of Mathematics and Statistics,
Dalhousie University, \\ Halifax, Nova Scotia, Canada  B3H 3J5}

\author{\textsc{A. Paliathanasis}}
\email{anpaliat@phys.uoa.gr}
\affiliation{Institute of Systems Science, Durban University of Technology, PO Box 1334,
Durban 4000, South Africa}
\affiliation{Departamento de Matem\'aticas, Universidad Cat\'olica del Norte, \\ Avda. Angamos 0610, Casilla 1280 Antofagasta, Chile}

\author{\textsc{Jonathan Tot}}
\email{jonathan.tot@dal.ca}
\affiliation{Department of Mathematics and Statistics,
Dalhousie University, \\ Halifax, Nova Scotia, Canada  B3H 3J5}

\author{\textsc{Balkar Yildirim}}
\email{bl541771@dal.ca}
\affiliation{Department of Mathematics and Statistics,
Dalhousie University, \\ Halifax, Nova Scotia, Canada  B3H 3J5}

\begin{abstract}
We comprehensively analyze the dynamics for the gravitational field equations for the Chiral-Quintom theory in a Friedman-Lemaître-Robertson-Walker cosmology with an additional matter source. We consider a new set of dimensionless variables and write the field equations in the equivalent form of an algebraic-differential system. Specifically, we consider two families of quintom models where the two scalar fields interact in the kinetic sector. We mathematically focus on the dynamical effect of spatial curvature. Physically, we find two periods of inflation related to the Universe's early and late-time acceleration phases. 
\end{abstract}


\maketitle
\tableofcontents

\section{Introduction}

Scalar fields have many applications in cosmology. They have been introduced to describe the source related to the acceleration of the Universe. The early-time acceleration phase of the Universe, known as inflation, is attributed to the inflaton field, which is usually described by a scalar field minimally coupled to gravity. Additionally, a scalar field has been introduced to describe the acceleration of the late Universe, which is related to dark energy (DE). In the vast literature on the subject, numerous models have arguments in their favour but without representing the definitive answer \cite{Sahni:2006pa, Copeland:2006wr}. Current astrophysical observations are consistent with the cosmological constant model ($w=-1$), where $w$ is the effective Equation of State Parameter (EoS) of matter; however, some favour the case when the equation of state parameter crosses the phantom divide line $w<-1$. 
 
 Various proposed DE models by cosmologists can cross the phantom barrier. A single scalar field with canonical (quintessence) or non-canonical kinetic energy does not give the desired result \cite{Melchiorri:2002ux, Vikman:2004dc, Sen:2005ra}. Nevertheless, the result is possible if these models were considered within theories with non-minimal couplings \cite{Curbelo:2005dh}. Beyond the single scalar field model, cosmologists have explicitly studied the quintom models \cite{Feng:2004ad, Wei:2005fq, Wei:2005si, Wei:2005nw, Guo:2004fq, Zhang:2005eg, Feng:2004ff, Wu:2005ap, Xia:2004rw, Zhao:2005vj, Zhang:2005kj, Lazkoz:2006pa, Lazkoz:2007mx, MohseniSadjadi:2006hb, Alimohammadi:2006tw, Elizalde:2004mq, Lazkoz:2006pa, Lazkoz:2007mx, Elizalde:2008yf, Cai:2009zp, Leon:2012vt, Leon:2018lnd}. 
 
The quintom models belong to the family of multi-scalar field models. Specifically, it is a two-scalar field model where one scalar field is a quintessence and the second is a phantom field with negative kinetic energy. Such a model provides a cosmological history where the quintessence dominates in early times with $w>-1$, and the phantom field dominates later, with $w<-1$. The global behaviour of the quintom model is different from the behaviour provided by the individual scalar fields, that is, the quintessence and the phantom scalar field models. An essential characteristic of the quintom model is that the EoS parameter  $w$ can cross the phantom divide line $-1$ more than once. Quintom-like behaviour (with a $w=-1$ crossing) has also been found in the context of holographic dark energy \cite{Zhang:2005yz, Zhang:2005hs, Zhang:2006qu}.
 
Quintom cosmologies with an exponential potential and dark matter (DM) were investigated from a dynamical systems perspective in \cite{Lazkoz:2006pa}. The novel behaviour of quintom models admitting either tracking attractors or phantom attractors was observed. A similar analysis with a more general potential function was studied in \cite{Lazkoz:2007mx}. Moreover, in \cite{Zhang:2005eg, Lazkoz:2006pa}, the interaction between the conventional scalar and phantom fields was considered. It has been proven that in the absence of interactions, the solution dominated by the phantom field should act as the system's global attractor, and the energy exchange does not affect its attractor behaviour.
Furthermore, the case in which the interaction term dominates against the mixed terms of the potential was studied in \cite{Lazkoz:2006pa}. A set of dynamical variables better adapted than those in \cite{Zhang:2005eg} was devised. Notably, it was proved that the hypothesis in \cite{Zhang:2005eg} is correct only in the cases in which the phantom phase excludes the presence of scaling attractors (in which the energy density of the quintom field and the energy density of DM are proportional). A similar result was attained in \cite{Wei:2005fq}, where it was shown that in the context of the hessence model (a non-canonical complex scalar field with an internal degree of freedom rather than two independent real scalar fields) in which the equation-of-state parameter can cross the phantom divide \cite{Wei:2005nw}, assuming the exponential and (inverse) power-law potentials, some stable attractors could exist, which are either scaling solutions or hessence-dominated solutions with EoS larger than or equal to $-1$. The analysis in \cite{Lazkoz:2007mx} support the latter results.

The case where radiation and dark matter (as described by perfect fluids) were included was studied in \cite{Leon:2012vt}. A complete dynamical analysis was presented in \cite{Leon:2018lnd} for a class of exponential potentials. In particular, the unstable and centre manifold of the massless scalar field cosmology motivated by the numerical results given in \cite{Lazkoz:2006pa} were constructed. The role of the spatial curvature on the dynamics was also investigated, and several monotonic functions were defined on relevant invariant sets for the Quintom cosmology. 

In  \cite{Leon:2018lnd}, a Friedmann–Lemaître–Robertson–Walker (FLRW) model for the geometry without restricting to the case of flat space-like slices was considered. 
Some additional arguments were presented to support the results of references \cite{Zhang:2005eg, Guo:2004fq, Lazkoz:2006pa, Lazkoz:2007mx}. The authors emphasized the importance of determining well-defined monotone functions to rule out periodic, recurrent, or homoclinic orbits. The analysis goes beyond the simple study of critical points and, hence, is more general than the analysis of those references. For exponential potentials, it reinforces the fact, first noticed in reference \cite{Lazkoz:2006pa}, that the dynamics are dominated by critical points on the boundary of some invariant sets and by heteroclinic orbits joining them. 

The Chiral generalized cosmology is a related model whose dynamics were studied in \cite{Paliathanasis:2020wjl}. In \cite{Paliathanasis:2020sfe}, a detailed analysis of the dynamics of four chiral-like cosmological models with phantom terms in a spatially flat FLRW background space was presented. When the second scalar field is phantom, the parameter of the equation of state of the cosmological fluid crosses the phantom divide line twice without the appearance of ghosts. This is the quintom model that we shall study here.

This paper extends the analysis presented in \cite{Tot:2022dpr}. In this study, we shall show the complete analysis of curvature-dominated equilibrium points and scaling solutions for two different quintom-chiral models, and we will include matter. 

The paper is organized as follows. \S\ref{sec2} and \S\ref{sec3} include the principal analysis of this study, where we present a global analysis for the phase space of the field equations. The case where matter is included is also considered. Finally, \S\ref{sec4} summarizes our results, and we present some conclusions.

\section{First model}
\label{sec2}

We consider the two-scalar field model with action integral
\begin{equation}
	S=\int\sqrt{-g}dx^{4}\left(  R-\frac{1}{2}g^{\mu\nu}\nabla_{\mu}\phi \nabla_{\nu}\phi + \frac{1}{2}g^{\mu\nu}e^{\kappa\; \phi}\nabla_{\mu}\psi \nabla_{\nu}\psi-V\left(  \phi\right)  \right) + S_{\text{matter}}, \label{sp.01}
\end{equation}
where the two scalar fields $\phi\left(x^{\mu}\right)$, which is a quintessence-like field, and $\psi\left(x^{\nu}\right) $, which is a phantom-like field, have kinetic terms which lie on a two-dimensional manifold, and $V(\phi)$ is the quintessence potential. For the background space, we assume the FLRW metric
\begin{equation}
	ds^2 = -dt^2 +a^2(t)\left( \frac{ dr^2}{1-Kr^2} +r^2\left( d\theta^2 +\sin^2\theta\,d\varphi^2 \right)\right),\label{sp.02}
\end{equation}
where $K$ is the spatial curvature for the three-dimensional hypersurface. For $K=0$, we have a spatially flat universe; for $K=1$, we have a closed universe; for $K=-1$, the line element (\ref{sp.02}) describes an open universe. The quantity $S_{\text{matter}}$ corresponds to the matter action considered as a perfect fluid, with energy density $\rho_m$, pressure $p_m$ with the constant equation of state parameter $p_m= w \rho_m$.

Now we study the cosmological model with action integral (\ref{sp.01}) for the line element (\ref{sp.02}). The gravitational field equations are
\begin{align}
& -3H^2 +\frac{1}{2}\dot{\phi}^2 -\frac{1}{2}e^{\kappa \phi}\dot{\psi}^2 +V\!\left( \phi \right) + \rho_m -\frac{3K}{a^2}=0, \label{0sp.04}\\
& 2\dot{H} +3H^2 +\frac{1}{2}\dot{\phi}^2 -\frac{1}{2}e^{\kappa  \phi}\dot{\psi}^2 -V\!\left( \phi \right) + w \rho_m +  \frac{K}{a^2}=0, \label{0sp.05}\\
& \ddot{\phi}+3H\dot{\phi} +\frac{\kappa}{2}e^{\kappa\, \phi}\dot{\psi}^2 +V'\left( \phi \right) =0, \label{0sp.06} \\
& \ddot{\psi} + 3H\dot{\psi} +\kappa \dot{\phi}  \dot{\psi} =0, \label{0sp.07} \\
& \dot{\rho_m} + 3 H(1+w) \rho_m=0, \label{matter} 
\end{align}
where $H=\frac{\dot{a}}{a}$ is the Hubble function, and we consider the equation of state parameter satisfying $-1\leq w\leq 1$. 
Following \cite{Lazkoz:2006pa} we take an exponential potential $V(\phi)= V_0 e^{\lambda \phi}$.

\subsection{Dynamical systems formulation ($K=0$)}

We define 
\begin{align}
  &   \chi^2 = \frac{1}{2} \dot{\phi}^2 + V(\phi) + \rho_m, \label{Zerochi_lambda} \; \text{and}
\\
	&h^2= \frac{3 H^2}{\chi^2}, \; \eta^2= \frac{e^{\kappa \phi}\dot{\psi}^2}{2 \chi^2}, \;  \Phi^2 = \frac{\dot{\phi}^2}{2 \chi^2}, \; \Psi= \frac{V(\phi)}{\chi^2}, \; \Omega= \frac{\rho_m}{\chi^2},\label{Zerovars}
\end{align}
which satisfy
\begin{equation}
	h^2 + \eta^2= \Phi^2+ \Psi + \Omega=1.  \label{eq102}
\end{equation}
Using \eqref{eq102}, we have bounded variables
\begin{equation}
	0 \leq h^2, \Phi^2, \eta^2,\Psi,\Omega \leq 1.  
\end{equation}
Then, equations \eqref{0sp.05}, \eqref{0sp.06}, \eqref{0sp.07} become
\begin{align}
& \frac{\dot{H}}{\chi^2}=\frac{1}{2} \left(\eta ^2-h^2-\Phi ^2-w \Omega +\Psi \right), \label{flathubble} 	\\
& \frac{\ddot{\phi}}{\chi^2}=-  \kappa\,\eta^2 -\sqrt{6} h \Phi -\lambda  \Psi,  \\
& \frac{\ddot{\psi}}{\dot{\psi}\chi}=-\frac{\sqrt{6} h+2 \kappa  \Phi }{\sqrt{2}}, \\
& \frac{\dot{\rho_m}}{\chi^2}=-\sqrt{3} (1+w) h\,\Omega\,\chi. 
\end{align}

Introducing the time derivative $d\tau=\chi d t$ and taking the derivatives of the variables with respect to the new time variable (i.e., $f^{\prime}=df/d\tau$), the field equations become
\begin{align}
	&\Phi^\prime= -\frac{ \kappa\,\eta^2}{\sqrt{2}} -\sqrt{3}h\Phi-\frac{ \lambda\Psi}{\sqrt{2}} -\frac{ \Phi\dot{\chi}}{\chi^2}, \\
	&h^\prime= \frac{1}{2}\sqrt{3}\left( \eta^2-h^2-\Phi^2 +\Psi\right) -\frac{ h\dot{\chi}}{\chi^2}, \\
	&\eta^\prime= -\frac{ \kappa\,\eta\,\Phi}{\sqrt{2}} -\sqrt{3}\eta h -\frac{ \eta\dot{\chi}}{\chi^2}, \\
	&\Psi^\prime= \sqrt{2}\lambda\Phi\Psi -\frac{ 2\Psi\dot{\chi}}{\chi^2}, \\
	& \Omega^{\prime}= -\sqrt{3} (1+w)h\,\Omega -\frac{2 \Omega\,\dot{\chi}}{\chi
   ^2}. 
\end{align}
Next, by substituting
\begin{align}
	&\frac{\dot{\chi}}{\chi^2}= -\frac{1}{2} \left(\sqrt{2}\kappa\,\eta^2 \Phi +\sqrt{3} h \left[2 \Phi ^2 +(1+w)\Omega \right]\right), \label{flatchi_de}\\
	&\eta^2= 1-h^2 , \\
	&\Psi= 1 - \Omega -\Phi^2 ,
\end{align}
 we obtain
\begin{align}
    \Phi^\prime &= \frac{1}{2}\left(\sqrt{2} \left(\lambda  \left(\Phi ^2+\Omega -1\right)-\left(h^2-1\right) \kappa  \left(\Phi^2 - 1\right)\right) +\sqrt{3}h\Phi\left[2\Phi^2 + (1+w)\Omega -2\right]\right),\label{eq54}\\
    h^{\prime}  &= \frac{1}{2}\left(h^2 - 1\right) \left(\sqrt{3} \left[2 \Phi ^2 +(1+w)\Omega -2\right]-\sqrt{2}\kappa\, h\Phi \right), \label{eq55}\\
    \Omega^{\prime} & =\Omega\left(\sqrt{3} h \left[2 \Phi^2+(1+w)(\Omega -1) \right]-\sqrt{2} \left(h^2-1\right) \kappa\,\Phi \right), \label{eq56}
\end{align}
and the auxiliary equations
\begin{align}
    \eta^{\prime} &= \frac{1}{2} h \sqrt{1-h^2} \left(\sqrt{3} \left[2 \Phi^2 +(1+w)\Omega -2\right]-\sqrt{2}\kappa\,h\Phi\right),\\
    \Psi^{\prime} &= \left(1 -\Phi^2 -\Omega \right)\left( -\sqrt{2} \kappa\,h^2\Phi +\sqrt{3}h\left[2 \Phi^2 +(1+w)\Omega \right] +\sqrt{2}(\kappa +\lambda)\Phi  \right). 
\end{align}

\begin{table}[]
    \centering
    \resizebox{\textwidth}{!}{   
    \begin{tabular}{|ccccccc|}
\hline 
Label & $\Phi$ & $h$ & $\Omega$ & $k_1$ & $k_2$ & $k_3$ \\\hline
$A$ & $ 1$ & $1$ & $0$ & $-\sqrt{3} (w-1)$ & $-\sqrt{2} \kappa$  & $\sqrt{2} \lambda +2 \sqrt{3}$ \\\hline
$B$ &  $-1$ & $1$ & $0$ & $-\sqrt{3} (w-1)$ & $\sqrt{2} \kappa$  & $2 \sqrt{3}-\sqrt{2} \lambda$  \\\hline
$C$ &  $1$ & $-1$ & $0$ & $\sqrt{3} (w-1)$ &$ -\sqrt{2} \kappa$  & $\sqrt{2} \lambda -2 \sqrt{3}$ \\\hline
$D$ &  $-1$ & $-1$ & $0$ & $\sqrt{3} (w-1)$ & $\sqrt{2} \kappa $ & $-\sqrt{2} \lambda -2 \sqrt{3}$ \\\hline
$E$ & $-\frac{\lambda }{\sqrt{6}}$ & $1 $& $0$ & $\frac{\lambda ^2-3 w-3}{\sqrt{3}}$ & $\frac{\lambda ^2-6}{2 \sqrt{3}} $&  $ \frac{\lambda  (\kappa +\lambda )-6}{\sqrt{3}}$ \\\hline
$F$ & $\frac{\lambda }{\sqrt{6}}$ & $-1$ & $0$ & $\frac{-\lambda ^2+3(w+1)}{\sqrt{3}}$ &$ -\frac{\lambda ^2-6}{2 \sqrt{3}}$ &  $ \frac{6-\lambda  (\kappa +\lambda )}{\sqrt{3}} $\\\hline
$G$ & $\frac{1}{\sqrt{\frac{1}{6} \kappa  (\kappa +\lambda )+1}}$ & $-\frac{\kappa +\lambda }{\sqrt{\kappa  (\kappa +\lambda )+6}}$ & $0$ 
& $\frac{\kappa  (1+w)+\lambda (w-1)}{\sqrt{\frac{1}{3} \kappa  (\kappa +\lambda )+2}}$ & $\frac{\sqrt{3} \kappa -\sqrt{-\kappa\left(4 \left(\kappa ^2-6\right) \lambda +8 \kappa  \lambda ^2-27 \kappa +4 \lambda ^3\right)}}{2
   \sqrt{\kappa  (\kappa +\lambda )+6}}$ & $\frac{\sqrt{3} \kappa +\sqrt{-\kappa\left(4 \left(\kappa ^2-6\right) \lambda +8 \kappa  \lambda ^2-27 \kappa +4 \lambda ^3\right)}}{2 \sqrt{\kappa (\kappa +\lambda )+6}}$ \\\hline
$H$ & $-\frac{1}{\sqrt{\frac{1}{6} \kappa  (\kappa +\lambda )+1}}$ & $\frac{\kappa +\lambda }{\sqrt{\kappa  (\kappa    +\lambda )+6}}$ & $0$ & $-\frac{\kappa  (1+w)+\lambda (w-1)}{\sqrt{\frac{1}{3} \kappa  (\kappa +\lambda )+2}}$ & $ -\frac{\sqrt{3} \kappa -\sqrt{-\kappa\left(4 \left(\kappa ^2-6\right) \lambda +8 \kappa  \lambda ^2-27 \kappa +4 \lambda ^3\right)}}{2
   \sqrt{\kappa  (\kappa +\lambda )+6}}$ & $-\frac{\sqrt{3} \kappa +\sqrt{-\kappa\left(4 \left(\kappa ^2-6\right) \lambda +8 \kappa  \lambda ^2-27 \kappa +4 \lambda ^3\right)}}{2 \sqrt{\kappa (\kappa +\lambda )+6}}$ \\\hline
$O^{-}$ &  $-1$ &$ 0$ & $0$ & $-\sqrt{2} \kappa$  &$ -\frac{\kappa }{\sqrt{2}}$ & $-\sqrt{2} (\kappa +\lambda ) $\\\hline
$O^{+}$ &  $1$ & $0$ & $0$ & $\sqrt{2} \kappa$  & $\frac{\kappa }{\sqrt{2}}$  & $\sqrt{2} (\kappa +\lambda ) $\\\hline
$I$ & $0$ & $-1$ & $1$ & $-\sqrt{3} (w-1)$ & $-\frac{1}{2} \sqrt{3} (w-1)$ &$ -\sqrt{3} (w+1)$ \\\hline
$J$ &  $0$ & $1$ & $1$  &$\sqrt{3} (w-1)$ & $\frac{1}{2} \sqrt{3} (w-1)$ & $\sqrt{3} (w+1) $\\\hline
$K$ & $\frac{\sqrt{\frac{3}{2}} (w+1)}{\lambda }$ & $-1$ & $1-\frac{3 (w+1)}{\lambda ^2} $ & $-\frac{\sqrt{3} (\kappa  (1+w)+\lambda  (w-1))}{\lambda }$ & $\frac{\sqrt{3} \left(\lambda (1-w)-\sqrt{(w-1)(\lambda ^2 (9 w+7)-24 (1+w)^2)}\right)}{4 \lambda }$ & $\frac{\sqrt{3} \left(\lambda (1-w) +\sqrt{(w-1)(\lambda ^2 (9 w+7)-24 (1+w)^2)}\right)}{4 \lambda }$\\\hline
$L$ & $-\frac{\sqrt{\frac{3}{2}} (w+1)}{\lambda }$ & $1 $& $1-\frac{3 (w+1)}{\lambda ^2}$ & $-\frac{\sqrt{3} (\kappa  (1+w)+\lambda  (w-1))}{\lambda }$ & $\frac{\sqrt{3} \left(\lambda (1-w)-\sqrt{(w-1)(\lambda ^2 (9 w+7)-24 (1+w)^2)}\right)}{4 \lambda }$ & $\frac{\sqrt{3} \left(\lambda (1-w) +\sqrt{(w-1)(\lambda ^2 (9 w+7)-24 (1+w)^2)}\right)}{4 \lambda }$\\\hline
$M$ & $\frac{w-1}{\sqrt{\frac{2 \kappa ^2}{3}+(w-1)^2}}$ & $\frac{\kappa }{\sqrt{\kappa ^2+\frac{3}{2} (w-1)^2}}$ & $  \frac{2 \kappa ^2}{2 \kappa ^2+3 (w-1)^2}$ &$ -\frac{\mu_{11}}{2 \left(2 \kappa ^2+3
   (w-1)^2\right)^{7/2}}$ & $-\frac{\mu_{12}}{2 \left(2 \kappa ^2+3 (w-1)^2\right)^{7/2}}$& $-\frac{\mu_{13}}{2 \left(2 \kappa ^2+3 (w-1)^2\right)^{7/2}} $\\\hline
$N$ & $-\frac{w-1}{\sqrt{\frac{2 \kappa ^2}{3}+(w-1)^2}}$ & $-\frac{\kappa }{\sqrt{\kappa ^2+\frac{3}{2} (w-1)^2}}$ & $ \frac{2 \kappa ^2}{2 \kappa ^2+3 (w-1)^2} $& $\frac{\mu_{11}}{2 \left(2 \kappa ^2+3
   (w-1)^2\right)^{7/2}} $& $\frac{\mu_{12}}{2 \left(2 \kappa ^2+3 (w-1)^2\right)^{7/2}}$ & $  \frac{ \mu_{13}}{2 \left(2 \kappa ^2+3 (w-1)^2\right)^{7/2}}$ \\\hline
    \end{tabular}}
    \caption{Equilibrium points  of the system \eqref{eq54}, \eqref{eq55} and \eqref{eq56} and their eigenvalues.}
    \label{tab:0}
\end{table}

The main focus of the analysis is to investigate whether different orbits inflate. The condition for a point in the phase space to be inflationary, using the deceleration parameter, $q$, is defined by
\begin{align}
	&q = -\frac{\dot{H}}{H^2}-1 < 0, \label{q.def}
\end{align}
which, using equation \eqref{Zerovars}, in terms of the dimensionless variables, we can find the inflationary condition with these variables:
\begin{align}
    q = 2+  \frac{6 \Phi ^2+3 (w+1) \Omega -6}{2 h^2} <0.
\end{align}
   
The equilibrium points   $(\Phi,h, \Omega)$  of the system \eqref{eq54}, \eqref{eq55} and \eqref{eq56} are (see Tab.  \ref{tab:0}): 

\begin{itemize}

\item 
 $A: (1,1,0)$, with eigenvalues $ \left\{-\sqrt{3} (w-1),-\sqrt{2} \kappa ,\sqrt{2} \lambda +2 \sqrt{3}\right\}$.  Therefore, it is a source for $\{w<1, \kappa<0, \lambda >-\sqrt{6}\}$. It is a saddle for $\{w<1, \kappa>0, \lambda <-\sqrt{6}\}$, or $ \{w<1, \kappa>0, \lambda >-\sqrt{6}\} $, or $ \{w<1, \kappa<0, \lambda <-\sqrt{6}\}$. The deceleration parameter is $q=2$. Therefore, it represents a stiff fluid solution. It never inflates.
 
 \item 
 $B: (-1,1,0)$, with eigenvalues $\left\{-\sqrt{3} (w-1),\sqrt{2} \kappa ,2 \sqrt{3}-\sqrt{2} \lambda \right\}$.  Therefore, it is a source for $\{w<1, \kappa>0,\lambda<\sqrt{6}\}$. It is is a saddle for $\{w<1, \kappa<0, \lambda >\sqrt{6}\} $, or $ \{w<1, \kappa<0, \lambda <\sqrt{6}\} $, or $ \{w<1, \kappa>0, \lambda >\sqrt{6}\}$. The deceleration parameter is $q=2$. Therefore, it represents a stiff fluid solution. It never inflates.

\item 
 $C: (1,-1,0)$, with eigenvalues $ \left\{\sqrt{3} (w-1),-\sqrt{2} \kappa ,\sqrt{2} \lambda -2 \sqrt{3}\right\}$. Therefore, it is a sink for $\{w<1, \kappa>0, \lambda<\sqrt{6}\}$. It is  a saddle for $\{w<1, \kappa>0, \lambda >\sqrt{6}\} $, or $ \{w<1, \kappa<0, \lambda <\sqrt{6}\}$, or $\{w<1, \kappa<0, \lambda >\sqrt{6}\}$. The deceleration parameter is $q=2$. Therefore, it represents a stiff fluid solution. It never inflates.
 
\item  
$D: (-1,-1,0)$, with eigenvalues  $\left\{\sqrt{3} (w-1),\sqrt{2} \kappa ,-\sqrt{2} \lambda -2 \sqrt{3}\right\}$. Therefore, it is a sink for $\{w<1, \kappa<0, \lambda >-\sqrt{6}\}$. It is a saddle for $\{w<1, \kappa<0, \lambda <-\sqrt{6}\} $, or $ \{w<1, \kappa>0, \lambda >-\sqrt{6}\}$, or $\{w<1, \kappa>0, \lambda <-\sqrt{6}\}$. The deceleration parameter is $q=2$. Therefore, it represents a stiff fluid solution. It never inflates.

\item 
 $E: \left(-\frac{\lambda }{\sqrt{6}},1,0\right)$, with  eigenvalues $\left\{\frac{\lambda ^2-3 w-3}{\sqrt{3}},\frac{\lambda ^2-6}{2 \sqrt{3}},\frac{\lambda  (\kappa +\lambda )-6}{\sqrt{3}}\right\}$ exists for $-\sqrt{6}\leq \lambda \leq \sqrt{6}$. It is nonhyperbolic for $\lambda ^2-3 w-3=0$, or $\lambda ^2=6$ or $\lambda  (\kappa +\lambda )=6$. It is a sink for \newline  $\left\{-\sqrt{6}<\lambda <0,  \frac{1}{3} \left(\lambda ^2-3\right)<w\leq 1,  \kappa >\frac{6-\lambda ^2}{\lambda }\right\}$, 
 or $ \{\lambda =0,  -1\leq w\leq 1\}$, or \newline $
   \left\{0<\lambda <\sqrt{6},  \frac{1}{3} \left(\lambda ^2-3\right)<w\leq 1,  \kappa <\frac{6-\lambda ^2}{\lambda }\right\}$. It is a saddle otherwise. The deceleration parameter is $q=\frac{1}{2} \left(\lambda ^2-2\right)$, therefore, the solution is inflationary for $\lambda^2<2$.

\item 
$F: \left(\frac{\lambda }{\sqrt{6}},-1,0\right)$, with  eigenvalues $\left\{\frac{-\lambda ^2+3(w+1)}{\sqrt{3}},-\frac{\lambda ^2-6}{2 \sqrt{3}},\frac{6-\lambda  (\kappa +\lambda )}{\sqrt{3}}\right\}$ exists for $-\sqrt{6}\leq \lambda \leq \sqrt{6}$. It is nonhyperbolic for $\lambda ^2-3 w-3=0$, or $\lambda ^2=6$ or $\lambda  (\kappa +\lambda )=6$. It is a source for $\left\{-\sqrt{6}<\lambda <0,  \frac{1}{3} \left(\lambda ^2-3\right)<w<1,  \kappa >\frac{6-\lambda ^2}{\lambda }\right\}$, or $ \{\lambda =0,  -1<w<1\}$ , or $
   \left\{0<\lambda <\sqrt{6},  \frac{1}{3} \left(\lambda ^2-3\right)<w<1,  \kappa <\frac{6-\lambda ^2}{\lambda }\right\}$. It is a saddle otherwise. The deceleration parameter is $q=\frac{1}{2} \left(\lambda ^2-2\right)$, therefore, the solution is inflationary for $\lambda^2<2$.

\item 
$G: \left(\frac{1}{\sqrt{\frac{1}{6} \kappa  (\kappa +\lambda )+1}},-\frac{\kappa +\lambda }{\sqrt{\kappa  (\kappa +\lambda )+6}},0\right)$, with  eigenvalues \newline $ \left\{\frac{\kappa  (1+w)+\lambda 
   (w-1)}{\sqrt{\frac{1}{3} \kappa  (\kappa +\lambda )+2}},\frac{\sqrt{3} \kappa -\sqrt{-\kappa\left(4 \left(\kappa ^2-6\right) \lambda +8 \kappa  \lambda ^2-27 \kappa +4 \lambda ^3\right)}}{2
   \sqrt{\kappa  (\kappa +\lambda )+6}},\frac{\sqrt{3} \kappa +\sqrt{-\kappa\left(4 \left(\kappa ^2-6\right) \lambda +8 \kappa  \lambda ^2-27 \kappa +4 \lambda ^3\right)}}{2 \sqrt{\kappa 
   (\kappa +\lambda )+6}}\right\}$. 
  Exists for $\kappa <0, -\frac{\sqrt{\kappa ^2+24}}{2}-\frac{\kappa }{2}<\lambda <-\kappa$, or $\kappa >0, -\kappa <\lambda <\frac{\sqrt{\kappa ^2+24}}{2}-\frac{\kappa }{2}$. It is a nonhyperbolic for $\kappa =0, \lambda =-\sqrt{6}$, or $\kappa =0,  \lambda =\sqrt{6}$, or $\kappa \neq 0, \lambda =-\kappa$, or $w-1\neq 0, \sqrt{2} \kappa  \sqrt{-\frac{\kappa ^2-3(w-1)}{w-1}}, 0,
   \lambda =\frac{\kappa  (-w)-\kappa }{w-1}$, or $\lambda \neq 0, w=1, \kappa =0$, or $\kappa =0$, or
   $\lambda =\frac{1}{2} \left(-\sqrt{\kappa ^2+24}-\kappa \right)$, or $\lambda =\frac{1}{2} \left(\sqrt{\kappa
   ^2+24}-\kappa \right)$,  
  or a saddle otherwise. The deceleration parameter is $q=2-\frac{3 \kappa }{\kappa +\lambda }$, therefore, the solution inflates for $\frac{\kappa}{\kappa+\lambda}>\frac{2}{3}$. 
   
\item 
$H: \left(-\frac{1}{\sqrt{\frac{1}{6} \kappa  (\kappa +\lambda )+1}},\frac{\kappa +\lambda }{\sqrt{\kappa  (\kappa +\lambda )+6}},0\right)$, with  eigenvalues \newline $\left\{-\frac{\kappa  (1+w)+\lambda 
   (w-1)}{\sqrt{\frac{1}{3} \kappa  (\kappa +\lambda )+2}},-\frac{\sqrt{3} \kappa -\sqrt{-\kappa\left(4 \left(\kappa ^2-6\right) \lambda +8 \kappa  \lambda ^2-27 \kappa +4 \lambda ^3\right)}}{2
   \sqrt{\kappa  (\kappa +\lambda )+6}}, -\frac{\sqrt{3} \kappa +\sqrt{-\kappa\left(4 \left(\kappa ^2-6\right) \lambda +8 \kappa  \lambda ^2-27 \kappa +4 \lambda ^3\right)}}{2 \sqrt{\kappa 
   (\kappa +\lambda )+6}}\right\}$. Exists for $\kappa <0, -\frac{\sqrt{\kappa ^2+24}}{2}-\frac{\kappa }{2}<\lambda <-\kappa$, or $\kappa >0, -\kappa <\lambda <\frac{\sqrt{\kappa ^2+24}}{2}-\frac{\kappa }{2}$. It is a nonhyperbolic for $\kappa =0, \lambda =-\sqrt{6}$, or $\kappa =0,  \lambda =\sqrt{6}$, or $\kappa \neq 0, \lambda =-\kappa$, or $w-1\neq 0, \sqrt{2} \kappa  \sqrt{-\frac{\kappa ^2-3(w-1)}{w-1}}, 0,
   \lambda =\frac{\kappa  (-w)-\kappa }{w-1}$, or $\lambda \neq 0, w=1, \kappa =0$, or $\kappa =0$, or
   $\lambda =\frac{1}{2} \left(-\sqrt{\kappa ^2+24}-\kappa \right)$, or $\lambda =\frac{1}{2} \left(\sqrt{\kappa
   ^2+24}-\kappa \right)$,  
  or a saddle otherwise. The deceleration parameter is $q=2-\frac{3 \kappa }{\kappa +\lambda }$, therefore, the solution inflates for $\frac{\kappa}{\kappa+\lambda}>\frac{2}{3}$. 

\item 
 $O^-: (-1,0,0)$, with eigenvalues $\left\{-\sqrt{2} \kappa ,-\frac{\kappa }{\sqrt{2}},-\sqrt{2} (\kappa +\lambda )\right\}$. It always exists. $O^{-}$ is a sink for $\{\kappa >0, \lambda >-\kappa\}$, a saddle for $\{\kappa <0, \lambda >-\kappa \}$, or $\{\kappa >0, \lambda <-\kappa \}$, a source for  $\kappa <0, \lambda <-\kappa$. The deceleration parameter is $q=2$. Therefore, it represents a stiff fluid solution. It never inflates.
 
\item 
$O^{+}: (1,0,0)$, with eigenvalues $\left\{\frac{\kappa }{\sqrt{2}},\sqrt{2} \kappa ,\sqrt{2} (\kappa +\lambda )\right\}$. It always exists. $O+$ is a sink for $\{\kappa <0, \lambda <-\kappa\}$, a saddle for $\{\kappa <0, \lambda >-\kappa \}$, or $ \{\kappa >0, \lambda <-\kappa\}$ , a source for $\kappa >0, \lambda >-\kappa$. The deceleration parameter is $q=2$. Therefore, it represents a stiff fluid solution. It never inflates.

The previous cases correspond to the matterless case. 

Now we consider matter-dominated and scaling solutions: 
 
 \item 
$I: (0,-1,1)$, with eigenvalues $\left\{-\sqrt{3} (w-1),-\frac{1}{2} \sqrt{3} (w-1),-\sqrt{3} (1+w)\right\}$. It always exists. It is nonhyperbolic for $w\in\{-1,1\}$. The point is a saddle for $-1<w<1$. The deceleration parameter is $q=\frac{1}{2} (3 w+1)$. It is an inflationary solution for $w<-\frac{1}{3}$. 

\item 
$J: (0,1,1)$, with eigenvalues $\left\{\frac{1}{2} \sqrt{3} (w-1),\sqrt{3} (w-1),\sqrt{3} (1+w)\right\}$. It always exists. It is nonhyperbolic for $w\in\{-1,1\}$. The point is a saddle for $-1<w<1$. The deceleration parameter is $q=\frac{1}{2} (3 w+1)$. It is an inflationary solution for $w<-\frac{1}{3}$.

\item 
$K: \left(\frac{\sqrt{\frac{3}{2}} (1+w)}{\lambda },-1,1-\frac{3 (1+w)}{\lambda ^2}\right)$, with  eigenvalues \newline 
\begin{footnotesize}
$\left\{-\frac{\sqrt{3} (\kappa  (1+w)+\lambda  (w-1))}{\lambda }, \frac{\sqrt{3} \left(\lambda (1-w)
   -\sqrt{(w-1)(\lambda ^2 (9 w+7)-24 (1+w)^2)}\right)}{4 \lambda }, \frac{\sqrt{3} \left(\lambda (1-w) +\sqrt{(w-1)(\lambda ^2 (9 w+7)-24 (1+w)^2)}\right)}{4 \lambda }\right\}$.
\end{footnotesize}
Exists for  $\left\{\lambda \leq -\sqrt{6}, -1\leq w\leq 1\right\} $, or $ \left\{-\sqrt{6}<\lambda <0, -1\leq w\leq \frac{1}{3}\left(\lambda ^2-3\right)\right\}$ \newline 
    $ \left\{0<\lambda <\sqrt{6}, -1\leq w\leq \frac{1}{3} \left(\lambda ^2-3\right)\right\} $, or $ \left\{\lambda \geq \sqrt{6}, -1\leq w\leq 1\right\}$. Nonhyperbolic for \newline $\left\{\lambda \neq 0, w=-1\right\}$, or $  \left\{-1<w<1,  \lambda =\sqrt{3(w+1)}\right\}$, or $  \left\{-1<w<1,  \lambda =-\sqrt{3(w+1)}\right\}$, or $  \left\{-1<w<1,  \lambda <-\sqrt{3(w+1)},  \kappa =\frac{\lambda(1-w)}{w+1}\right\}$, or $ 
   \left\{-1<w<1,  \lambda >\sqrt{3(w+1)},  \kappa =\frac{\lambda(1-w)}{w+1}\right\} $, or $  \left\{w=1, 
   \lambda \geq \sqrt{6}\right\}$, or $  \left\{w=1,  \lambda \leq -\sqrt{6}\right\}$.  
   
   It is a source for $\left\{-1<w<1, \lambda <-\sqrt{3 w+3}, \kappa
   >\frac{\lambda(1-w)}{w+1}\right\}$,  or \newline
   $\left\{-1<w<1, \lambda >\sqrt{3 w+3}, \kappa
   <\frac{\lambda(1-w)}{w+1}\right\}$. It is a saddle otherwise. The deceleration parameter is $q=\frac{1}{2} (3 w+1)$. It is an inflationary solution for $w<-\frac{1}{3}$. 
    
\item 
$L:  \left(-\frac{\sqrt{\frac{3}{2}} (1+w)}{\lambda },1,1-\frac{3 (1+w)}{\lambda ^2}\right)$, with  eigenvalues \newline 
\begin{footnotesize}
$\left\{\frac{\sqrt{3} (\kappa  (1+w)+\lambda  (w-1))}{\lambda }, -\frac{\sqrt{3} \left(\lambda (1-w)
   -\sqrt{(w-1)(\lambda ^2 (9 w+7)-24 (1+w)^2)}\right)}{4 \lambda }, -\frac{\sqrt{3} \left(\lambda (1-w) +\sqrt{(w-1)(\lambda ^2 (9 w+7)-24 (1+w)^2)}\right)}{4 \lambda }\right\} $.
   \end{footnotesize}
Exists for $\left\{\lambda \leq -\sqrt{6}, -1\leq w\leq 1\right\} $, or $ \left\{-\sqrt{6}<\lambda <0, -1\leq w\leq \frac{1}{3}\left(\lambda ^2-3\right)\right\}$ \newline 
    $ \left\{0<\lambda <\sqrt{6}, -1\leq w\leq \frac{1}{3} \left(\lambda ^2-3\right)\right\} $, or $ \left\{\lambda \geq \sqrt{6}, -1\leq w\leq 1\right\}$.  Nonhyperbolic for \newline $\left\{\lambda \neq 0, w=-1\right\}$, or $  \left\{-1<w<1,  \lambda =\sqrt{3(w+1)}\right\}$, or $  \left\{-1<w<1,  \lambda =-\sqrt{3(w+1)}\right\}$, or $  \left\{-1<w<1,  \lambda <-\sqrt{3(w+1)},  \kappa =\frac{\lambda(1-w)}{w+1}\right\}$, or $ 
   \left\{-1<w<1,  \lambda >\sqrt{3(w+1)},  \kappa =\frac{\lambda(1-w)}{w+1}\right\} $, or $  \left\{w=1, 
   \lambda \geq \sqrt{6}\right\}$, or $  \left\{w=1,  \lambda \leq -\sqrt{6}\right\}$.   
   
   It is a sink for  $\left\{-1<w<1, \lambda <-\sqrt{3 w+3}, \kappa
   >\frac{\lambda(1-w)}{w+1}\right\}$,  or \newline
   $\left\{-1<w<1, \lambda >\sqrt{3 w+3}, \kappa
   <\frac{\lambda(1-w)}{w+1}\right\}$. It is a saddle otherwise. The deceleration parameter is $q=\frac{1}{2} (3 w+1)$. It is an inflationary solution for $w<-\frac{1}{3}$.

\item $M, N: \left(\pm\frac{w-1}{\sqrt{\frac{2 \kappa ^2}{3}+(w-1)^2}}, \pm \frac{\kappa }{\sqrt{\kappa ^2+\frac{3}{2} (w-1)^2}},\frac{2 \kappa ^2}{2 \kappa ^2+3 (w-1)^2}\right)$, exist for $\{\kappa <0, -1\leq w\leq 1\}$ or $\{\kappa =0, -1\leq w<1\}$ or $\{\kappa >0, -1\leq w\leq 1\}$. The  eigenvalues are \newline $\left\{\mp \frac{\mu_{11}}{2 \left(2 \kappa ^2+3
   (w-1)^2\right)^{7/2}}, \mp \frac{\mu_{12}}{2 \left(2 \kappa ^2+3 (w-1)^2\right)^{7/2}}, \mp \frac{\mu_{13}}{2 \left(2 \kappa ^2+3 (w-1)^2\right)^{7/2}}\right\}$. We denote by $\mu_{1 j}, j\in\{1,2,3\}$  the $j$-th root of the  polynomial \newline 
$P_1(\mu):= \mu ^3+\sqrt{6}
   \mu ^2 \left(2 \kappa ^2+3 (w-1)^2\right)^3 (\kappa +3 \kappa  w+2 \lambda  (w-1))+12 \kappa  \mu  (w-1)
   \left(2 \kappa ^2+3 (w-1)^2\right)^6 (2 \kappa +\lambda  (w-1))-24 \sqrt{6} \kappa ^2 (w-1)^2 \left(2
   \kappa ^2+3 (w-1)^2\right)^9 (\kappa  (w+1)+\lambda  (w-1))$.    
 The deceleration parameter is $q=\frac{1}{2} (3 w+1)$. It is an inflationary solution for $w<-\frac{1}{3}$.   
Finally, the stability of the points $M$ and $N$ (which have opposite dynamical behaviours) is examined numerically. For simplicity, we consider the case of dust matter, $w=0$. Fig.  \ref{fig:MN} presents the real parts of the complicated eigenvalues related to $N$ compared with zero for dust. This diagram shows its saddle nature.

\begin{figure}
    \centering
    \includegraphics[scale=0.6]{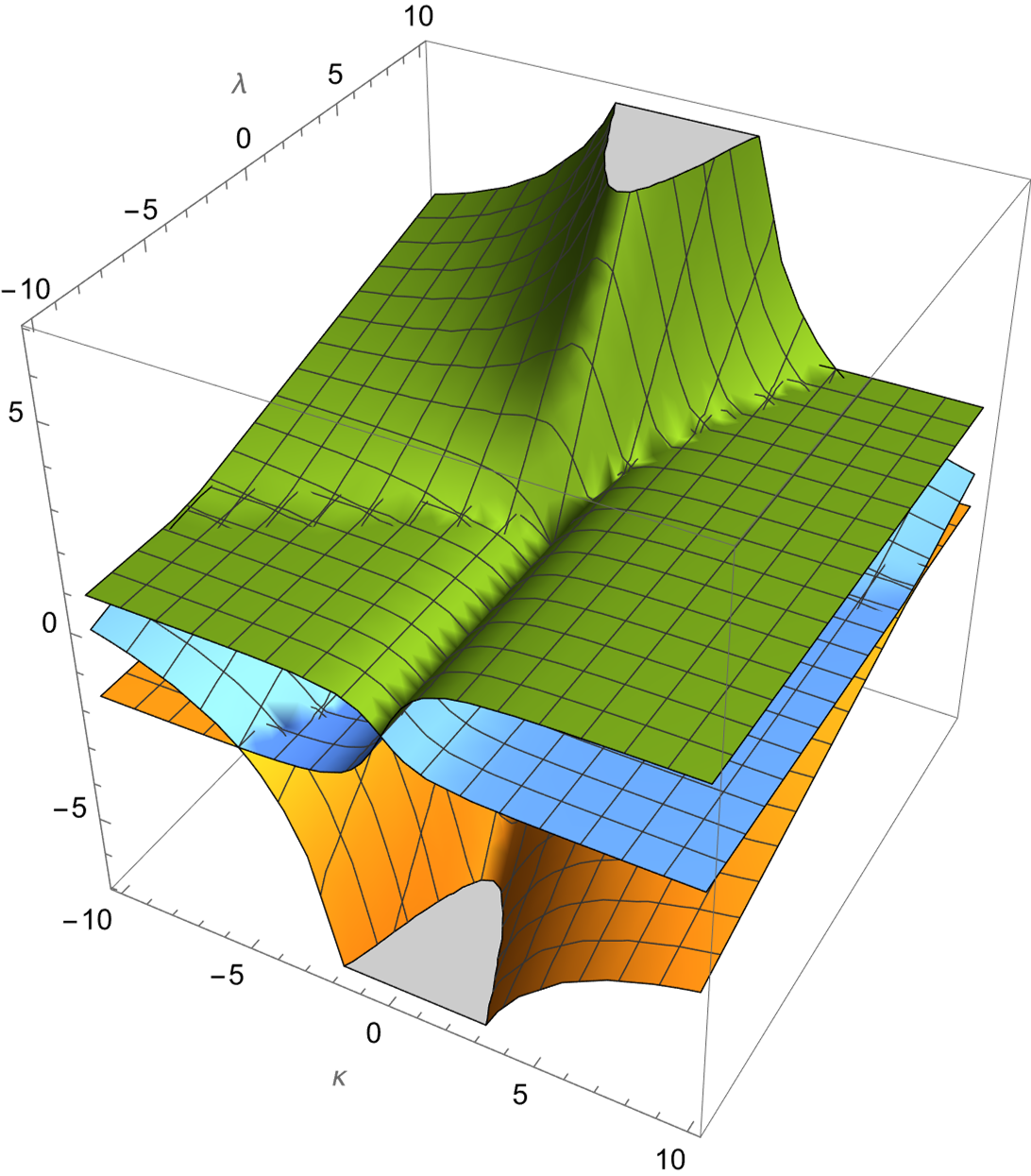}
    \caption{Real parts of the eigenvalues $\left\{\frac{\mu_{11}}{2 \left(2 \kappa ^2+3
   (w-1)^2\right)^{7/2}}, \frac{\mu_{12}}{2 \left(2 \kappa ^2+3 (w-1)^2\right)^{7/2}},  \frac{ \mu_{13}}{2 \left(2 \kappa ^2+3 (w-1)^2\right)^{7/2}}\right\}$, where $\mu_{1j}, j=1,2,3$ are the
roots of the  polynomial 
$P_1(\mu):= \mu ^3+\sqrt{6}
   \mu ^2 \left(2 \kappa ^2+3 (w-1)^2\right)^3 (\kappa +3 \kappa  w+2 \lambda  (w-1))+12 \kappa  \mu  (w-1)
   \left(2 \kappa ^2+3 (w-1)^2\right)^6 (2 \kappa +\lambda  (w-1))-24 \sqrt{6} \kappa ^2 (w-1)^2 \left(2
   \kappa ^2+3 (w-1)^2\right)^9 (\kappa  (w+1)+\lambda  (w-1))$ related to $N$, compared with zero for dust. This diagram shows its saddle nature.}
    \label{fig:MN}
\end{figure}

\end{itemize}
\subsubsection{Flat ($K=0$) vacuum model ($\rho_m=0$)}
\label{Subsect:II1}
When we consider the vacuum solution $\rho_m=0$, $\Omega=0$, the variables \eqref{Zerovars} satisfy
\begin{equation}
    h^2 + \eta^2= \Phi^2+ \Psi=1. \label{flateq102}
\end{equation}

Using  \eqref{flateq102}, we have bounded variables
\begin{equation}
    0 \leq h^2, \Phi^2, \eta^2,\Psi \leq 1.  
\end{equation}

Again, introducing the time derivative $d\tau=\chi d t$, $f^{\prime}=df/d\tau$,  the field equations becomes
\begin{align}
    & \Phi^{\prime}= -\frac{\eta ^2 \kappa }{\sqrt{2}}-\sqrt{3} h \Phi -\frac{\lambda  \Psi }{\sqrt{2}}-\frac{\Phi  \dot{\chi}}{\chi^2},\\
    &  h^{\prime}= \frac{1}{2} \sqrt{3} \left(\eta ^2-h^2-\Phi ^2+\Psi \right)-\frac{h \dot{\chi}}{\chi^2}, \\
    & \eta^{\prime}= -\frac{\kappa\,\eta\Phi}{\sqrt{2}}-\sqrt{3} \eta  h-\frac{\eta  \dot{\chi}}{\chi^2},\\
    & \Psi^{\prime}= \sqrt{2} \lambda  \Phi  \Psi -\frac{2\Psi  \dot{\chi}}{\chi^2}. 
\end{align}
Next, by substituting
\begin{align}
    & \dot{\chi}= -\frac{\Phi  \chi ^2 \left(\left(1-h^2\right) \kappa +\sqrt{6} h \Phi \right)}{\sqrt{2}}, \\
    & \eta^2=1-h^2,\\
    & \Psi= 1- \Phi^2,
\end{align}
we obtain
\begin{align}
  & \Phi^{\prime} =-  \frac{\sqrt{2}\left(1-\Phi ^2\right)}{2} \left(\lambda +\sqrt{6} h \Phi + \kappa \left(1-h^2\right)  \right), \label{systA}\\
 & h^{\prime}= \frac{1}{2} \left(1-h^2\right) \left(\sqrt{2} \kappa h \Phi +2 \sqrt{3} \left(1-\Phi
   ^2\right)\right), \label{systB}
\end{align}
with the auxiliary equations
\begin{align}
      & \eta^{\prime}= \frac{1}{2} h \sqrt{1-h^2} \left(2 \sqrt{3} \left(\Phi ^2-1\right)-\sqrt{2} \kappa h \Phi
   \right),\\
   & \Psi^{\prime}= -\sqrt{2} \Phi  \left(\Phi ^2-1\right) \left(\kappa \left(1-h^2\right) +\sqrt{6} h \Phi +\lambda
   \right).
\end{align}

Hence, we investigate the reduced dynamical system for the vector $(\Phi, h)$ given by \eqref{systA}-\eqref{systB} defined in the phase-plane $(\Phi, h) \in [-1,1]\times [-1,1]$. 

The main focus of the analysis is to investigate whether different orbits inflate. The condition for a point in the phase space to be inflationary, using the deceleration parameter, $q$, is defined by \eqref{q.def}, which, in terms of the dimensionless variables, becomes
\begin{align}
	&q = \frac{3}{h^2}\left(h^2+\Phi^2-1\right)-1 < 0.\label{q.exp}
\end{align}

The equilibrium points are $(\Phi,h)$: 

\begin{itemize}
    \item $A: (1,1 )$, which exists for all parameter values. The eigenvalues of the linearization are $\left\{-\sqrt{2} \kappa ,\sqrt{2} \lambda +2 \sqrt{3}\right\}$. Therefore, $A$ is a sink for $\{\kappa>0, \lambda <-\sqrt{6}\}$, a saddle for $\{\kappa>0, \lambda >-\sqrt{6}\} $, or $ \{\kappa<0, \lambda <-\sqrt{6}\}$ or a source for  $\{\kappa<0, \lambda >-\sqrt{6}\}$. It is a solution dominated by the kinetic term of the quintessence field. The deceleration parameter is $q=2$. Therefore, it never inflates.

    \item $B: (-1,1 )$, which exists for all parameter values. The eigenvalues of the linearization are $\left\{\sqrt{2} \kappa ,2 \sqrt{3}-\sqrt{2} \lambda \right\}$. Therefore, $B$ is a sink for $\{\kappa<0, \lambda >\sqrt{6}\}$, a saddle for $\{\kappa<0, \lambda <\sqrt{6}\} $, or $ \{\kappa>0, \lambda >\sqrt{6}\}$ or a source for  $\{\kappa>0, \lambda <\sqrt{6}\}$. It is a solution dominated by the kinetic term of quintessence field. The deceleration parameter is $q=2$. Therefore, it never inflates.

    \item $C: (1,-1 )$, which exists for all parameter values. The eigenvalues of the linearization are $\left\{-\sqrt{2} \kappa ,\sqrt{2} \lambda -2 \sqrt{3}\right\}$. Therefore, $C$ is a sink for $\{\kappa>0, \lambda <\sqrt{6}\}$, a saddle for $\{\kappa>0, \lambda >\sqrt{6}\} $, or $ \{\kappa<0, \lambda <\sqrt{6}\}$ or a source for  $\{\kappa<0, \lambda >\sqrt{6}\}$. It is a solution dominated by the kinetic term of quintessence field. The deceleration parameter is $q=2$. Therefore, it never inflates.
    
    \item $D: (-1,-1 )$, which exists for all parameter values. The eigenvalues of the linearization are $\left\{\sqrt{2} \kappa ,-\sqrt{2} \lambda -2 \sqrt{3}\right\}$. Therefore, $D$ is  a sink for $\{\kappa<0, \lambda >-\sqrt{6}\}$, a saddle for $\{\kappa<0, \lambda <-\sqrt{6}\} $, or $ \{\kappa>0, \lambda >-\sqrt{6}\}$ or a source for  $\{\kappa>0, \lambda <-\sqrt{6}\}$. It is a solution dominated by the kinetic term of quintessence field. The deceleration parameter is $q=2$. Therefore, it never inflates.
    
    \item $E: \left(-\frac{\lambda}{\sqrt{6}}, 1 \right)$. Exists  for $-\sqrt{6}\leq \lambda\leq \sqrt{6}$. The  eigenvalues of the linearization are $\left\{\frac{\lambda ^2-6}{2 \sqrt{3}},\frac{\lambda (\kappa + \lambda) -6}{\sqrt{3}}\right\}$.  Therefore, $E$ is a sink for $\kappa \in \mathbb{R}$ and $\left\{-\sqrt{6}<\lambda <0, \kappa >\frac{6-\lambda^2}{\lambda }\right\}$, or $ \{\lambda =0\} $, or $ \left\{0<\lambda <\sqrt{6}, \kappa   <\frac{6-\lambda ^2}{\lambda }\right\}$, or a saddle for $\left\{-\sqrt{6}<\lambda <0, \kappa <\frac{6-\lambda ^2}{\lambda }\right\}$, \newline or $ \left\{0<\lambda <\sqrt{6}, \kappa >\frac{6-\lambda ^2}{\lambda }\right\}$. It is a quintessence dominated solution. It is an inflationary solution for $-\sqrt{2}<\lambda <\sqrt{2}$. 
   
    \item $F: \left(\frac{\lambda}{\sqrt{6}},  -1 \right)$. Exists  for $-\sqrt{6}\leq \lambda\leq\sqrt{6}$. The  eigenvalues of the linearization are $\left\{-\frac{\lambda ^2-6}{2 \sqrt{3}},\frac{6-\lambda  (\kappa +\lambda)}{\sqrt{3}}\right\}$.  Therefore, $F$  a saddle for $\left\{-\sqrt{6}<\lambda <0,  \kappa <\frac{6-\lambda ^2}{\lambda }\right\}$, or $ \left\{0<\lambda <\sqrt{6},  \kappa >\frac{6-\lambda ^2}{\lambda }\right\}$ or a source for $\kappa \in \mathbb{R}$ and $\left\{-\sqrt{6}<\lambda <0, \kappa >\frac{6-\lambda^2}{\lambda }\right\}$, or $ \{\lambda =0\} $, or $ \left\{0<\lambda <\sqrt{6}, \kappa <\frac{6-\lambda ^2}{\lambda }\right\}$. It is a quintessence dominated solution. It is an inflationary solution for $-\sqrt{2}<\lambda <\sqrt{2}$. 
   
    \item $G: \left(\frac{\sqrt{6}}{\sqrt{\kappa^2 + \kappa \lambda +6}}, -\frac{\kappa + \lambda}{\sqrt{\kappa^2 + \kappa \lambda +6}} \right)$. Exists  for $\left\{\kappa  (\kappa +\lambda )>0,\kappa  \lambda +\lambda ^2-6<0\right\}$, which leads to $\left\{\kappa <0, -\frac{\sqrt{\kappa ^2+24}}{2}-\frac{\kappa }{2}<\lambda <-\kappa \right\}$, or $ \left\{\kappa >0, -\kappa <\lambda <\frac{\sqrt{\kappa^2 +24}}{2}-\frac{\kappa }{2}\right\}$. The eigenvalues of the linearization are \newline $\left\{\frac{\sqrt{3} \kappa -\sqrt{-\kappa\left(4 \left(\kappa ^2-6\right) \lambda +8\kappa\lambda ^2-27 \kappa +4 \lambda ^3\right)}}{2 \sqrt{\kappa(\kappa +\lambda)+6}},\frac{\sqrt{3} \kappa +  \sqrt{-\kappa\left(4 \left(\kappa ^2-6\right) \lambda +8\kappa\lambda ^2-27 \kappa +4 \lambda ^3\right)}}{2 \sqrt{\kappa  (\kappa +\lambda)+6}}\right\}$. It is a saddle for $\left\{\kappa >0, -\kappa <\lambda <\frac{\sqrt{ \kappa^2+24}}{2}-\frac{\kappa}{2}\right\}$, or $ \left\{\kappa <0, -\frac{\sqrt{\kappa ^2+24}}{2}-\frac{\kappa}{2}<\lambda <-\kappa \right\}$. It is an inflationary solution for $\left\{\kappa  (\kappa +\lambda )>0,\kappa  \lambda +\lambda^2 -6<0, \frac{2}{3}\leq \frac{\kappa}{\kappa +\lambda} \right\}$, that is, for \newline $\left\{\kappa \leq -2 \sqrt{2},  -\frac{\sqrt{\kappa ^2+24}}{2}-\frac{\kappa }{2}<\lambda<-\kappa \right\}$, or $ \left\{-2 \sqrt{2}<\kappa <0,  \frac{\kappa }{2}<\lambda <-\kappa\right\}$, \newline or $ \left\{0<\kappa \leq 2 \sqrt{2},  -\kappa <\lambda <\frac{\kappa}{2}\right\}$,  or $ \left\{\kappa >2 \sqrt{2},  -\kappa <\lambda <\frac{\sqrt{\kappa^2+24}}{2}-\frac{\kappa }{2}\right\}$. 
   
    \item $H: \left(-\frac{\sqrt{6}}{\sqrt{\kappa^2 + \kappa \lambda +6}}, \frac{\kappa + \lambda}{\sqrt{\kappa^2 + \kappa \lambda +6}} \right)$. Exists  for $\left\{\kappa <0, -\frac{\sqrt{\kappa ^2+24}}{2}-\frac{\kappa }{2}<\lambda <-\kappa\right\}$, \newline or $ \left\{\kappa >0, -\kappa <\lambda <\frac{\sqrt{\kappa^2+24}}{2}-\frac{\kappa }{2}\right\}$. The eigenvalues of the linearization are \newline  $\left\{\frac{-\sqrt{3} \kappa -\sqrt{-\kappa\left(4 \left(\kappa ^2-6\right) \lambda +8\kappa\lambda ^2-27 \kappa +4 \lambda ^3\right)}}{2 \sqrt{\kappa  (\kappa +\lambda)+6}},\frac{-\sqrt{3} \kappa +\sqrt{-\kappa\left(4 \left(\kappa ^2-6\right) \lambda +8\kappa\lambda ^2-27 \kappa +4 \lambda ^3\right)}}{2 \sqrt{\kappa  (\kappa +\lambda)+6}}\right\}$. It is a saddle for $\left\{\kappa >0, -\kappa <\lambda <\frac{\sqrt{\kappa ^2+24}}{2}-\frac{\kappa}{2}\right\}$, or $ \left\{\kappa <0, -\frac{\sqrt{\kappa^2+24}}{2} -\frac{\kappa}{2}<\lambda<-\kappa \right\}$. It is an inflationary solution for the same cases as $G$. 
   
    \item $O^{\pm}:(\pm 1, 0)$, with eigenvalues $\left\{\pm\frac{\kappa }{\sqrt{2}}, \pm\sqrt{2} (\kappa +\lambda )\right\}$. $O+$ is a sink for $\kappa <0, \lambda <-\kappa$, a saddle for $\{\kappa <0, \lambda >-\kappa \}$, or $ \{\kappa >0, \lambda <-\kappa\}$ , a source for $\kappa >0, \lambda >-\kappa$. $O^{-}$ is a sink for $\kappa >0, \lambda >-\kappa$, a saddle for $\{\kappa <0, \lambda >-\kappa \}$, or $\{\kappa >0, \lambda <-\kappa \}$, a source for  $\kappa <0, \lambda <-\kappa$. The deceleration parameter is $q=2$, representing stiff fluid solutions that never inflate.

\end{itemize}

\subsubsection{Discussion}
\begin{figure}[h!]
    \centering
    \includegraphics[scale=0.6]{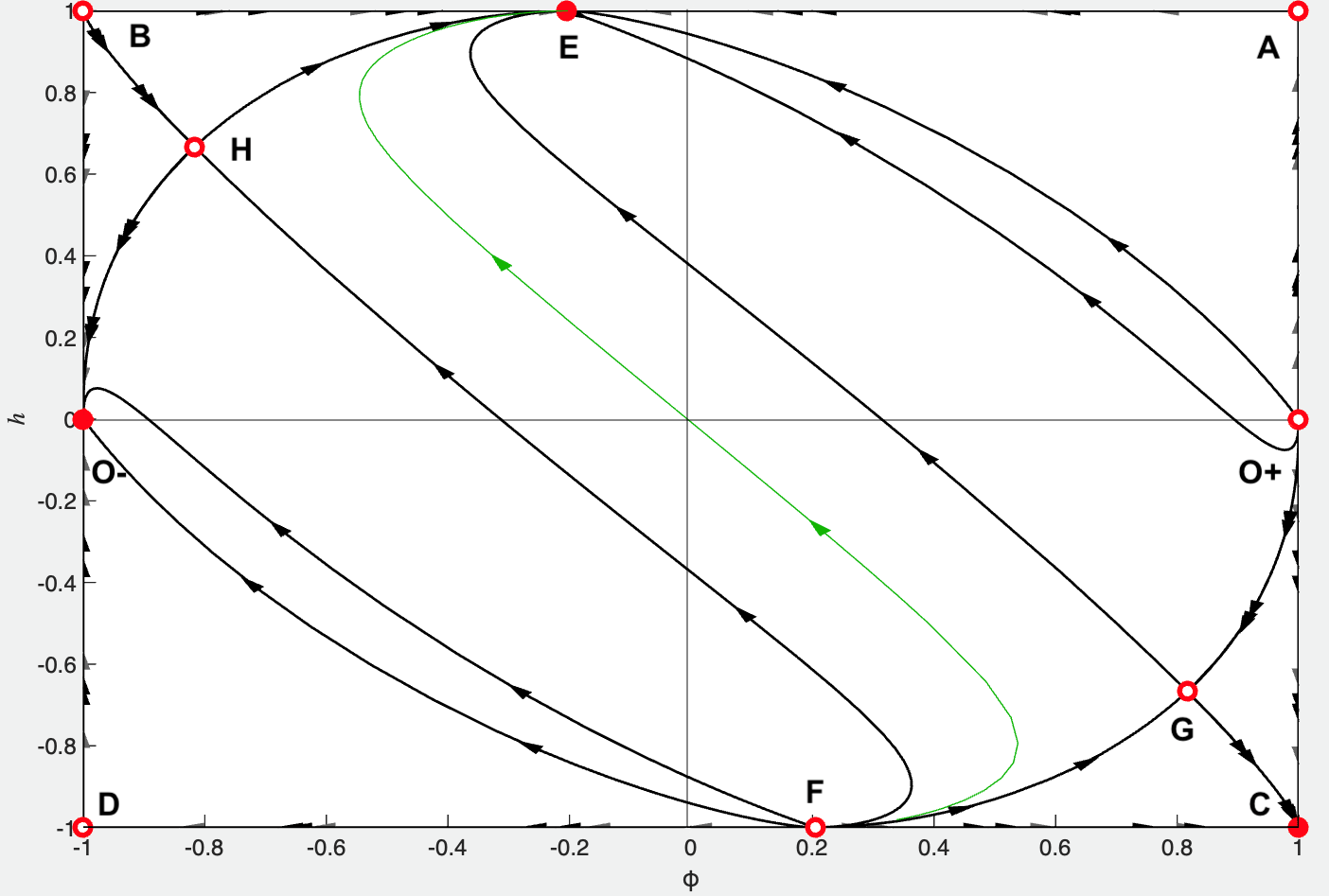}
    \caption{This figure shows a skeleton of the orbits of the dynamical system \eqref{systA}-\eqref{systB} within the phase-space $[-1,1]\times [-1,1]$, for parameter values $\lambda=0.5,\kappa=1.5$.  The stable and unstable manifolds of the saddles $G$ and $H$ are shown, as well as orbits that are tangent to the eigenvectors of nodes (sinks $E, O^+, C$ and sources $F, O^-, B$). Shown in green is the solution which passes through the origin. This solution is completely time-symmetric. All orbits in the region bounded by the manifolds from $F\rightarrow H\rightarrow E$ and $F\rightarrow G\rightarrow E$ represent bouncing cosmologies, the bounce occurring at the $h$-axis ($H=0$).}
    \label{skeleton}
\end{figure}

The main focus of the analysis is to investigate whether different orbits inflate. The condition for a point in the phase space to be inflationary, using the deceleration parameter,  is \eqref{q.exp}. The results are presented in Tab.  \ref{ep_table}. We see a parameter space region allowing for two periods of inflation. The expanding region of the phase diagram contains the saddle $H$ in the interior region and a sink $E$ on the boundary. Both points are inflationary (as are $G$ and $F$). At $H$, both fields contribute to inflation, while $E$ corresponds to standard single-field inflation. We can see that there exist orbits that approach a saddle and inflate for an arbitrarily long time, then approach the sink and inflate asymptotically again. This is shown in more detail in Fig. \ref{skeleton}.

\begin{table}[!ht]
	\centering
	\begin{tabular}{|c|c|c|c|c|c|}
		\hline 
		Label  & Existence & Sink & Saddle & Source & Inflation  \\\hline
		A & $\forall \lambda, k$  & $\kappa>0, \lambda<-\sqrt{6}$ & $\begin{array}{c}
			\kappa>0, \lambda>-\sqrt{6},\\
			\text{or} \\ \kappa<0, \lambda<-\sqrt{6}
		\end{array}$ & $\kappa<0, \lambda>-\sqrt{6}$ & N/A \\\hline 
		B & $\forall \lambda, k$  & $\kappa<0, \lambda>\sqrt{6}$ & $\begin{array}{c}
			\kappa<0, \lambda<\sqrt{6}, \\
			\text{or} \\  \kappa>0, \lambda>\sqrt{6}
		\end{array}$ & $\kappa>0, \lambda<\sqrt{6}$ & N/A \\\hline 
		C & $\forall \lambda, k$  & $\kappa>0, \lambda<\sqrt{6}$ & Same as $B$ & $\kappa<0, \lambda>\sqrt{6}$ & N/A \\ \hline 
		D & $\forall \lambda, k$  & $\kappa<0, \lambda>-\sqrt{6}$ & Same as $A$ & $\kappa>0, \lambda<-\sqrt{6}$ & N/A \\ \hline 
		$O^+$ & $\forall \lambda, k$  & $\kappa<0, \lambda<-\kappa$ & $\begin{array}{c}
			\kappa>0, \lambda<-\kappa\\ 
			\text{or} \\ \kappa<0, \lambda>-\kappa
		\end{array}$ & $\kappa>0, \lambda>-\kappa$ &  N/A \\\hline 
		$O^-$ & $\forall \lambda, k$  & $\kappa>0, \lambda>-\kappa$ & Same as $O^+$ & $\kappa<0, \lambda<-\kappa$ &  N/A \\\hline 
		$E$ & $\abs{\lambda}<\sqrt{6}$ & $\begin{array}{c}
			\abs{\lambda}<\sqrt{6},  \\
			\kappa \lambda +\lambda^2 -6 <0 
		\end{array}$ & $\begin{array}{c}
			\abs{\lambda}<\sqrt{6},  \\
			\kappa \lambda +\lambda^2 -6 >0 
		\end{array}$ & DNE & $\abs{\lambda}<\sqrt{2}$\\\hline
		$F$ & Same as $E$ & DNE & Same as $E$ &  $\begin{array}{c}
			\abs{\lambda}<\sqrt{6},  \\
			\kappa \lambda +\lambda^2 -6 <0 
		\end{array}$ & Same as $E$ \\\hline 
		$G$, $H$ & $\begin{array}{c}
			0<\kappa(\kappa+ \lambda) \\
			\kappa \lambda +\lambda^2 -6 <0
		\end{array}$ & DNE & $\begin{array}{c}
			\kappa>0,  \\
			\lambda( 6-\lambda^2)\qquad\qquad\quad \\
			\qquad>\kappa (\kappa \lambda +2 \lambda^2-6) \\
			\text{or}\\
			\kappa<0,  \\
			\lambda( 6-\lambda^2)\qquad\qquad\quad \\
			\qquad<\kappa (\kappa \lambda +2 \lambda^2-6)  
		\end{array}$ & DNE & $\frac{2}{3}<\frac{\kappa}{\kappa + \lambda}$ \\\hline 
	\end{tabular}
	
	\caption{Existence and stability of the equilibrium points of the reduced  dynamical system for the vector $(\Phi, h)$ given by \eqref{systA}-\eqref{systB}.}
	\label{ep_table}
\end{table}

This model exhibits bouncing cosmologies as reported in \cite{Tot:2022dpr}. In the dynamical system \eqref{systA}-\eqref{systB}, these are seen as the solutions bounded by the manifolds $F\rightarrow H\rightarrow E$ and $F\rightarrow G\rightarrow E$, as shown in Fig. \ref{skeleton}. However, this is not the only possible behaviour of orbits asymptotic to $F$ at early times. For example, in one region of parameter space, the stable manifold of $H$ connects to $B$ and $O^+$. In comparison, the unstable manifold of $G$ connects to $C$ and $O^-$. In this case, all orbits from $F$ are in the basin of attraction of $O^-$. All orbits in the basin of attraction of $E$ have $O^+$ as their early-time attractor. The solution that passes through the origin goes from $O^+$ at early times to $O^-$ at late times.

\begin{figure}[h!]
    \centering
    \includegraphics[scale=0.7]{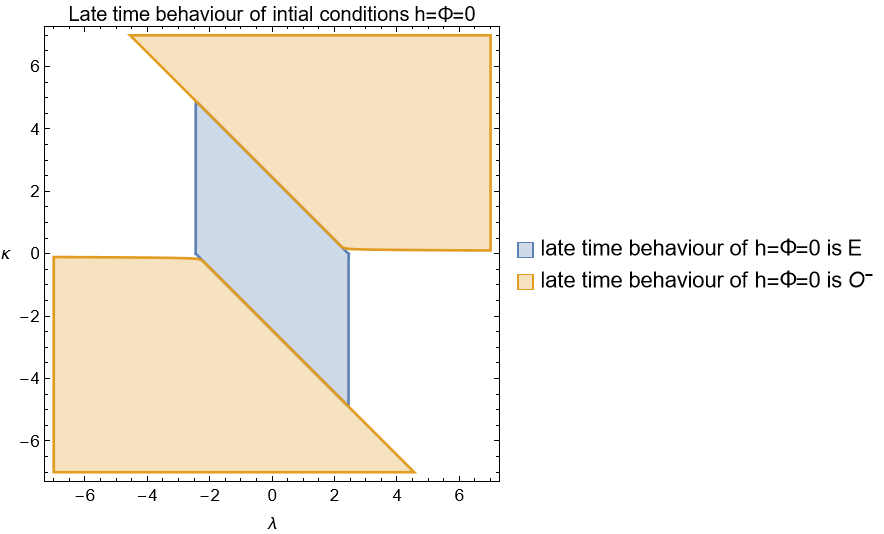}
    \caption{Regions of the parameters space $(\lambda,\kappa)$ for which the late time behaviour of initial conditions $\Phi=h=0$ is either $E$ or $O^-$.}
    \label{regions}
\end{figure}

In the bouncing case, when the origin is in the basis of attraction of $E$, then $h\rightarrow 1^-$ as $t\rightarrow\infty$. In the latter case, we have $h\rightarrow0^+$ as $t\rightarrow\infty$. Thus by integrating from the origin across parameter space and taking the value of $h$ at a sufficiently late time, we can determine which case the given parameter values are in Fig. \ref{regions} shows these regions of parameters space.

\begin{figure}[h!]
    \centering
    \includegraphics[scale=0.4]{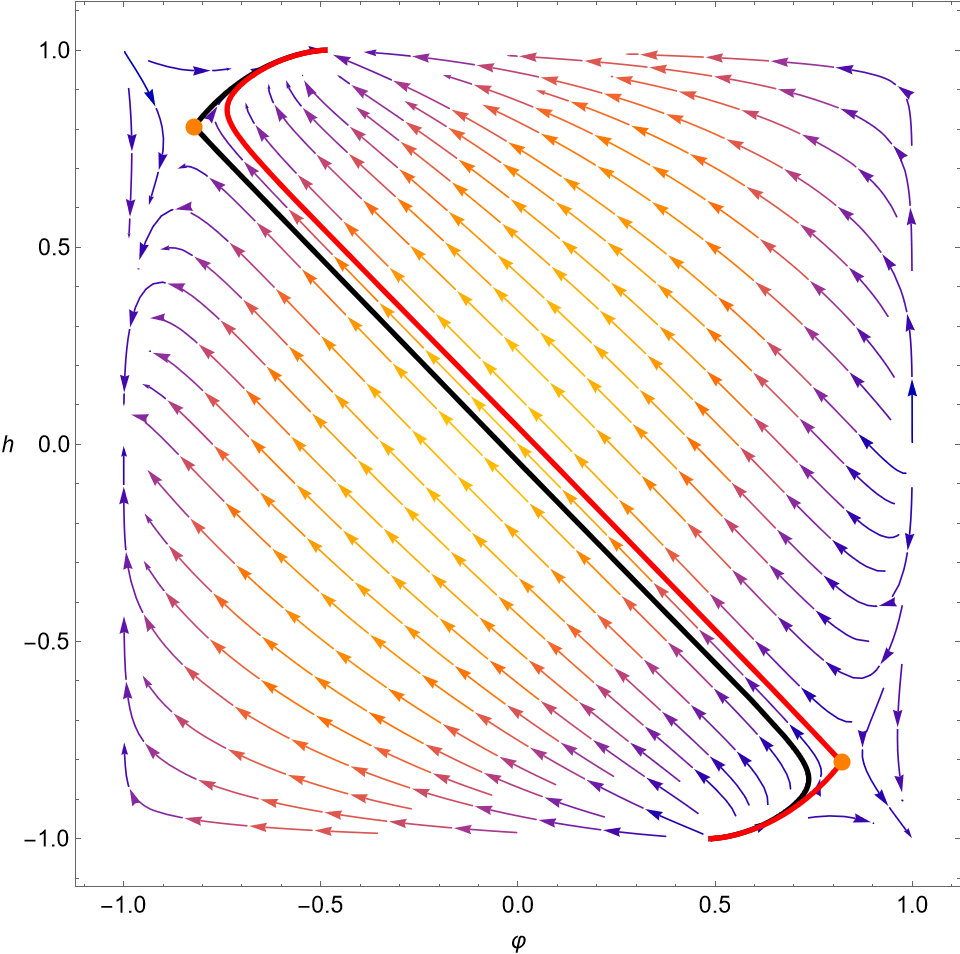}
    \caption{A phase portrait of \eqref{systA}-\eqref{systB} for $\lambda=\kappa=1.2$, showing the manifold $F\rightarrow H\rightarrow E$ in black and $F\rightarrow H\rightarrow E$ in red.}
    \label{near_bif}
\end{figure}

Helpfully, this bifurcation can be found explicitly. Near the boundary in Fig. \ref{regions} (within the blue region), the manifolds mentioned above come very near to each other, and their central portions approach the straight line connecting $G$ and $H$, as shown in Fig. \ref{near_bif}. The bifurcation is a saddle-connection bifurcation, as the unstable manifold of $G$ overlaps with the stable manifold of $H$. From \eqref{systA}-\eqref{systB}, it can be seen that the solution from the origin is tangent to the line $\Phi+\frac{\kappa+\lambda}{\sqrt{6}}h=0$. Thus we seek parameter values for which this line is an invariant set.

It can readily be seen that 
\begin{align}
    &\left(\Phi+\frac{\kappa+\lambda}{\sqrt{6}}h\right)^\prime= \nonumber \\
    \left(\Phi+\frac{\kappa+\lambda}{\sqrt{6}}h\right)&\cdot\sqrt{3}h\left[\Phi^2- \kappa\,h\Phi/\sqrt{6} +(\kappa^2+\kappa\lambda-6)/6\right]+\sqrt{2}h\cdot\kappa\left(6-(\kappa+\lambda)^2\right)/12 \label{flat_poly_div}
\end{align}

For $\kappa=0$ or $\kappa+\lambda=\pm\sqrt{6}$, we have $G^\prime=0$ where $G=\Phi+\frac{\kappa+\lambda}{\sqrt{6}}h$ vanishes, so $G=0$ is an invariant set.  The conditions $\kappa+\lambda=\pm\sqrt{6}$ make $G=\Phi\pm h$.

\subsection{Model with positive curvature ($K=+1$)}
\label{Subsect:IIB}
Defining 
\begin{equation}
    \chi^2 = \frac{1}{2} \dot{\phi}^2 + V(\phi) + \rho_m, \label{Positivechi_lambda}
\end{equation}
and
\begin{align}
	&h^2= \frac{3 H^2}{\chi^2}, \; \eta^2= \frac{e^{\kappa \phi}\dot{\psi}^2}{2 \chi^2}, \;  \Phi^2 = \frac{\dot{\phi}^2}{2 \chi^2}, \; \Psi= \frac{V(\phi)}{\chi^2}, \; \Omega= \frac{\rho_m}{\chi^2},  \; \Omega_K= \frac{3}{\chi^2 a^2}, \label{Positivevars}
\end{align}
which satisfy
\begin{equation}
	h^2 + \eta^2+ \Omega_K= \Phi^2+ \Psi + \Omega=1, \label{Positiveeq102}
\end{equation}
using \eqref{Positiveeq102}, we have bounded variables
\begin{equation}
	0 \leq h^2, \Phi^2, \eta^2,\Psi,\Omega, \Omega_K \leq 1.  
\end{equation}
Introducing the time derivative $d\tau=\chi d t$, and taking the derivatives of the variables for the new time variable  (i.e., $f^{\prime}=df/d\tau$), the field equations \eqref{0sp.04}, \eqref{0sp.05},\eqref{0sp.06}, \eqref{0sp.07} and \eqref{matter} become 
\begin{align}
  \Phi^{\prime} & =   -\frac{\eta ^2 \kappa }{\sqrt{2}}-\sqrt{3} h \Phi -\frac{\lambda  \Psi }{\sqrt{2}}-\frac{\Phi  \dot{\chi}}{\chi   ^2},\\
  h^{\prime} & = -\frac{3 \left(-\eta ^2+h^2+\Phi ^2+w \Omega -\Psi \right)+K \Omega_K}{2 \sqrt{3}}-\frac{h \dot{\chi}}{\chi ^2},\\
  \eta^{\prime} & = -\frac{\eta  \kappa  \Phi }{\sqrt{2}}-\sqrt{3} \eta  h-\frac{\eta  \dot{\chi}}{\chi ^2},\\
  \Phi^{\prime}  & = \sqrt{2} \lambda  \Phi  \Psi -\frac{2 \Psi  \dot{\chi}}{\chi ^2},\\
  \Omega^{\prime} &  =-\sqrt{3} h (w+1) \Omega -\frac{2 \Omega  \dot{\chi}}{\chi^2},\\
  \Omega_K^{\prime} & = -\frac{2 h \Omega_K}{\sqrt{3}}-\frac{2 \Omega_K \dot{\chi}}{\chi ^2}
\end{align}
Next, by substituting,  
\begin{align}
\dot{\chi} &= -\frac{1}{2} \chi ^2 \left(\sqrt{2} \eta ^2 \kappa  \Phi +\sqrt{3} h \left(2 \Phi ^2+(w+1)\Omega \right)\right),\\
    \Psi & =1- \Phi^2-\Omega,\\
    \eta^2 & =1-h^2-\Omega_K, 
\end{align}
we obtain the main equations 
\begin{align}
   \Phi^{\prime} & =    \frac{1}{2} \left(\sqrt{2} \left(\lambda  \left(\Phi ^2+\Omega -1\right)-\kappa  \left(\Phi ^2-1\right)
   \left(h^2+\Omega_K-1\right)\right)+\sqrt{3} h \Phi  \left(2 \Phi ^2+(w+1)\Omega  -2\right)\right), \label{Full1}\\
     h^{\prime} & = \frac{1}{6} \Big(-3 \sqrt{2} h^3 \kappa  \Phi +3 \sqrt{3} h^2 \left(2 \Phi ^2+(w+1)\Omega
   -2\right)-3 \sqrt{2} h \kappa  \Phi  (\Omega_K-1) \nonumber \\
   & \;\;\;\;\;\;\;\;  -\sqrt{3} \left(6 \Phi ^2+3 (w+1) \Omega +4 \Omega_K-6\right)\Big), \label{Full2}\\
   \Omega^{\prime} & = \Omega  \left(\sqrt{3} h \left(2 \Phi ^2+w (\Omega -1)+\Omega -1\right)-\sqrt{2}
   \kappa  \Phi  \left(h^2+\Omega_K-1\right)\right), \label{Full3}\\
    \Omega_K^{\prime} & = \Omega_K  \left( \frac{h\left(6 \Phi ^2+3 (w+1) \Omega -2\right)}{\sqrt{3}}-\sqrt{2} \kappa  \Phi \left(h^2+\Omega_K-1\right)\right), \label{Full4}
\end{align}
and the auxiliary ones
\begin{align}
 \eta^{\prime} & =   \frac{1}{2} \sqrt{1-h^2-\Omega_K} \left(\sqrt{3} h \left(2 \Phi ^2+(w+1)\Omega -2\right)-\sqrt{2} \kappa  \Phi  \left(h^2+\Omega_K\right)\right),\\ 
  \Psi^{\prime}& = \left(-\Phi ^2-\Omega +1\right)\left(-\sqrt{2} \kappa  \Phi  \left(h^2+\Omega_K-1\right)+\sqrt{3} h \left(2 \Phi ^2+(w+1)\Omega
   \right)+\sqrt{2} \lambda  \Phi \right).
\end{align}

\begin{table}[]
    \centering
\resizebox{\textwidth}{!}{   
  \begin{tabular}{|ccccccccc|}\hline 
Label & $\Phi$ & $h$ & $\Omega$ & $\Omega_K$ & $k_1$ & $k_2$ & $k_3$ & $k_4$\\\hline 
$A$ & $ 1$ & $1$ & $0$ & $0$ & $\frac{4}{\sqrt{3}}$ &$-\sqrt{3} (w-1)$ &$ -\sqrt{2} \kappa $ & $\sqrt{2} \lambda +2 \sqrt{3} $\\\hline
$B$ & $ -1 $& $1$ & $0$ & $0$ &$ \frac{4}{\sqrt{3}}$ & $-\sqrt{3} (w-1)$ & $\sqrt{2} \kappa$  & $2 \sqrt{3}-\sqrt{2} \lambda$  \\\hline
$C$ & $1$ & $-1$ & $0$ & $0$ & $-\frac{4}{\sqrt{3}}$ & $\sqrt{3} (w-1)$ & $-\sqrt{2} \kappa$  &$ \sqrt{2} \lambda -2 \sqrt{3}$ \\\hline
$D$ &  $-1$ &$-1$ & $0$ & $0$ & $-\frac{4}{\sqrt{3}}$ & $\sqrt{3} (w-1)$ &$ \sqrt{2} \kappa$  & $-\sqrt{2} \lambda -2 \sqrt{3} $\\\hline
$E$ & $ -\frac{\lambda }{\sqrt{6}}$ & $1$ & $0$ & $0$ & $\frac{\lambda ^2-3 w-3}{\sqrt{3}}$ & $\frac{\lambda ^2-6}{2 \sqrt{3}}$  & $\frac{\lambda ^2-2}{\sqrt{3}}$ & $\frac{\lambda  (\kappa +\lambda )-6}{\sqrt{3}}$\\\hline
$F$ & $ \frac{\lambda }{\sqrt{6}}$ & $-1$ &$ 0$ &$ 0$ & $\frac{-\lambda ^2+3(w+1)}{\sqrt{3}}$ & $-\frac{\lambda ^2-6}{2 \sqrt{3}}$ & $-\frac{\lambda ^2-2}{\sqrt{3}}$ & $\frac{6-\lambda  (\kappa +\lambda )}{\sqrt{3}} $\\\hline
$G$ & $ \frac{1}{\sqrt{\frac{1}{6} \kappa  (\kappa +\lambda )+1}}$ & $-\frac{\kappa +\lambda }{\sqrt{\kappa  (\kappa +\lambda )+6}}$ & $0$ & $0$ & $\frac{2 (\kappa -2 \lambda )}{\sqrt{3} \sqrt{\kappa  (\kappa +\lambda )+6}}$ & $\frac{\kappa  (w+1)+\lambda  (w-1)}{\sqrt{\frac{1}{3} \kappa  (\kappa +\lambda )+2}}$ & $ \frac{\sqrt{3} \kappa -\sqrt{-\kappa\left(4 \left(\kappa ^2-6\right) \lambda +8 \kappa  \lambda ^2-27 \kappa +4 \lambda ^3\right)}}{2 \sqrt{\kappa  (\kappa +\lambda )+6}}$ & $\frac{\sqrt{3} \kappa +\sqrt{-\kappa\left(4 \left(\kappa ^2-6\right) \lambda +8 \kappa  \lambda ^2-27 \kappa +4 \lambda ^3\right)}}{2 \sqrt{\kappa  (\kappa
   +\lambda )+6}}$ \\\hline
$H$ & $ -\frac{1}{\sqrt{\frac{1}{6} \kappa  (\kappa +\lambda )+1}} $& $\frac{\kappa +\lambda }{\sqrt{\kappa  (\kappa+\lambda )+6}}$ & $0$ & $0$ & $-\frac{2 (\kappa -2 \lambda )}{\sqrt{3} \sqrt{\kappa  (\kappa +\lambda )+6}}$ & $-\frac{\kappa  (w+1)+\lambda  (w-1)}{\sqrt{\frac{1}{3} \kappa  (\kappa +\lambda )+2}}$ & $-\frac{\sqrt{3} \kappa -\sqrt{-\kappa\left(4 \left(\kappa ^2-6\right) \lambda +8 \kappa  \lambda ^2-27 \kappa +4 \lambda ^3\right)}}{2 \sqrt{\kappa  (\kappa +\lambda )+6}}$ & $-\frac{\sqrt{3} \kappa +\sqrt{-\kappa\left(4 \left(\kappa ^2-6\right) \lambda +8 \kappa  \lambda ^2-27 \kappa +4 \lambda ^3\right)}}{2 \sqrt{\kappa  (\kappa
   +\lambda )+6}}$\\\hline
$O^{-}$ & $ -1$ & $0$ & $0$ & $0$ & $-\sqrt{2} \kappa$  & $-\sqrt{2} \kappa$  & $-\frac{\kappa }{\sqrt{2}}$ & $-\sqrt{2} (\kappa +\lambda )$ \\\hline
$O^{+}$ & $ 1$ &$ 0 $& $0$ & $0$ & $\frac{\kappa }{\sqrt{2}}$ & $\sqrt{2} \kappa $ & $\sqrt{2} \kappa$  & $\sqrt{2} (\kappa +\lambda )$  \\\hline
$I$ &  $0 $& $-1$ & $1$ &$0$ & $-\sqrt{3} (w-1)$ & $-\frac{1}{2} \sqrt{3} (w-1)$ & $-\sqrt{3} (w+1) $& $-\frac{3 w+1}{\sqrt{3}}$\\\hline
$J$ & $ 0$ & $1$ &$ 1$ & $0 $& $\frac{1}{2} \sqrt{3} (w-1)$ & $\sqrt{3} (w-1)$ &$ \sqrt{3} (w+1)$ & $\frac{3 w+1}{\sqrt{3}}$ \\\hline
$K$ & $\frac{\sqrt{\frac{3}{2}} (w+1)}{\lambda }$ & $-1$ & $1-\frac{3 (w+1)}{\lambda ^2}$ & $0 $& $-\frac{3 w+1}{\sqrt{3}}$ & $-\frac{\sqrt{3} (\kappa  (w+1)+\lambda  (w-1))}{\lambda }$ & $\frac{\sqrt{3} \left( -\sqrt{w-1}\sqrt{\lambda ^2 (9 w+7)-24 (w+1)^2}+\lambda  (1-w)\right)}{4 \lambda }$ &$ \frac{\sqrt{3} \left(\lambda +\sqrt{w-1} \sqrt{\lambda ^2 (9 w+7)-24 (w+1)^2}+\lambda  (-w)\right)}{4 \lambda }$ \\\hline
$L$ & $ -\frac{\sqrt{\frac{3}{2}} (w+1)}{\lambda }$ & $1$ & $1-\frac{3 (w+1)}{\lambda ^2}$ & $0$ &$ \frac{3 w+1}{\sqrt{3}}$ &$  \frac{\sqrt{3} (\kappa  (w+1)+\lambda  (w-1))}{\lambda }$ & $-\frac{\sqrt{3} \left(\lambda +\sqrt{w-1} \sqrt{\lambda ^2 (9 w+7)-24 (w+1)^2}+\lambda  (-w)\right)}{4 \lambda }$ & $\frac{\sqrt{3} \left(\sqrt{w-1} \sqrt{\lambda ^2 (9 w+7)-24 (w+1)^2}+\lambda  (w-1)\right)}{4 \lambda }$ \\\hline
$M$ & $ \frac{w-1}{\sqrt{\frac{2 \kappa ^2}{3}+(w-1)^2}}$ & $\frac{\kappa }{\sqrt{\kappa ^2+\frac{3}{2} (w-1)^2}}$ & $ \frac{2 \kappa ^2}{2 \kappa ^2+3 (w-1)^2} $& $0 $& $\frac{\kappa  (3 w+1)}{\sqrt{3 \kappa ^2+\frac{9}{2}(w-1)^2}}$ &  $ -\frac{\mu_{11}}{6 \left(2 \kappa ^2+3 (w-1)^2\right)^{7/2}}$ & $-\frac{\mu_{12}}{6  \left(2 \kappa ^2+3 (w-1)^2\right)^{7/2}}$ & $-\frac{\mu_{13}}{6 \left(2 \kappa ^2+3
   (w-1)^2\right)^{7/2}}$ \\\hline
$N$ &  $-\frac{w-1}{\sqrt{\frac{2 \kappa ^2}{3}+(w-1)^2}}$ & $-\frac{\kappa }{\sqrt{\kappa ^2+\frac{3}{2} (w-1)^2}}$ & $  \frac{2 \kappa ^2}{2 \kappa ^2+3 (w-1)^2} $&$0$ & $-\frac{\kappa  (3 w+1)}{\sqrt{3 \kappa ^2+\frac{9}{2} (w-1)^2}}$ &$ \frac{\mu_{11}}{6 \left(2 \kappa ^2+3 (w-1)^2\right)^{7/2}}$ & $\frac{\mu_{12}}{6  \left(2 \kappa ^2+3 (w-1)^2\right)^{7/2}}$ & $\frac{\mu_{13}}{6 \left(2 \kappa ^2+3
   (w-1)^2\right)^{7/2}}$ \\\hline
$P$ & $-\frac{1}{\sqrt{3}}$ & $\frac{\lambda }{\sqrt{2}}$ & $0$ & $1-\frac{\lambda ^2}{2}$ & $\sqrt{\frac{2}{3}} (\kappa -2\lambda )$ & $-\frac{\lambda  (3 w+1)}{\sqrt{6}}$ &$ -\frac{\sqrt{8-3 \lambda ^2}+\lambda }{\sqrt{6}} $&
  $\frac{\sqrt{8-3 \lambda ^2}-\lambda }{\sqrt{6}}$ \\\hline
$Q$ & $ \frac{1}{\sqrt{3}}$ & $-\frac{\lambda }{\sqrt{2}}$ & $0$ & $1-\frac{\lambda ^2}{2}$ & $-\sqrt{\frac{2}{3}} (\kappa    -2 \lambda )$ & $\frac{\lambda  (3 w+1)}{\sqrt{6}} $& $\frac{\lambda -\sqrt{8-3 \lambda^2}}{\sqrt{6}}$ &  $  \frac{\sqrt{8-3 \lambda ^2}+\lambda }{\sqrt{6}} $\\\hline
$MC$ & $0$& $0$ & $\frac{2 (\kappa -2 \lambda )}{3 \kappa  (w+1)-4 \lambda }$ & $\frac{3 (\kappa  (w+1)+\lambda(w-1))}{3 \kappa  (w+1)-4 \lambda }$ &$ -\mu_{1} $& $\mu_{1}$ &$ -\mu_{2}$ &$\mu_{2}$\\
\hline 
\end{tabular}}
    \caption{Equilibrium points of the system \eqref{Full1}, \eqref{Full2}, \eqref{Full3} and \eqref{Full4} and their corresponding eigenvalues.}
    \label{tab:2}
\end{table}
We can find the inflationary condition with these variables where $q$, the deceleration parameter, is negative:
\begin{align}
    q = 2+ \frac{6 \Phi ^2+3 (w+1) \Omega +4  \Omega_{K}-6}{2 h^2} <0.
\end{align}
   
The equilibrium points $\left(\Phi, h, \Omega, \Omega_K\right)$ of the system \eqref{Full1}, \eqref{Full2}, \eqref{Full3} and \eqref{Full4} are the following (see Tab.  \ref{tab:2}): 
\begin{itemize}
\item  $A: =\left( 1, 1, 0, 0\right)$, with eigenvalues  $\left\{\frac{4}{\sqrt{3}},-\sqrt{3} (w-1), -\sqrt{2} \kappa, \sqrt{2} \lambda +2 \sqrt{3} \right)$. It is a saddle for $\{w<1, \kappa>0, \lambda <-\sqrt{6}\}$, or $ \{w<1, \kappa>0, \lambda >-\sqrt{6}\} $, or $ \{w<1, \kappa<0, \lambda <-\sqrt{6}\}$ or a source for  $\{w<1, \kappa<0, \lambda >-\sqrt{6}\}$. It is a solution dominated by the kinetic term of quintessence field. The deceleration parameter is $q=2$. Therefore, it never inflates.
    
\item $B: \left( -1, 1, 0, 0\right)$, with eigenvalues $\left\{\frac{4}{\sqrt{3}},-\sqrt{3} (w-1),\sqrt{2} \kappa, 2 \sqrt{3}-\sqrt{2} \lambda\right\}$. It is is a saddle for $\{w<1, \kappa<0, \lambda >\sqrt{6}\} $, or $ \{w<1, \kappa<0, \lambda <\sqrt{6}\} $, or $ \{w<1, \kappa>0, \lambda >\sqrt{6}\}$ or a source for  $\{w<1, \kappa>0, \lambda <\sqrt{6}\}$. It is a solution dominated by the kinetic term of quintessence field. The deceleration parameter is $q=2$. Therefore, it never inflates.

\item $C: \left(1, -1, 0, 0\right)$, with eigenvalues  $\left\{-\frac{4}{\sqrt{3}}, \sqrt{3} (w-1), -\sqrt{2} \kappa, \sqrt{2} \lambda -2 \sqrt{3}\right\}$. It is a sink for $\{w<1, \kappa>0, \lambda <\sqrt{6}\}$, a saddle for $\{w<1, \kappa>0, \lambda >\sqrt{6}\} $, or $ \{w<1, \kappa<0, \lambda <\sqrt{6}\}$, or $\{w<1, \kappa<0, \lambda >\sqrt{6}\}$. It is a solution dominated by the kinetic term of quintessence field. The deceleration parameter is $q=2$. Therefore, it never inflates.

\item $D: \left(-1, -1, 0,0\right)$, with eigenvalues  $\left\{-\frac{4}{\sqrt{3}},\sqrt{3} (w-1), \sqrt{2} \kappa,-\sqrt{2} \lambda -2 \sqrt{3}\right\}$. It is  a sink for $\{w<1, \kappa<0, \lambda >-\sqrt{6}\}$, a saddle for $\{w<1, \kappa<0, \lambda <-\sqrt{6}\} $, or $ \{w<1, \kappa>0, \lambda >-\sqrt{6}\}$, or $\{w<1, \kappa>0, \lambda <-\sqrt{6}\}$. It is a solution dominated by the kinetic term of quintessence field. The deceleration parameter is $q=2$. Therefore, it never inflates.

\item $E: \left(-\frac{\lambda }{\sqrt{6}}, 1, 0, 0\right)$, with eigenvalues $\left\{\frac{\lambda ^2-3 w-3}{\sqrt{3}}, \frac{\lambda ^2-6}{2 \sqrt{3}}, \frac{\lambda ^2-2}{\sqrt{3}},\frac{\lambda  (\kappa +\lambda )-6}{\sqrt{3}}\right\}$ exists for $-\sqrt{6}\leq \lambda \leq \sqrt{6}$.  It is nonhyperbolic for $\lambda ^2-3 w-3=0$, or $\lambda ^2=6$, or $\lambda^2=2$, or $\lambda  (\kappa +\lambda )=6$. It is a sink for $\kappa \in \mathbb{R}$ and $\left\{-\sqrt{2}<\lambda <0,  \kappa >\frac{6-\lambda
   ^2}{\lambda }\right\}$, or $ \{\lambda =0\} $, or $ \left\{0<\lambda <\sqrt{2},  \kappa
   <\frac{6-\lambda ^2}{\lambda }\right\}$. It is a saddle otherwise. 

\item $F: \left(\frac{\lambda }{\sqrt{6}}, -1,  0,  0\right)$, with eigenvalues $\left\{\frac{-\lambda ^2+3(w+1)}{\sqrt{3}}, -\frac{\lambda ^2-6}{2 \sqrt{3}}, -\frac{\lambda ^2-2}{\sqrt{3}}, \frac{6-\lambda  (\kappa +\lambda )}{\sqrt{3}} \right\}$ exists for $-\sqrt{6}\leq \lambda \leq \sqrt{6}$.  It is nonhyperbolic for $\lambda ^2-3 w-3=0$, or $\lambda ^2=6$, or $\lambda^2=2$, or $\lambda  (\kappa +\lambda )=6$. It is a source for $\kappa \in \mathbb{R}$ and $\left\{-\sqrt{2}<\lambda <0, \kappa >\frac{6-\lambda
   ^2}{\lambda }\right\}$, or $ \{\lambda =0\}$, or $ \left\{0<\lambda <\sqrt{2}\land \kappa
   <\frac{6-\lambda ^2}{\lambda }\right\}$.  It is a saddle otherwise. 

\item $G: \left(\frac{1}{\sqrt{\frac{1}{6} \kappa  (\kappa +\lambda )+1}}, -\frac{\kappa +\lambda }{\sqrt{\kappa  (\kappa +\lambda )+6}}, 0, 0\right)$, with eigenvalues \newline 
\begin{footnotesize}
$\left\{\frac{2 (\kappa -2 \lambda )}{\sqrt{3} \sqrt{\kappa  (\kappa +\lambda )+6}}, \frac{\kappa  (w+1)+\lambda  (w-1)}{\sqrt{\frac{1}{3} \kappa  (\kappa +\lambda )+2}}, \frac{\sqrt{3} \kappa -\sqrt{-\kappa\left(4 \left(\kappa ^2-6\right) \lambda +8 \kappa  \lambda ^2-27 \kappa +4 \lambda ^3\right)}}{2 \sqrt{\kappa  (\kappa +\lambda )+6}}, \frac{\sqrt{3} \kappa +\sqrt{-\kappa\left(4 \left(\kappa ^2-6\right) \lambda +8 \kappa  \lambda ^2-27 \kappa +4 \lambda ^3\right)}}{2 \sqrt{\kappa  (\kappa +\lambda )+6}}\right\}$.  
\end{footnotesize}
Exists for $\kappa <0, -\frac{\sqrt{\kappa ^2+24}}{2}-\frac{\kappa }{2}<\lambda <-\kappa$, or $\kappa >0, -\kappa <\lambda <\frac{\sqrt{\kappa ^2+24}}{2}-\frac{\kappa }{2}$. It is a nonhyperbolic for $\kappa=2\lambda$, or $\kappa =0, \lambda =-\sqrt{6}$, or $\kappa =0,  \lambda =\sqrt{6}$, or $\kappa \neq 0, \lambda =-\kappa$, or $w-1\neq 0, \sqrt{2} \kappa  \sqrt{-\frac{\kappa ^2-3(w-1)}{w-1}}, 0,
   \lambda =\frac{\kappa  (-w)-\kappa }{w-1}$, or $\lambda \neq 0, w=1, \kappa =0$, or $\kappa =0$, or
   $\lambda =\frac{1}{2} \left(-\sqrt{\kappa ^2+24}-\kappa \right)$, or $\lambda =\frac{1}{2} \left(\sqrt{\kappa
   ^2+24}-\kappa \right)$,  
  or a saddle otherwise.

\item $H: \left(-\frac{1}{\sqrt{\frac{1}{6} \kappa  (\kappa +\lambda )+1}}, \frac{\kappa +\lambda }{\sqrt{\kappa  (\kappa+\lambda )+6}}, 0, 0\right)$, with eigenvalues \newline 
\begin{footnotesize}
$\left\{\frac{4 \lambda -2 \kappa }{\sqrt{3} \sqrt{\kappa  (\kappa +\lambda )+6}},-\frac{\kappa  (w+1)+\lambda  (w-1)}{\sqrt{\frac{1}{3} \kappa  (\kappa +\lambda )+2}}, \frac{-\sqrt{3} \kappa -\sqrt{-\kappa\left(4 \left(\kappa ^2-6\right) \lambda +8 \kappa  \lambda ^2-27 \kappa +4 \lambda ^3\right)}}{2 \sqrt{\kappa  (\kappa +\lambda )+6}}, \frac{-\sqrt{3} \kappa +\sqrt{-\kappa\left(4 \left(\kappa ^2-6\right) \lambda +8 \kappa  \lambda ^2-27 \kappa +4 \lambda ^3\right)}}{2 \sqrt{\kappa  (\kappa
   +\lambda )+6}}\right\}$.
\end{footnotesize}
Exists for $\kappa <0, -\frac{\sqrt{\kappa ^2+24}}{2}-\frac{\kappa }{2}<\lambda <-\kappa$, or $\kappa >0, -\kappa <\lambda <\frac{\sqrt{\kappa ^2+24}}{2}-\frac{\kappa }{2}$. It is a nonhyperbolic for $\kappa=2\lambda$, or $\kappa =0, \lambda =-\sqrt{6}$, or $\kappa =0,  \lambda =\sqrt{6}$, or $\kappa \neq 0, \lambda =-\kappa$, or $w-1\neq 0, \sqrt{2} \kappa  \sqrt{-\frac{\kappa ^2-3(w-1)}{w-1}}, 0,
   \lambda =\frac{\kappa  (-w)-\kappa }{w-1}$, or $\lambda \neq 0, w=1, \kappa =0$, or $\kappa =0$, or
   $\lambda =\frac{1}{2} \left(-\sqrt{\kappa ^2+24}-\kappa \right)$, or $\lambda =\frac{1}{2} \left(\sqrt{\kappa
   ^2+24}-\kappa \right)$,  
  or a saddle otherwise. 
   
\item $O^{-}: \left( -1, 0, 0, 0\right)$, with eigenvalues $\left\{-\sqrt{2} \kappa, -\sqrt{2} \kappa, -\frac{\kappa }{\sqrt{2}}, -\sqrt{2} (\kappa +\lambda )\right\}$.  $O^{-}$ is a sink for $\kappa >0, \lambda >-\kappa$, a saddle for $\{\kappa <0, \lambda >-\kappa \}$, or $\{\kappa >0, \lambda <-\kappa \}$, a source for  $\kappa <0, \lambda <-\kappa$.  The deceleration parameter is $q=2$. Therefore, it represents a stiff fluid solution. It never inflates. 

\item $O^{+}: \left(1, 0, 0, 0\right)$, with eigenvalues  $\left\{\frac{\kappa }{\sqrt{2}}, \sqrt{2} \kappa , \sqrt{2} \kappa, \sqrt{2} (\kappa +\lambda )\right\}$. $O^+$ is a sink for $\kappa <0, \lambda <-\kappa$, a saddle for $\{\kappa <0, \lambda >-\kappa \}$, or $ \{\kappa >0, \lambda <-\kappa\}$ , a source for $\kappa >0, \lambda >-\kappa$.  The deceleration parameter is $q=2$. Therefore, it represents a stiff fluid solution. It never inflates.

\item $I: \left(0, -1, 1,0\right)$, with eigenvalues $\left\{-\sqrt{3} (w-1), -\frac{1}{2} \sqrt{3} (w-1), -\sqrt{3} (w+1), -\frac{3 w+1}{\sqrt{3}}\right\}$.  It always exists. It is nonhyperbolic for $w\in\{-1,1\}$. The point is a saddle for $-1<w<1$. The deceleration parameter is $q=\frac{1}{2} (3 w+1)$. It is an inflationary solution for $w<-\frac{1}{3}$. 

\item $J: \left(0, 1, 1, 0 \right)$, with eigenvalues $\left\{\frac{1}{2} \sqrt{3} (w-1), \sqrt{3} (w-1), \sqrt{3} (w+1), \frac{3 w+1}{\sqrt{3}}\right\}$.  It always exists. It is nonhyperbolic for $w\in\{-1,1\}$. The point is a saddle for $-1<w<1$. The deceleration parameter is $q=\frac{1}{2} (3 w+1)$. It is an inflationary solution for $w<-\frac{1}{3}$.

\item $K: \left(\frac{\sqrt{\frac{3}{2}} (w+1)}{\lambda }, -1, 1-\frac{3 (w+1)}{\lambda ^2},0\right)$, with eigenvalues \newline 
\begin{footnotesize}
$\Big\{-\frac{3 w+1}{\sqrt{3}}, -\frac{\sqrt{3} (\kappa  (w+1)+\lambda  (w-1))}{\lambda }, \frac{\sqrt{3} \left(\sqrt{w-1} \sqrt{\lambda ^2 (9 w+7)-24 (w+1)^2}+\lambda  (1-w)\right)}{4 \lambda }, -\frac{\sqrt{3} \left(\sqrt{w-1} \sqrt{\lambda ^2 (9 w+7)-24 (w+1)^2}+\lambda  (w-1)\right)}{4 \lambda }\Big\}$.
\end{footnotesize}
Exists for $\left\{\lambda \leq -\sqrt{6}, -1\leq w\leq 1\right\} $,  \newline or $ \left\{-\sqrt{6}<\lambda <0, -1\leq w\leq \frac{1}{3}\left(\lambda ^2-3\right)\right\}$, or 
    $ \left\{0<\lambda <\sqrt{6}, -1\leq w\leq \frac{1}{3} \left(\lambda ^2-3\right)\right\} $, or $ \left\{\lambda \geq \sqrt{6}, -1\leq w\leq 1\right\}$. Nonhyperbolic for $w=-\frac{1}{3}$, or $\left\{\lambda \neq 0, w=-1\right\}$, \newline or $  \left\{-1<w<1,  \lambda =\sqrt{3(w+1)}\right\}$, or $  \left\{-1<w<1,  \lambda =-\sqrt{3(w+1)}\right\}$,  \newline or $  \left\{-1<w<1,  \lambda <-\sqrt{3(w+1)},  \kappa =\frac{\lambda(1-w)}{w+1}\right\}$, or $ 
   \left\{-1<w<1,  \lambda >\sqrt{3(w+1)},  \kappa =\frac{\lambda(1-w)}{w+1}\right\} $, or $  \left\{w=1, 
   \lambda \geq \sqrt{6}\right\}$, or $  \left\{w=1,  \lambda \leq -\sqrt{6}\right\}$. 
   
   It is a source for $\left\{-1<w<-\frac{1}{3},  \lambda <-\sqrt{3 w+3},  \kappa
   >\frac{\lambda(1-w)}{w+1}\right\}$, or \newline $\left\{-1<w<-\frac{1}{3}, \lambda >\sqrt{3 w+3}, \kappa <\frac{\lambda(1-w)}{w+1}\right\}$. It is a saddle otherwise. The deceleration parameter is $q=\frac{1}{2} (3 w+1)$. It is an inflationary solution for $w<-\frac{1}{3}$. 

   \item $L: \left(-\frac{\sqrt{\frac{3}{2}} (w+1)}{\lambda }, 1, 1-\frac{3 (w+1)}{\lambda ^2}, 0\right)$, with eigenvalues \newline 
\begin{footnotesize}
$\Big\{\frac{3 w+1}{\sqrt{3}},  \frac{\sqrt{3} (\kappa  (w+1)+\lambda  (w-1))}{\lambda }, -\frac{\sqrt{3} \left(\sqrt{w-1} \sqrt{\lambda ^2 (9 w+7)-24 (w+1)^2}+\lambda  (1-w)\right)}{4 \lambda }, \frac{\sqrt{3} \left(\sqrt{w-1} \sqrt{\lambda ^2 (9 w+7)-24 (w+1)^2}+\lambda  (w-1)\right)}{4 \lambda }\Big\}$.
\end{footnotesize}
Exists for $\left\{\lambda \leq -\sqrt{6}, -1\leq w\leq 1\right\} $,  or $ \left\{-\sqrt{6}<\lambda <0, -1\leq w\leq \frac{1}{3}\left(\lambda ^2-3\right)\right\}$, \newline  or 
    $ \left\{0<\lambda <\sqrt{6}, -1\leq w\leq \frac{1}{3} \left(\lambda ^2-3\right)\right\} $, or $ \left\{\lambda \geq \sqrt{6}, -1\leq w\leq 1\right\}$. Nonhyperbolic for $w=-\frac{1}{3}$, or $\left\{\lambda \neq 0, w=-1\right\}$,  or $  \left\{-1<w<1,  \lambda =\sqrt{3(w+1)}\right\}$, \newline or $  \left\{-1<w<1,  \lambda =-\sqrt{3(w+1)}\right\}$,  or $  \left\{-1<w<1,  \lambda <-\sqrt{3(w+1)},  \kappa =\frac{\lambda(1-w)}{w+1}\right\}$, \newline  or $ 
   \left\{-1<w<1,  \lambda >\sqrt{3(w+1)},  \kappa =\frac{\lambda(1-w)}{w+1}\right\} $, or $  \left\{w=1, 
   \lambda \geq \sqrt{6}\right\}$, or $  \left\{w=1,  \lambda \leq -\sqrt{6}\right\}$. 
   
   It is a sink for $\left\{-1<w<-\frac{1}{3}, \lambda <-\sqrt{3 w+3}, \kappa
   >\frac{\lambda(1-w)}{w+1}\right\}$,  or \newline
   $\left\{-1<w<-\frac{1}{3}, \lambda >\sqrt{3 w+3}, \kappa
   <\frac{\lambda(1-w)}{w+1}\right\}$. It is a saddle otherwise. The deceleration parameter is $q=\frac{1}{2} (3 w+1)$. It is an inflationary solution for $w<-\frac{1}{3}$.

\item $M, N: \left(\pm\frac{w-1}{\sqrt{\frac{2 \kappa ^2}{3}+(w-1)^2}}, \pm\frac{\kappa }{\sqrt{\kappa ^2+\frac{3}{2} (w-1)^2}}, \frac{2 \kappa ^2}{2 \kappa ^2+3 (w-1)^2}, 0 \right)$. 
$M$ and $N$ exist for $\{\kappa <0, -1\leq w\leq 1\}$ or $\{\kappa =0, -1\leq w<1\}$ or $\{\kappa >0, -1\leq w\leq 1\}$.  The eigenvalues are \newline $\left\{\pm \frac{\kappa  (3 w+1)}{\sqrt{3 \kappa ^2+\frac{9}{2}(w-1)^2}}, \mp\frac{\mu_{11}}{6 \left(2 \kappa ^2+3 (w-1)^2\right)^{7/2}}, \mp\frac{\mu_{12}}{6  \left(2 \kappa ^2+3 (w-1)^2\right)^{7/2}}, \mp\frac{\mu_{13}}{6 \left(2 \kappa ^2+3
   (w-1)^2\right)^{7/2}}\right\}$, where we define  $\mu_{1 j}, j\in\{1,2,3\}$ as the $j$-th root of the   polynomial \newline 
$P_1(\mu):= \mu ^3+3 \sqrt{6} \mu ^2 \left(2 \kappa ^2+3 (w-1)^2\right)^3 (\kappa +3 \kappa  w+2 \lambda 
   (w-1))+108 \kappa  \mu  (w-1) \left(2 \kappa ^2+3 (w-1)^2\right)^6 (2 \kappa +\lambda  (w-1))-648 \sqrt{6}
   \kappa ^2 (w-1)^2 \left(2 \kappa ^2+3 (w-1)^2\right)^9 (\kappa  (w+1)+\lambda  (w-1)$. 
 
The stability of the points $M$ and $N$ (which have opposite dynamical behaviours) is examined numerically. For dust matter ($w=0$), the signs of the real parts of the three complicated eigenvalues of $M$ are the same as in Fig.  \ref{fig:MN}. Therefore, for dust, the points are saddles. 

\item $P: \left(-\frac{1}{\sqrt{3}}, \frac{\lambda }{\sqrt{2}}, 0, 1-\frac{\lambda ^2}{2}\right)$, with eigenvalues  $\left\{ \sqrt{\frac{2}{3}} (\kappa -2\lambda ), -\frac{\lambda  (3 w+1)}{\sqrt{6}}, -\frac{\sqrt{8-3 \lambda ^2}+\lambda }{\sqrt{6}}, \frac{\sqrt{8-3 \lambda ^2}-\lambda }{\sqrt{6}}\right\}$. Exists for $-\sqrt{2}\leq \lambda \leq \sqrt{2}$. It is nonhyperbolic for $w=-\frac{1}{3}$, or $\kappa=2\lambda$, or  $\lambda^2=2$. It is a saddle otherwise.

\item $Q: \left(\frac{1}{\sqrt{3}}, -\frac{\lambda }{\sqrt{2}}, 0, 1-\frac{\lambda ^2}{2}\right)$, with eigenvalues $\left\{-\sqrt{\frac{2}{3}} (\kappa    -2 \lambda ), \frac{\lambda  (3 w+1)}{\sqrt{6}}, \frac{\lambda -\sqrt{8-3 \lambda^2}}{\sqrt{6}}, \frac{\sqrt{8-3 \lambda ^2}+\lambda }{\sqrt{6}}\right\}$.  Exists for $-\sqrt{2}\leq \lambda \leq \sqrt{2}$. It is nonhyperbolic for $w=-\frac{1}{3}$, or $\kappa=2\lambda$, or  $\lambda^2=2$. It is a saddle otherwise.

\item $MC: \left(0, 0, \frac{2 (\kappa -2 \lambda )}{3 \kappa  (w+1)-4 \lambda }, \frac{3 (\kappa  (w+1)+\lambda(w-1))}{3 \kappa  (w+1)-4 \lambda }\right)$, with eigenvalues $\left\{-\mu_{1}, \mu_{1}, -\mu_{2}, \mu_{2}\right\}$, where \newline  
\begin{footnotesize}
   $\mu_1=\frac{\sqrt{3 w+1} \sqrt{-\kappa  \lambda ^2-\lambda  \left(\kappa ^2+2 w-2\right)-\sqrt{\kappa ^4 \lambda
   ^2+2 \kappa ^3 \lambda  \left(\lambda ^2+w+1\right)+\kappa ^2 \left(\lambda ^4-2 \lambda ^2
   (w+9)+(w+1)^2\right)-4 \kappa  \lambda  (w-1) \left(\lambda ^2+w+1\right)+4 \lambda ^2 (w-1)^2}+\kappa 
   (w+1)}}{\sqrt{6 \kappa  (w+1)-8 \lambda }}$, \newline $\mu_2=\frac{\sqrt{3 w+1} \sqrt{-\kappa  \lambda ^2-\lambda  \left(\kappa ^2+2 w-2\right)+\sqrt{\kappa ^4 \lambda
   ^2+2 \kappa ^3 \lambda  \left(\lambda ^2+w+1\right)+\kappa ^2 \left(\lambda ^4-2 \lambda ^2
   (w+9)+(w+1)^2\right)-4 \kappa  \lambda  (w-1) \left(\lambda ^2+w+1\right)+4 \lambda ^2 (w-1)^2}+\kappa 
   (w+1)}}{\sqrt{6 \kappa  (w+1)-8 \lambda }}$. 
\end{footnotesize} 
Exists for $3 \kappa  (w+1)-4 \lambda \neq 0$. It is nonhyperbolic for $\kappa\in\left\{2 \lambda, -\lambda(w-1)/(w+1)\right\}$ or a saddle, otherwise.
\end{itemize}

\subsubsection{Vacuum model ($\rho_m=0$) with positive curvature ($K=+1$)}
\label{m1k+1}

When we consider the vacuum solution $\rho_m=0$, $\Omega=0$, the variables \eqref{Positivevars} satisfy
\begin{equation}
    h^2 + \eta^2 + \Omega_K= \Phi^2+ \Psi=1. \label{K+1eq10}
\end{equation}

Using  \eqref{K+1eq10}, we have bounded variables 
\begin{equation}
    0 \leq  {h}^2, {\Phi}^2,  {\eta}^2,  {\Psi},  {\Omega}_K \leq 1.  \label{k+1bounds}
\end{equation}

Eliminating $\Psi=1- \Phi^2$ and $\eta^2=1-h^2-\Omega_K$, we obtain
\begin{align}
    & \Phi^{\prime}= \frac{1}{2} \left(\Phi ^2-1\right) \left(-\sqrt{2} \kappa  \left(h^2+ \Omega_K-1\right)+2 \sqrt{3} h \Phi +\sqrt{2} \lambda \right),  \label{k+1sys1} \\
    & h^{\prime}= \frac{1}{6} \left(-3 \sqrt{2} h^3 \kappa  \Phi +6 \sqrt{3} h^2 \left(\Phi^2 - 1\right) -3\sqrt{2} h \kappa \Phi(\Omega_K-1)-\sqrt{3} \left(4 \Omega_K+6 \Phi^2-6\right)\right), \\
    & \Omega_K^{\prime}= \frac{1}{3} \Omega_K \left(-3 \sqrt{2} h^2 \kappa  \Phi +2 \sqrt{3} h \left(3 \Phi ^2-1\right)-3 \sqrt{2} \kappa\,\Phi(\Omega_K-1)\right). \label{k+1sys3}
\end{align}

We can find the inflationary condition with these variables where $q$, the deceleration parameter, is negative:
\begin{align}
    q = 2 + \frac{3 \Phi ^2+2  \Omega_{K}-3}{h^2} <0.
\end{align}

The equilibrium points are $(\Phi,h, \Omega_K)$: 

\begin{itemize}
    \item $A: (1,1,0)$, with eigenvalues $\left\{\frac{4}{\sqrt{3}},-\sqrt{2} \kappa ,\sqrt{2} \lambda +2 \sqrt{3}\right\}$. It is a saddle for $\{\kappa>0, \lambda <-\sqrt{6}\}$, or $ \{\kappa>0, \lambda >-\sqrt{6}\} $, or $ \{\kappa<0, \lambda <-\sqrt{6}\}$ or a source for  $\kappa<0, \lambda >-\sqrt{6}$.  It is a solution dominated by the kinetic term of quintessence field. The deceleration parameter is $q=2$. Therefore, it never inflates.

    \item $B: (-1,1,0)$, with eigenvalues $\left\{\frac{4}{\sqrt{3}},\sqrt{2} \kappa ,2 \sqrt{3}-\sqrt{2} \lambda \right\}$. It is is a saddle for $\{\kappa<0, \lambda >\sqrt{6}\} $, or $ \{\kappa<0, \lambda <\sqrt{6}\} $, or $ \{\kappa>0, \lambda >\sqrt{6}\}$ or a source for  $\kappa>0, \lambda <\sqrt{6}$.  It is a solution dominated by the kinetic term of quintessence field. The deceleration parameter is $q=2$. Therefore, it never inflates.

   \item $C: (1,-1,0)$, with eigenvalues $\left\{-\frac{4}{\sqrt{3}},-\sqrt{2} \kappa ,\sqrt{2} \lambda -2 \sqrt{3}\right\}$. It is a sink for $\kappa>0, \lambda <\sqrt{6}$, a saddle for $\{\kappa>0, \lambda >\sqrt{6}\} $, or $ \{\kappa<0, \lambda <\sqrt{6}\}$, or $\{\kappa<0, \lambda >\sqrt{6}\}$.  It is a solution dominated by the kinetic term of quintessence field. The deceleration parameter is $q=2$. Therefore, it never inflates.

   \item $D: (-1,-1,0)$, with eigenvalues $\left\{-\frac{4}{\sqrt{3}},\sqrt{2} \kappa ,-\sqrt{2} \lambda -2 \sqrt{3}\right\}$. It is  a sink for $\kappa<0, \lambda >-\sqrt{6}$, a saddle for $\{\kappa<0, \lambda <-\sqrt{6}\} $, or $ \{\kappa>0, \lambda >-\sqrt{6}\}$, or $\{\kappa>0, \lambda <-\sqrt{6}\}$. It is a solution dominated by the kinetic term of quintessence field. The deceleration parameter is $q=2$. Therefore, it never inflates.

   \item $E: \left(-\frac{\lambda}{\sqrt{6}}, 1,0\right)$, with  eigenvalues $\left\{\frac{\lambda ^2-2}{\sqrt{3}},\frac{\lambda ^2-6}{2 \sqrt{3}},\frac{\lambda  (\kappa
   +\lambda )-6}{\sqrt{3}}\right\}$ exists for $-\sqrt{6}\leq \lambda \leq \sqrt{6}$. It is a sink for $\kappa \in \mathbb{R}$ and $\left\{-\sqrt{2}<\lambda <0,  \kappa >\frac{6-\lambda
   ^2}{\lambda }\right\}$, or $ \{\lambda =0\} $, or $ \left\{0<\lambda <\sqrt{2},  \kappa
   <\frac{6-\lambda ^2}{\lambda }\right\}$. It is a saddle for $\left\{-\sqrt{6}<\lambda <-\sqrt{2} \land  \left(\kappa <\frac{6-\lambda ^2}{\lambda } \lor 
   \kappa >\frac{6-\lambda ^2}{\lambda }\right)\right\}$, or $ \left\{-\sqrt{2}<\lambda <0, 
   \kappa <\frac{6-\lambda ^2}{\lambda }\right\}$, or $ \left\{0<\lambda <\sqrt{2},  \kappa
   >\frac{6-\lambda ^2}{\lambda }\right\}$, or $ \left\{\sqrt{2}<\lambda <\sqrt{6}, \kappa <\frac{6-\lambda ^2}{\lambda }\right\}$, or $ \left\{\sqrt{2}<\lambda <\sqrt{6}, \kappa >\frac{6-\lambda ^2}{\lambda
   }\right\}$.

   \item $F: \left(\frac{\lambda}{\sqrt{6}},  -1,0\right)$, with  eigenvalues $\left\{-\frac{\lambda ^2-2}{\sqrt{3}},-\frac{\lambda ^2-6}{2 \sqrt{3}},\frac{6-\lambda 
   (\kappa +\lambda )}{\sqrt{3}}\right\}$ exists for $-\sqrt{6}\leq \lambda \leq \sqrt{6}$. It is a source for $\kappa \in \mathbb{R}$ and $\left\{-\sqrt{2}<\lambda <0, \kappa >\frac{6-\lambda
   ^2}{\lambda }\right\}$, or $ \{\lambda =0\}$, or $ \left\{0<\lambda <\sqrt{2}, \kappa
   <\frac{6-\lambda ^2}{\lambda }\right\}$.  It is a saddle for $\left\{-\sqrt{6}<\lambda <-\sqrt{2} \land  \left(\kappa <\frac{6-\lambda ^2}{\lambda } \lor 
   \kappa >\frac{6-\lambda ^2}{\lambda }\right)\right\}$, or $ \left\{-\sqrt{2}<\lambda <0, 
   \kappa <\frac{6-\lambda ^2}{\lambda }\right\}$, or $ \left\{0<\lambda <\sqrt{2},  \kappa
   >\frac{6-\lambda ^2}{\lambda }\right\}$, or \newline $ \left\{\sqrt{2}<\lambda <\sqrt{6} \land 
   \left(\kappa <\frac{6-\lambda ^2}{\lambda }\lor \kappa >\frac{6-\lambda ^2}{\lambda
   }\right)\right\}$.
      
   \item $G: \left(\frac{\sqrt{6}}{\sqrt{\kappa^2 + \kappa \lambda +6}}, -\frac{\kappa + \lambda}{\sqrt{\kappa^2 + \kappa \lambda +6}},0\right)$, with  eigenvalues \newline $\left\{\frac{2 (\kappa -2 \lambda )}{\sqrt{3} \sqrt{\kappa  (\kappa +\lambda)+6}},\frac{\sqrt{3} \kappa -\sqrt{-\kappa\left(4 \left(\kappa ^2-6\right) \lambda +8\kappa\lambda ^2-27 \kappa +4 \lambda ^3\right)}}{2 \sqrt{\kappa  (\kappa +\lambda)+6}},\frac{\sqrt{3} \kappa +\sqrt{-\kappa\left(4 \left(\kappa ^2-6\right) \lambda +8\kappa\lambda ^2-27 \kappa +4 \lambda ^3\right)}}{2 \sqrt{\kappa  (\kappa +\lambda)+6}}\right\}$. It a saddle for  $\left\{0<\kappa \leq 2 \sqrt{2},  -\kappa <\lambda <\frac{\kappa }{2}\right\}$, or $\left\{\kappa >2 \sqrt{2},  -\kappa <\lambda <\frac{\sqrt{\kappa ^2+24}}{2}-\frac{\kappa}{2}\right\}$, or $ \left\{-2 \sqrt{2}<\kappa <0,  -\frac{\sqrt{\kappa^2+24}}{2}-\frac{\kappa }{2}<\lambda <\frac{\kappa }{2}\right\}$, or $ \left\{0<\kappa <2\sqrt{2},  \frac{\kappa }{2}<\lambda <\frac{\sqrt{\kappa ^2+24}}{2}-\frac{\kappa}{2}\right\}$, or $ \left\{\kappa \leq -2 \sqrt{2},  -\frac{\sqrt{\kappa^2+24}}{2}-\frac{\kappa }{2}<\lambda <-\kappa \right\}$, or $ \left\{-2 \sqrt{2}<\kappa <0, \frac{\kappa }{2}<\lambda <-\kappa \right\}$,  \newline or $ \left\{0<\kappa \leq 2 \sqrt{2}, -\kappa<\lambda <\frac{\kappa}{2}\right\}$, or $ \left\{\kappa >2 \sqrt{2}, -\kappa <\lambda<\frac{\sqrt{\kappa ^2+24}}{2}-\frac{\kappa }{2}\right\}$,  \newline or $ \left\{-2 \sqrt{2}<\kappa<0, -\frac{\sqrt{\kappa ^2+24}}{2}-\frac{\kappa}{2}<\lambda <\frac{\kappa }{2}\right\}$.
   
   \item $H: \left(-\frac{\sqrt{6}}{\sqrt{\kappa^2 + \kappa \lambda +6}}, \frac{\kappa + \lambda}{\sqrt{\kappa^2 + \kappa \lambda +6}},0\right)$, with  eigenvalues \newline $\left\{\frac{4 \lambda -2 \kappa }{\sqrt{3} \sqrt{\kappa  (\kappa +\lambda)+6}},\frac{-\sqrt{3} \kappa -\sqrt{-\kappa\left(4 \left(\kappa ^2-6\right) \lambda +8\kappa\lambda ^2-27 \kappa +4 \lambda ^3\right)}}{2 \sqrt{\kappa  (\kappa +\lambda)+6}},\frac{-\sqrt{3} \kappa +\sqrt{-\kappa\left(4 \left(\kappa ^2-6\right) \lambda +8\kappa\lambda ^2-27 \kappa +4 \lambda ^3\right)}}{2 \sqrt{\kappa  (\kappa +\lambda)+6}}\right\}$. It is a saddle for $\left\{0<\kappa <2 \sqrt{2}, \frac{\kappa }{2}<\lambda <\frac{\sqrt{\kappa^2+24}}{2}-\frac{\kappa }{2}\right\}$, or $ \left\{\kappa \leq -2 \sqrt{2},-\frac{\sqrt{\kappa ^2+24}}{2}-\frac{\kappa }{2}<\lambda <-\kappa \right\}$, or $ \left\{-2\sqrt{2}<\kappa <0, \frac{\kappa }{2}<\lambda <-\kappa \right\}$, or $ \left\{0<\kappa \leq
   2 \sqrt{2}, -\kappa <\lambda <\frac{\kappa }{2}\right\}$,  \newline or $ \left\{\kappa >2\sqrt{2}, -\kappa <\lambda <\frac{\sqrt{\kappa ^2+24}}{2}-\frac{\kappa }{2}\right\}$, or $\left\{-2 \sqrt{2}<\kappa <0, -\frac{\sqrt{\kappa ^2+24}}{2}-\frac{\kappa }{2}<\lambda<\frac{\kappa }{2}\right\}$ , or $ \left\{0<\kappa <2 \sqrt{2}, \frac{\kappa }{2}<\lambda
   <\frac{\sqrt{\kappa ^2+24}}{2}-\frac{\kappa }{2}\right\}$, or $ \left\{\kappa \leq -2\sqrt{2}, -\frac{\sqrt{\kappa ^2+24}}{2}-\frac{\kappa }{2}<\lambda <-\kappa\right\}$, or $\left\{-2 \sqrt{2}<\kappa <0,\frac{\kappa }{2}<\lambda <-\kappa \right\}$.
   
   \item $O^{\pm}:(\pm 1, 0,0)$, with eigenvalues $\left\{\pm\frac{\kappa }{\sqrt{2}},\pm\sqrt{2} \kappa , \pm\sqrt{2} (\kappa +\lambda )\right\}$. $O^+$ is a sink for $\kappa <0, \lambda <-\kappa$, a saddle for $\{\kappa <0, \lambda >-\kappa \}$, or $ \{\kappa >0, \lambda <-\kappa\}$ , a source for $\kappa >0, \lambda >-\kappa$. $O^{-}$ is a sink for $\kappa >0, \lambda >-\kappa$, a saddle for $\{\kappa <0, \lambda >-\kappa \}$, or $\{\kappa >0, \lambda <-\kappa \}$, a source for  $\kappa <0, \lambda <-\kappa$.

    Points $A$, $B$, $C$, $D$, $E$, $F$, $G$, $H$ and $O^\pm$  have the same physical interpretation as for the flat case. Existence conditions are the same, but the stability conditions slightly change. 
   
    For $-\sqrt{2}<\lambda <\sqrt{2}$, there are new points, which are the curvature scaling solutions.

    \item $P: \left(-\frac{1}{\sqrt{3}}, \frac{\lambda }{\sqrt{2}},  1-\frac{\lambda ^2}{2}\right)$, with  eigenvalues $\left\{\sqrt{\frac{2}{3}} (\kappa -2 \lambda ),-\frac{\sqrt{8-3 \lambda ^2}+\lambda}{\sqrt{6}},\frac{\sqrt{8-3 \lambda^2} -\lambda}{\sqrt{6}}\right\}$. It is a saddle for $\left\{-\sqrt{2}<\lambda <\sqrt{2}, \kappa >2 \lambda \right\}$, or $\left\{-\sqrt{2}<\lambda <\sqrt{2}, \kappa <2 \lambda \right\}$. The deceleration parameter is $q=0$. It never inflates. 
    
    \item $Q: \left(  \frac{1}{\sqrt{3}},  -\frac{\lambda }{\sqrt{2}}, 1-\frac{\lambda^2}{2}\right)$, with  eigenvalues $\left\{-\sqrt{\frac{2}{3}} (\kappa -2 \lambda ),\frac{\lambda -\sqrt{8-3 \lambda^2}}{\sqrt{6}},\frac{\sqrt{8-3 \lambda ^2}+\lambda }{\sqrt{6}}\right\}$. It is a saddle for $\left\{-\sqrt{2}<\lambda <\sqrt{2}, \kappa <2 \lambda \right\}$, or $\left\{-\sqrt{2}<\lambda<\sqrt{2}, \kappa >2 \lambda \right\}$. The deceleration parameter is $q=0$. It never inflates. 
 
\end{itemize}

\subsubsection{Discussion}
In this case, owing to \eqref{K+1eq10}-\eqref{k+1bounds}, the phase space is
\begin{equation}
    \lbrace(\Phi,h,\Omega_K)|\,\,-1\leq \Phi,h\leq1,\,0\leq\Omega_K\leq1-h^2\rbrace.\label{k+1space}
\end{equation}

\begin{figure}[h!]
    \centering
    \includegraphics[scale=0.4]{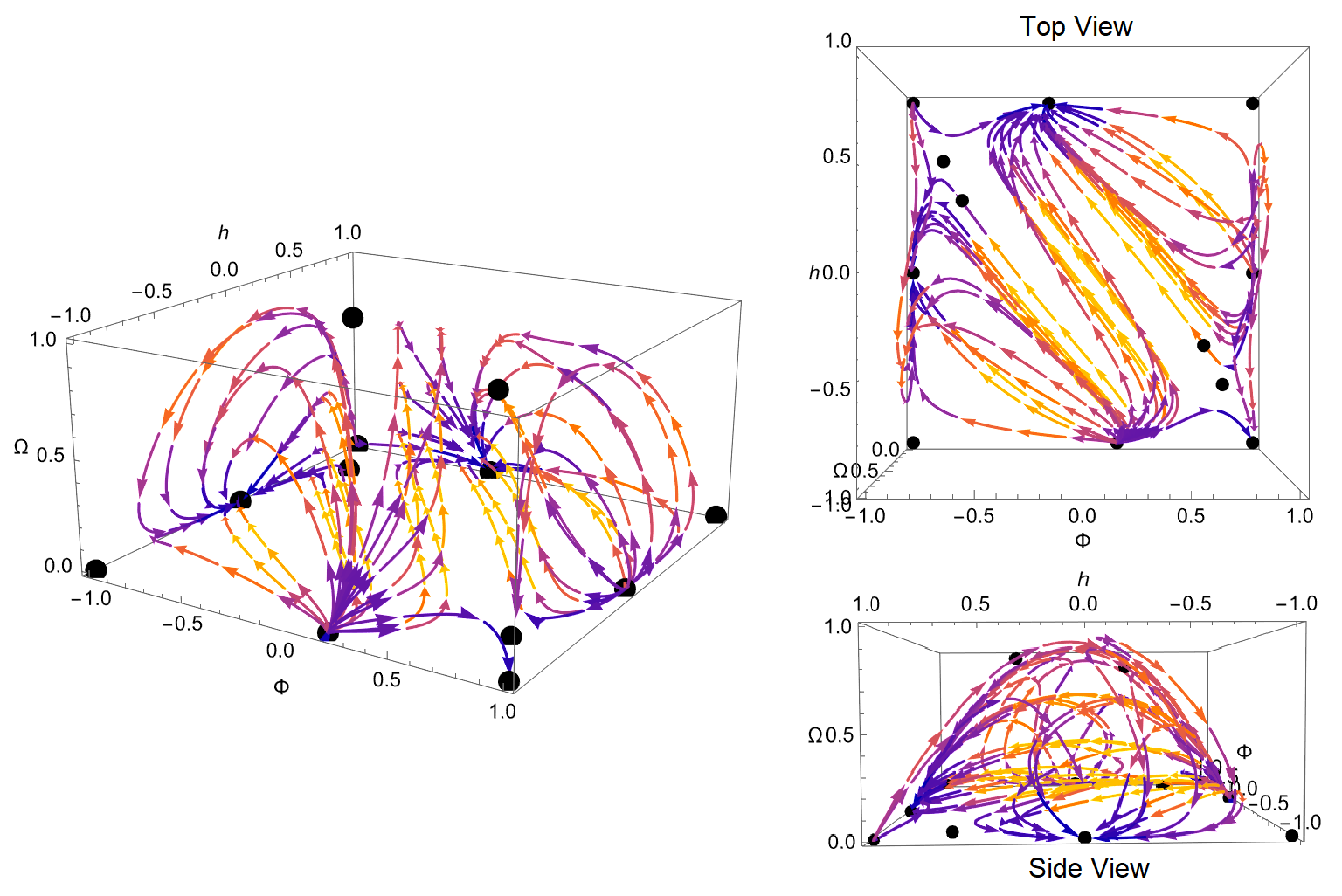}
    \caption{Three different views of the flows \eqref{k+1sys1}-\eqref{k+1sys3} for $\lambda=0.5,\kappa=1.5$.}
    \label{k+1M1flow}
\end{figure}

The boundaries of this set are invariant sets of \eqref{k+1sys1}-\eqref{k+1sys3}, defined by: $\Phi=\pm1,\,\Omega_K=0$ and $\Omega_K=1-h^2$ --- the last of these is confirmed by \eqref{invk+1}. The equilibria $P$ and $Q$ are points within the surface $\Omega_K=1-h^2$.  Fig. \ref{k+1M1flow} shows the flow for $\lambda=0.5,\kappa=1.5$.
\begin{equation}
    (h^2+\Omega_K-1)^\prime = 2\sqrt{3}(h^2+\Omega_K-1)\left[h(\Phi^2-1)-(\kappa/\sqrt{6})f(h^2+\Omega_K)\label{invk+1} \right].
\end{equation}

\begin{figure}[h!]
    \centering
    \includegraphics[scale=0.6]{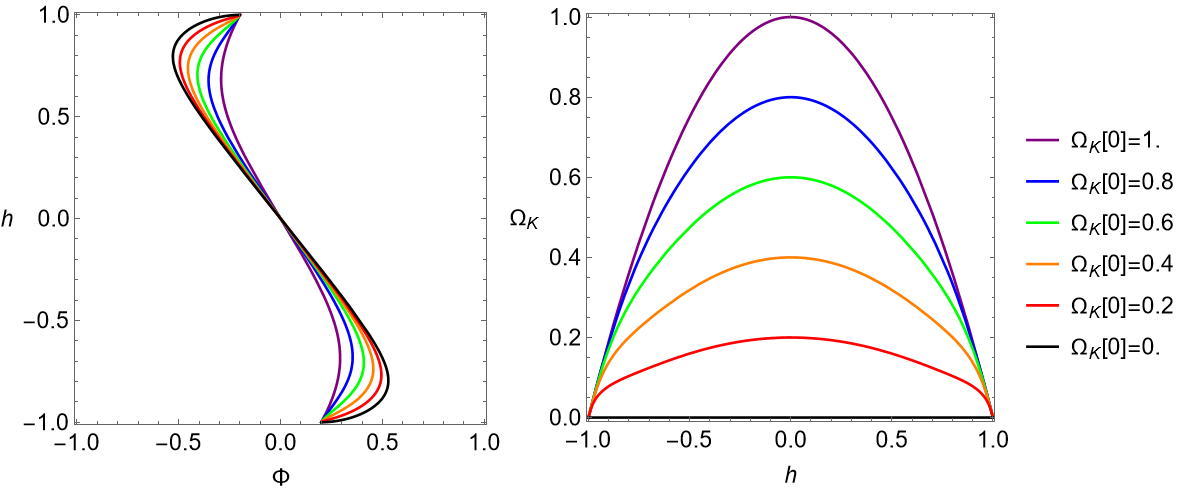}
    \caption{Time-symmetric solutions of \eqref{k+1sys1}-\eqref{k+1sys3}, with varying initial values of $\Omega_K(0)$, and $\lambda=0.5,\,\kappa=1.5$ }
    \label{k+1series}
\end{figure}

Fig. \ref{k+1series} highlights a handful of solutions with initial conditions $\Phi=h=0$ and varying $\Omega_K$, for parameters $\lambda=0.5,\,\kappa=1.5$.  These time-symmetric solutions correspond to bouncing cosmologies, varying the curvature value at the bounce (where $h=0)$.

\begin{figure}[h!]
	\centering
	\subfigure[\label{k+1bouncing}]{\includegraphics[scale=0.45]{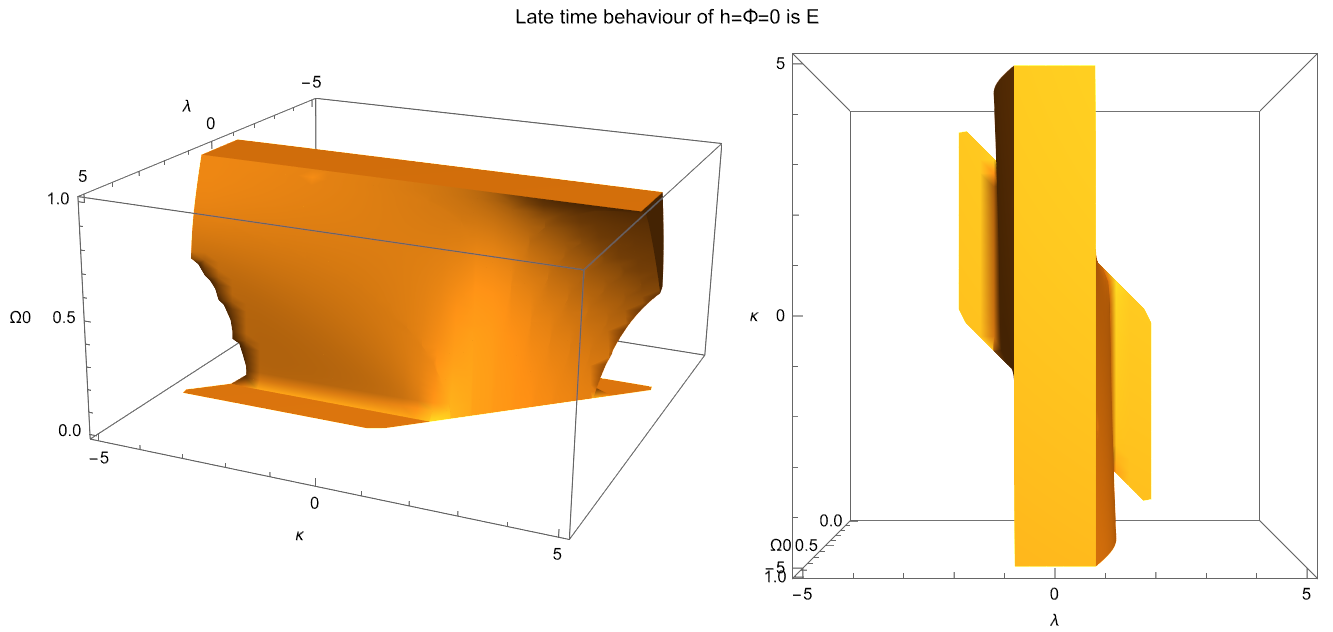}}
	\subfigure[\label{k+1crunching}]{\includegraphics[scale=0.4]{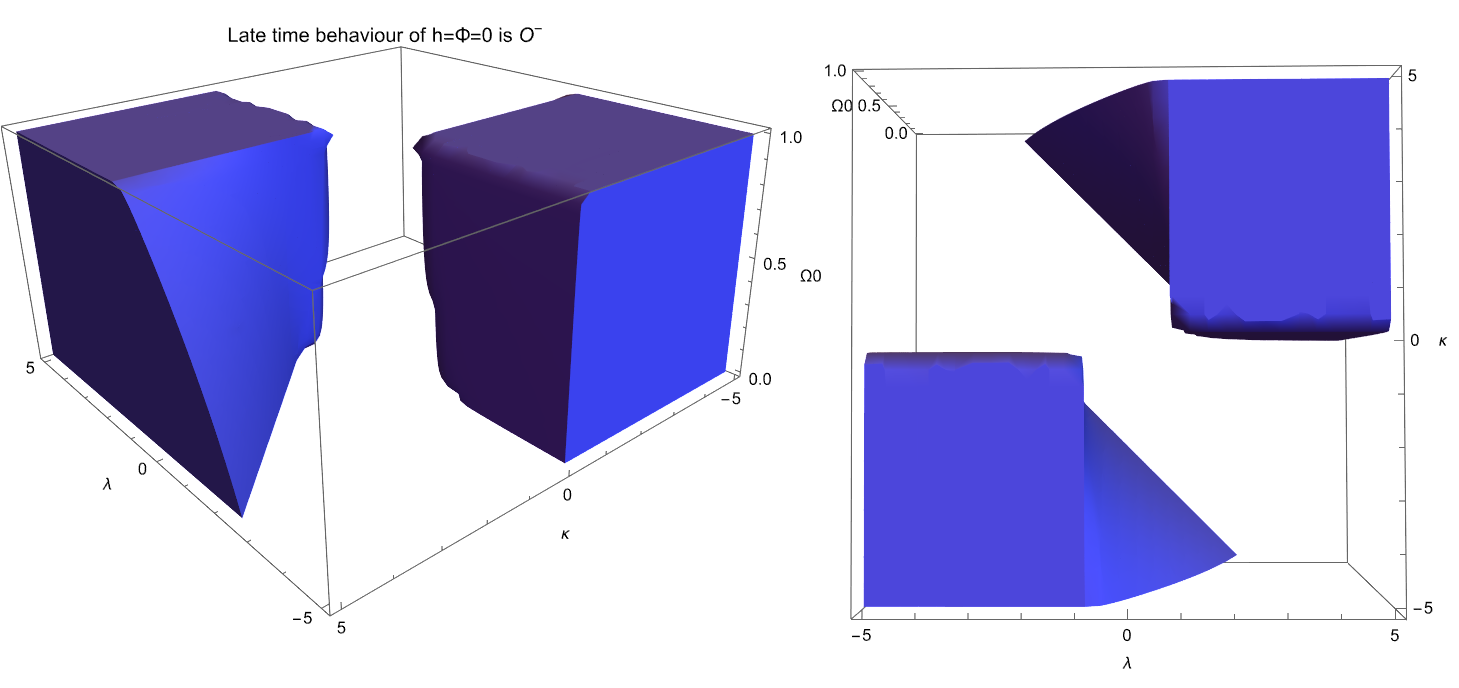}}
	\caption{Parameter values for which the late time cosmology of $h=\Phi=0$ is either $E$ (a) or $O^-$ (b), depending on varying initial value of $\Omega_K=\Omega0$.}
	\label{FIG3}
\end{figure}

Corresponding to the analysis in the flat case (see Fig. \ref{regions}) of the bifurcation from the flow $E\rightarrow F$ to $O^+\rightarrow O^-$ for the central region of the phase space, we make a similar computation by numerically integrating the equations \eqref{k+1sys1}-\eqref{k+1sys3} to sufficiently late times, and then checking whether either: $h\rightarrow 1$ and $\Phi+\lambda\rightarrow 0$, or $h\rightarrow 0$. This produces the regions in Fig. \ref{FIG3}.

\begin{figure}[h!]
	\centering
	\subfigure[\label{k+1comp2}]{\includegraphics[scale=0.55]{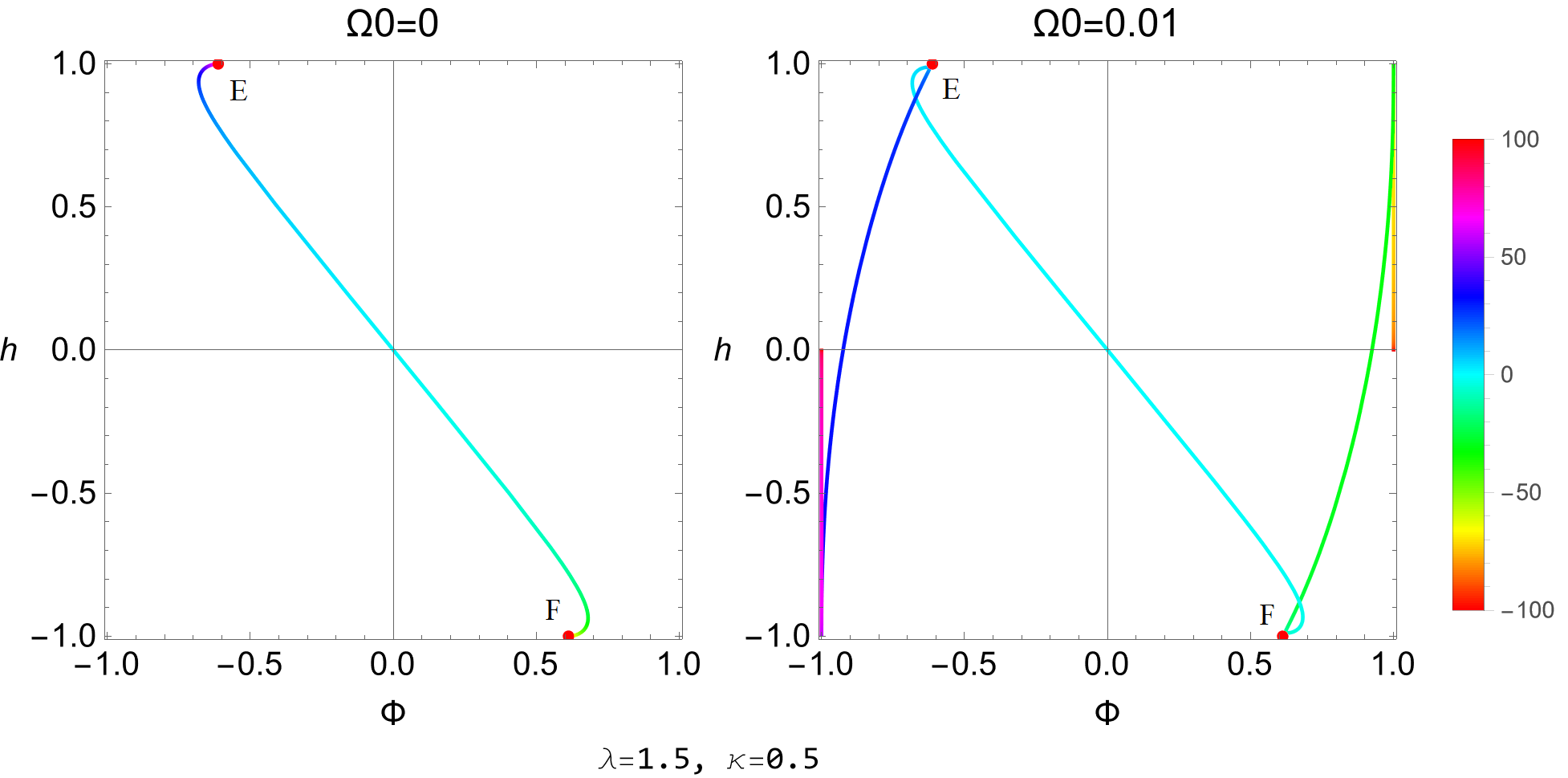}}
	\subfigure[\label{k+1comp1}]{\includegraphics[scale=0.4]{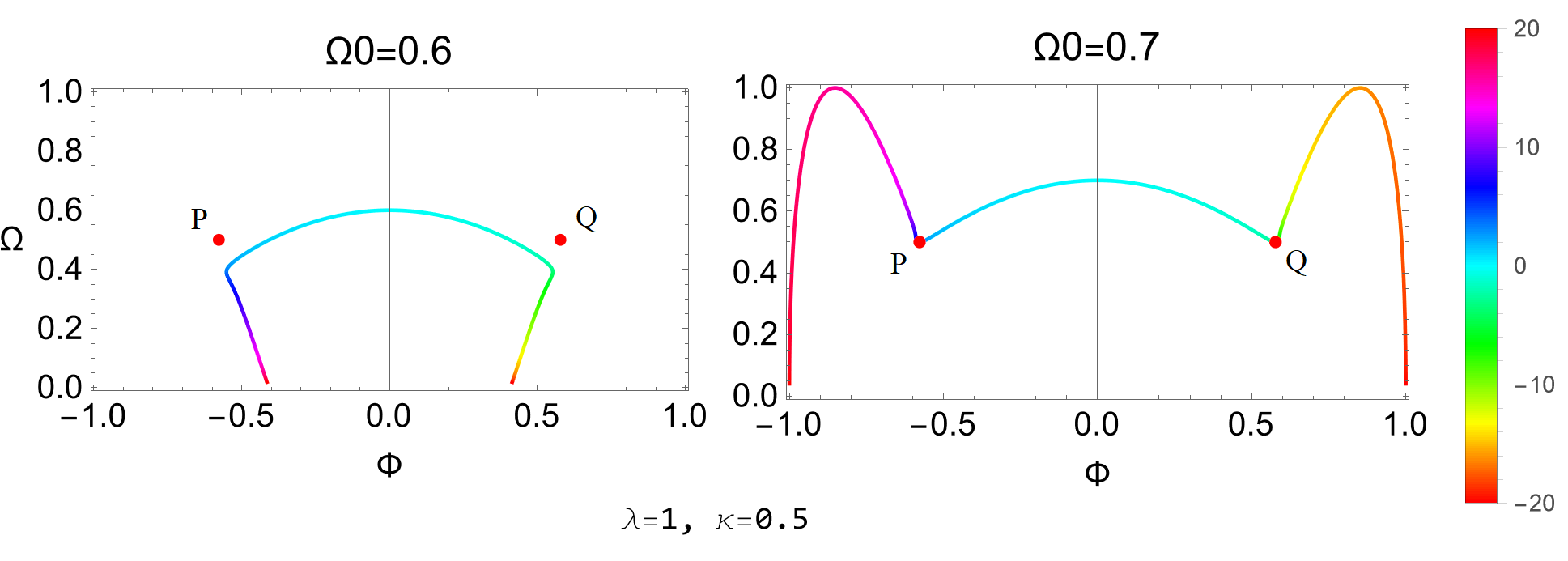}}
	\caption{Solutions of \eqref{k+1sys1}-\eqref{k+1sys3} demonstrating the interaction with saddles $E,F$, $P$ and $Q$.  In Fig. \ref{k+1comp2} solutions are shown for parameter values $\lambda=0.1.5,\kappa=0.5$; in the right panel with initial conditions $\Phi=h=\Omega_K=0$, and the solutions tend to $E (F)$ for early (late) times. However, as shown in the left panel, with $\Omega_K=0.01$ initially, $E$ and $F$ are saddle points in the three-dimensional dynamics, and solutions are deflected away, ultimately coming to the sink (source) $O^-(O^+)$ for early (late) times. In Fig. \ref{k+1comp1} parameters $\lambda=1,\kappa=0,5$ are used.  The left panel has initial conditions $\Phi=h=0$ and $\Omega_K=0.6$, and the saddles $P, Q$ deflect the solution downward towards $E(F)$ for early(late) times. The right panel has $\Omega_K=0.7$ initially, and now $Q, P$ deflect the solutions up, grazing the invariant surface ($\Omega_K=1-h^2$), before coming to $O^-(O^+)$ for early (late) times. The colour legend indicates the time-variable $\tau$ value along the solutions in each subplot.}
	\label{FIG3b}
\end{figure}

The flow in the invariant sets $\Omega_K=0,1-h^2$ can be investigated to reveal that the differences between $\Omega_K(0)=0$ and $\Omega_K(0)>0$. Substituting $\Omega_K=1-h^2$ into \eqref{k+1sys1}-\eqref{k+1sys3} gives the following projected system in the invariant surface
\begin{align}
    & \Phi^{\prime}= \sqrt{3} \left(\Phi ^2-1\right) \left( h \Phi + \lambda/\sqrt{6} \right), \label{max_curv_sys1} \\
    & h^{\prime}= (1-h^2)(1-2\Phi^2)/\sqrt{3} \label{max_curv_sys2}
\end{align}
which has equilibrium points at $(\Phi,h)=(\mp/\sqrt{3},\pm\lambda/\sqrt{2})$ (corresponding to $P,Q$), which are saddles for $\lambda^2<2$.  However, a similar polynomial division calculation as that done in the flat case (after line \eqref{flat_poly_div}) reveals that for $\Omega_K=1$, the saddle-connection bifurcation occurs at $\abs{\lambda}=\sqrt{2/3}$:
\begin{align}
    \left(\Phi/\sqrt{3}+\lambda h/\sqrt{2}\right)^\prime=& \nonumber \\
    \sqrt{3}\left(\Phi/\sqrt{3}+\lambda h/\sqrt{2}\right)&\left(-h + \Phi^2h - \sqrt{2/3}\lambda f + \lambda^2 h\right) +\frac{h^2}{\sqrt{6}}\lambda(2-3\lambda^2) \label{k+1_poly_div}
\end{align}
This corresponds to the vertical strip, which is the intersection of the region shown in Fig. \ref{k+1bouncing} with $\Omega=1$. For $\sqrt{2}<\abs{\lambda}<\sqrt{6}$, the region shown in Fig. \ref{k+1bouncing} is only for $\Omega_K(0)=0$ because, for these parameters, the equilibrium point $E$ is a sink in the sub-system $\Omega_K=0$, but it is a saddle in the three-dimensional dynamics. Thus for $\Omega_K(0)=0$, the late-time dynamics of $\Phi=h=0$ is towards $E$. Still, for any none-zero initial curvature $\Omega_K(0)=\Omega_0$, the saddle $E$ deflects the solution away, grazing the invariant surface $\Omega_K=1-h^2$ and coming near the saddle $D$, before being further deflected toward the sink $O^-$. This transition is shown in Fig. \ref{k+1comp2}. For $\abs{\lambda}<\sqrt{2}$, $E$ is a sink in the three-dimensional dynamics. The steep sides of the region's boundary in Fig. \ref{k+1bouncing} are due to interaction with the saddles $P, Q$, as shown in Fig. \ref{k+1comp1}.

\subsection{Dynamical systems formulation ($K=-1$)}
\label{Sect:IIC}
With $K=-1$, we  have that
\begin{equation}
    \bar{\chi}^2 = \frac{1}{2} \dot{\phi}^2 + V(\phi) + \frac{3}{a^2} + \rho_m=3 H^2+\frac{1}{2}e^{\kappa\,\phi}\dot{\psi}^2, \label{xi-lambda}
\end{equation}
and define dimensionless variables as 
\begin{align}
    & \bar{h}^2 = \frac{3 H^2}{\bar{\chi}^2}, \; \bar{\eta}^2= \frac{e^{\kappa \phi}\dot{\psi}^2}{2\bar{\chi}^2}, \;  \bar{\Phi}^2 = \frac{\dot{\phi}^2}{2 \bar{\chi}^2}, \; \bar{\Psi}= \frac{V(\phi)}{\bar{\chi}^2}, \; \bar{\Omega}_K=\frac{3}{a^2 \bar{\chi}^2}, \;  \bar{\Omega}= \frac{\rho_m}{\bar{\chi}^2}\label{NegativeNewvars}
\end{align}
which satisfy
\begin{equation}
    \bar{h}^2 + \bar{\eta}^2= \bar{\Phi}^2+ \bar{\Psi} + \bar{\Omega}_K +  \bar{\Omega}=1. \label{Negativeeq10}
\end{equation}
Using  \eqref{Negativeeq10}, we have bounded variables
\begin{equation}
    0 \leq \bar{h}^2, \bar{\Phi}^2, \bar{\eta}^2, \bar{\Psi}, \bar{\Omega}_K, \bar{\Omega} \leq 1.  
\end{equation}
Then, equations \eqref{0sp.05}, \eqref{0sp.06}, \eqref{0sp.07} and \eqref{matter} become
\begin{align}
	&\frac{\dot{H}}{\bar{\chi}^2}=\frac{1}{6} \left(\bar{\Omega}_K-3 \left(-\bar{\eta} ^2+\bar{h}^2+\bar{\Phi}^2+w \bar{\Omega} -\bar{\Psi}\right)\right), \label{IIIcurvhubble}\\
	&\frac{\ddot{\phi}}{\bar{\chi}^2}=-\kappa \bar{\eta}^2 -\sqrt{6} \bar{h}  \bar{\Phi} -\lambda \bar{\Psi},  \\
	&\frac{\ddot{\psi}}{\dot{\psi}\bar{\chi}}=-\sqrt{3} \bar{h}-\sqrt{2} \kappa  \bar{\Phi},\\
    &\frac{\dot{\rho_m}}{\bar{\chi}^2}=-\sqrt{3} (1+w) \bar{h} \bar{\chi}  \bar{\Omega}.
\end{align}
Now the new time derivative is again $d\bar{\tau}=\bar{\chi} d t$, and taking the derivatives of the variables with respect to $\bar{\tau}$ the field equations become
\begin{align}
	&{{\bar{\Phi}}}^\prime=-\frac{{\kappa\,\bar{\eta}} ^2 }{\sqrt{2}}-\sqrt{3} {\bar{h}} {\bar{\Phi}} -\frac{\lambda  \bar{\Psi} }{\sqrt{2}}-\frac{{\bar{\Phi}}  \dot{\bar{\chi} }}{\bar{\chi} ^2}, \\
	&{\bar{h}}^\prime=\frac{{\bar{\Omega}}_K-3 \left(-{\bar{\eta}} ^2+{\bar{h}}^2+{\bar{\Phi}} ^2+w {\bar{\Omega}} -\bar{\Psi} \right)}{2 \sqrt{3}}-\frac{{\bar{h}} \dot{\bar{\chi}}}{\bar{\chi} ^2}, \\
	&{\bar{\eta}}^\prime=-\frac{ \kappa\bar{\eta} {\bar{\Phi}} }{\sqrt{2}}-\sqrt{3} {\bar{\eta}}  {\bar{h}}-\frac{{\bar{\eta}}  \dot{\bar{\chi}}}{\bar{\chi} ^2},\\
	&\bar{\Psi}^\prime=\sqrt{2} \lambda  {\bar{\Phi}} 
   \bar{\Psi} -\frac{2 \bar{\Psi}  \dot{\bar{\chi} }}{\bar{\chi} ^2},\\
	&{\bar{\Omega}}_K^\prime=-\frac{2 {\bar{h}} {\bar{\Omega}}_K}{\sqrt{3}}-\frac{2 {\bar{\Omega}}_K \dot{\bar{\chi}}}{\bar{\chi} ^2},  \\
        &{\bar{\Omega}}^\prime=-\sqrt{3} (1+w) {\bar{h}} {\bar{\Omega}} -\frac{2 {\bar{\Omega}}  \dot{\bar{\chi} }}{\bar{\chi} ^2}.
\end{align}
Next, by substituting
\begin{align}
    & \dot{\bar{\chi}} =   -\sqrt{3} \bar{\chi} ^2 \left(\kappa\,{\bar{\eta}}^2{\bar{\Phi}}/\sqrt{6} +\bar{h} \left[{\bar{\Phi}}^2+ (1+w)\bar{\Omega}/2 +\bar{\Omega}_K/3\right]\right),\\
    & {\bar{\eta}}^2= 1- \bar{h}^2, \\
    & \bar{\Psi}  =1-{\bar{\Phi}}^2 -\bar{\Omega}-\bar{\Omega}_K,
\end{align}
we obtain 
\begin{align}
    {\bar{\Phi}}^\prime &=\kappa \bar{h}^2 (1-\bar{\Phi}^2)/\sqrt{2}+\sqrt{3}{\bar{h}}\bar{\Phi} \left({\bar{\Phi}}^2+ (1+w) {\bar{\Omega}}/2+ {\bar{\Omega}}_K/3-1\right) \nonumber \\
    & +\left[(\kappa +\lambda )({\bar{\Phi}} ^2-1)+\lambda ({\bar{\Omega}} + {\bar{\Omega}}_K)\right]/\sqrt{2}, \label{k-1sys1} \\ 
    {\bar{h}}^{\prime}  & =(\bar{h}^2-1)\left(-\kappa\bar{h}{\bar{\Phi}}/\sqrt{2} +\sqrt{3} \left[{\bar{\Phi}} ^2+(1+w) {\bar{\Omega}}/2 +{\bar{\Omega}}_K/3-1\right]\right), \label{k-1sys2}\\
    {\bar{\Omega}}_K^{\prime}  & = 2\bar{\Omega}_K\left( -\kappa(\bar{h}^2-1){\bar{\Phi}}/\sqrt{2}+\sqrt{3}{\bar{h}}\left[ {\bar{\Phi}} ^2 + (1+w) {\bar{\Omega}}/2 -{1}/{3}+\bar{\Omega}_K/3\right]\right), \label{k-1sys4} \\
    {\bar{\Omega}}^{\prime} & =2\bar{\Omega}\left(-\kappa( {\bar{h}}^2 -1){\bar{\Phi}}/\sqrt{2} +\sqrt{3}{\bar{h}} \left[ {\bar{\Phi}}^2 + (1+w)(\bar{\Omega}-1)/2 + {\bar{\Omega}}_K/3\right]\right), \label{k-1sys5}
\end{align}
with the auxiliary equations
\begin{align}
  {\bar{\eta}}^{\prime} &= \sqrt{1-{\bar{h}}^2-{\bar{\Omega}}_K} \left(\sqrt{3} {\bar{h}} \left[ {\bar{\Phi}}^2+ (1+w){\bar{\Omega}}/2 -1\right]- \kappa {\bar{\Phi}} \left({\bar{h}}^2 +{\bar{\Omega}}_K\right)/\sqrt{2}\right),\\
 \Psi^{\prime} &= 2\left(1-{\bar{\Phi}} ^2-{\bar{\Omega}-\bar{\Omega}_K} \right)
   \left( \left[\kappa+\lambda- \kappa \bar{h}^2\right]\bar{\Phi}/\sqrt{2} +\sqrt{3} {\bar{h}} \left[ {\bar{\Phi}}^2+(1+w) {\bar{\Omega}}/2+\bar{\Omega}_K/3\right]\right)  . 
\end{align}
We can find the inflationary condition with these variables where $q$, the deceleration parameter, is negative:
\begin{align}
    q = 2 + \frac{6 \bar{\Phi} ^2+3 (w+1) \bar{\Omega} +2 \bar{\Omega}_{K}-6}{2 \bar{h}^2}<0.
\end{align}

\begin{table}[]
    \centering
\resizebox{\textwidth}{!}{   
  \begin{tabular}{|ccccccccc|}\hline 
Label & $\bar{\Phi}$ & $\bar{h}$ & $\bar{\Omega}$ & $\bar{\Omega}_K$ & $k_1$ & $k_2$ & $k_3$ & $k_4$\\\hline 
$A$ & $1$ & $1$ & $0$ &  $0$ & $\frac{4}{\sqrt{3}}$ & $-\sqrt{3} (w-1)$ & $-\sqrt{2} \kappa $ & $\sqrt{2} \lambda +2 \sqrt{3}$ \\\hline
$B$ & $-1$ & $1$ & $0$ & $0$ &  $\frac{4}{\sqrt{3}}$ & $-\sqrt{3} (w-1)$ & $\sqrt{2} \kappa $ & $ 2 \sqrt{3}-\sqrt{2} \lambda$ \\\hline
$C$ & $1$ & $-1$ & $0$ & $0$ & $-\frac{4}{\sqrt{3}}$ & $\sqrt{3} (w-1)$ & $-\sqrt{2} \kappa$ & $ \sqrt{2} \lambda -2 \sqrt{3}$ \\\hline
$D$ & $-1$ & $-1$ & $0$ & $0$ & $-\frac{4}{\sqrt{3}}$ & $\sqrt{3} (w-1)$ & $\sqrt{2} \kappa$ & $-\sqrt{2} \lambda -2 \sqrt{3}$ \\\hline
$E$ & $-\frac{\lambda }{\sqrt{6}}$ & $1$ & $0$ & $0$ & $\frac{\lambda ^2-3 w-3}{\sqrt{3}}$ & $\frac{\lambda ^2-2}{\sqrt{3}}$ & $\frac{\lambda ^2-6}{2 \sqrt{3}}$ & $\frac{\lambda (\kappa +\lambda )-6}{\sqrt{3}}$ \\\hline
$F$ & $\frac{\lambda }{\sqrt{6}}$ & $-1$ & $0$ & $0$ & $\frac{-\lambda ^2+3 w+3}{\sqrt{3}}$ & $-\frac{\lambda ^2-2}{\sqrt{3}}$ & $-\frac{\lambda ^2-6}{2\sqrt{3}}$ & $\frac{6-\lambda  (\kappa +\lambda )}{\sqrt{3}}$\\\hline
$G$ & $\frac{1}{\sqrt{\frac{1}{6} \kappa  (\kappa +\lambda )+1}}$ & $-\frac{\kappa +\lambda }{\sqrt{\kappa  (\kappa +\lambda)+6}}$ & $ 0$ & $0$ & $\frac{2 (\kappa -2 \lambda )}{\sqrt{3} \sqrt{\kappa  (\kappa +\lambda )+6}}$ & $\frac{\kappa  (w+1)+\lambda (w-1)}{\sqrt{\frac{1}{3} \kappa  (\kappa +\lambda )+2}}$ & $\frac{\sqrt{3} \kappa -\sqrt{-\kappa \left(4 \left(\kappa ^2-6\right) \lambda +8 \kappa  \lambda ^2-27 \kappa +4 \lambda ^3\right)}}{2 \sqrt{\kappa  (\kappa +\lambda)+6}}$ & $\frac{\sqrt{3} \kappa +\sqrt{-\kappa \left(4 \left(\kappa ^2-6\right) \lambda +8 \kappa  \lambda ^2-27 \kappa +4 \lambda ^3\right)}}{2 \sqrt{\kappa  (\kappa +\lambda )+6}}$ \\\hline 
$H$ & $-\frac{1}{\sqrt{\frac{1}{6} \kappa  (\kappa +\lambda )+1}}$ & $\frac{\kappa +\lambda }{\sqrt{\kappa  (\kappa +\lambda)+6}}$ & $0$ & $0$ & $\frac{4 \lambda -2 \kappa }{\sqrt{3} \sqrt{\kappa  (\kappa +\lambda )+6}}$ & $-\frac{\kappa  (w+1)+\lambda 
   (w-1)}{\sqrt{\frac{1}{3} \kappa  (\kappa +\lambda )+2}}$ & $-\frac{\sqrt{3} \kappa +\sqrt{-\kappa \left(4\left(\kappa ^2-6\right) \lambda +8 \kappa  \lambda ^2-27 \kappa +4 \lambda ^3\right)}}{2 \sqrt{\kappa  (\kappa +\lambda)+6}}$ & $\frac{-\sqrt{3} \kappa +\sqrt{-\kappa \left(4 \left(\kappa ^2-6\right) \lambda +8 \kappa  \lambda ^2-27\kappa +4 \lambda ^3\right)}}{2 \sqrt{\kappa  (\kappa +\lambda )+6}}$ \\\hline
   $O^{+}$ & $1$ & $0$ & $0$ & $0$ & $\frac{\kappa }{\sqrt{2}}$ & $\sqrt{2} \kappa$ & $\sqrt{2} \kappa$ & $\sqrt{2} (\kappa +\lambda )$\\\hline
$O^{-}$ & $-1$ & $0$ & $0$ & $0$ & $-\sqrt{2} \kappa$ & $-\sqrt{2} \kappa$ & $-\frac{\kappa }{\sqrt{2}}$ & $-\sqrt{2} (\kappa +\lambda )$\\\hline
$I$ & $0$ & $-1$ & $0$ & $1$ & $-\sqrt{3} (w-1)$ & $-\frac{1}{2} \sqrt{3} (w-1)$ & $-\sqrt{3} (w+1)$ & $-\frac{3 w+1}{\sqrt{3}}$\\\hline
$J$ & $0$ & $1$ & $0$ & $1)$ & $\frac{1}{2} \sqrt{3} (w-1)$ & $\sqrt{3} (w-1)$ & $\sqrt{3} (w+1)$ & $\frac{3 w+1}{\sqrt{3}}$\\\hline
$K$ & $\frac{\sqrt{\frac{3}{2}} (w+1)}{\lambda }$ & $-1$ & $0$ & $1-\frac{3 (w+1)}{\lambda ^2}$ & $-\frac{3 w+1}{\sqrt{3}}$ & $-\frac{\sqrt{3} (\kappa  (w+1)+\lambda  (w-1))}{\lambda }$ & $\frac{\sqrt{3} \left(\lambda(1-w) +\sqrt{w-1} \sqrt{\lambda ^2 (9 w+7)-24 (w+1)^2}\right)}{4 \lambda }$ & $ -\frac{\sqrt{3} \left(\sqrt{w-1}\sqrt{\lambda ^2 (9 w+7)-24 (w+1)^2}+\lambda  (w-1)\right)}{4 \lambda }$\\\hline
$L$ & $-\frac{\sqrt{\frac{3}{2}} (w+1)}{\lambda }$ & $1$ & $0$ & $1-\frac{3 (w+1)}{\lambda ^2}$ & $\frac{3 w+1}{\sqrt{3}}$ & $\frac{\sqrt{3} (\kappa  (w+1)+\lambda  (w-1))}{\lambda }$ & $-\frac{\sqrt{3} \left(\lambda (1-w)  +\sqrt{w-1} \sqrt{\lambda ^2 (9 w+7)-24 (w+1)^2}\right)}{4 \lambda }$ & $\frac{\sqrt{3} \left(\sqrt{w-1}    \sqrt{\lambda ^2 (9 w+7)-24 (w+1)^2}+\lambda  (w-1)\right)}{4 \lambda }$\\\hline
$M$ & $\frac{w-1}{\sqrt{\frac{2 \kappa ^2}{3}+(w-1)^2}}$ & $\frac{\kappa }{\sqrt{\kappa ^2+\frac{3}{2} (w-1)^2}}$ & $0$ & $\frac{2 \kappa ^2}{2 \kappa ^2+3 (w-1)^2}$ & $\frac{\kappa  (3 w+1)}{\sqrt{3 \kappa ^2+\frac{9}{2} (w-1)^2}}$ & $\frac{\mu_{11}}{6 \left(2 \kappa ^2+3(w-1)^2\right)^3}$ & $\frac{\mu_{12}}{6 \left(2 \kappa ^2+3 (w-1)^2\right)^3}$ & $\frac{\mu_{13}}{6\left(2 \kappa ^2+3 (w-1)^2\right)^3}$ \\\hline
$N$ & $-\frac{w-1}{\sqrt{\frac{2 \kappa ^2}{3}+(w-1)^2}}$ & $-\frac{\kappa }{\sqrt{\kappa ^2+\frac{3}{2} (w-1)^2}}$ & $0$ & $\frac{2 \kappa ^2}{2 \kappa ^2+3 (w-1)^2}$ & $-\frac{\kappa  (3 w+1)}{\sqrt{3 \kappa ^2+\frac{9}{2} (w-1)^2}}$ & $ -\frac{\mu_{11}}{6 \left(2 \kappa ^2+3(w-1)^2\right)^3}$ & $-\frac{\mu_{12}}{6 \left(2 \kappa ^2+3 (w-1)^2\right)^3}$ & $-\frac{\mu_{13}}{6\left(2 \kappa ^2+3 (w-1)^2\right)^3}$ \\\hline
$P$ & $-\frac{\sqrt{\frac{2}{3}}}{\lambda }$ & $1$ & $1-\frac{2}{\lambda ^2}$ & $0$ & $-\frac{3 w+1}{\sqrt{3}}$ & $\frac{2 (\kappa -2 \lambda )}{\sqrt{3} \lambda }$ & $ -\frac{\lambda ^3+\sqrt{8 \lambda ^4-3 \lambda ^6}}{\sqrt{3} \lambda ^3}$ & $ \frac{\sqrt{8 \lambda ^4-3 \lambda ^6}-\lambda ^3}{\sqrt{3} \lambda^3}$\\ \hline
$Q$ & $\frac{\sqrt{\frac{2}{3}}}{\lambda }$ & $-1$ & $1-\frac{2}{\lambda ^2}$ & $0$ & $\frac{3 w+1}{\sqrt{3}}$ & $ -\frac{2 (\kappa -2 \lambda )}{\sqrt{3} \lambda }$ & $ \frac{\lambda ^3+\sqrt{8 \lambda  ^4-3 \lambda ^6}}{\sqrt{3} \lambda ^3}$ & $ \frac{\sqrt{8 \lambda ^4-3 \lambda ^6}-\lambda ^3}{\sqrt{3} \lambda^3}$ \\ \hline
$R$ & $-\frac{4}{\sqrt{6 \kappa ^2+16}}$ & $\frac{\kappa }{\sqrt{\kappa ^2+\frac{8}{3}}}$ & $\frac{3 \kappa ^2}{3 \kappa ^2+8}$ & $0$ & $-\frac{\kappa  (3 w+1)}{\sqrt{3 \kappa ^2+8}}$ & $\frac{\mu_{2 1}}{6 \left(3 \kappa ^2+8\right)^3}$ & $\frac{\mu_{2 2}}{6 \left(3 \kappa ^2+8\right)^3}$ & $\frac{\mu_{2 3}}{6 \left(3 \kappa2+8\right)^3}$\\\hline
$S$ & $\frac{4}{\sqrt{6 \kappa ^2+16}}$ & $ -\frac{\kappa }{\sqrt{\kappa ^2+\frac{8}{3}}}$ & $\frac{3 \kappa ^2}{3 \kappa ^2+8}$ & $ 0$ & $\frac{\kappa  (3 w+1)}{\sqrt{3 \kappa ^2+8}}$ & $-\frac{\mu_{2 1}}{6 \left(3 \kappa^2+8\right)^3}$ & $ -\frac{\mu_{2 2}}{6 \left(3 \kappa ^2+8\right)^3}$ & $ -\frac{\mu_{2 3}}{6 \left(3 \kappa ^2+8\right)^3}$\\\hline
$T$ & $0$ & $1$ & $1$ & $ 0$ & $-\frac{4}{\sqrt{3}}$ & $-\frac{2}{\sqrt{3}}$ & $\frac{2}{\sqrt{3}}$ & $-\frac{3 w+1}{\sqrt{3}}$\\\hline
$U$ & $0$ & $-1$ & $1$ & $0$ & $\frac{4}{\sqrt{3}}$ & $-\frac{2}{\sqrt{3}}$ & $\frac{2}{\sqrt{3}}$ & $\frac{3 w+1}{\sqrt{3}}$\\ \hline
$MC$ & $0$ & $ 0$ & $ \frac{3 (\kappa  (w+1)+\lambda  (w-1))}{\lambda +3 \lambda  w}$ & $ \frac{4 \lambda -2 \kappa }{\lambda +3\lambda  w}$ & $-\mu_1$ & $ \mu_1$ & $-\mu_2$ &  $\mu_2$\\\hline
\end{tabular}}
    \caption{Equilibrium points of the system \eqref{k-1sys1}, \eqref{k-1sys2}, \eqref{k-1sys4} and \eqref{k-1sys5} and their corresponding eigenvalues.}
    \label{tab:K=-1}
\end{table}

The equilibrium points of the system \eqref{k-1sys1}, \eqref{k-1sys2}, \eqref{k-1sys4} and \eqref{k-1sys5} are $(\bar{\Phi} , \bar{h}, \bar{\Omega} _K, \bar{\Omega})$ (see Tab.  \ref{tab:K=-1}):

\begin{itemize}
     
      \item $A: (1, 1, 0, 0)$, with eigenvalues $\left\{\frac{4}{\sqrt{3}},-\sqrt{3} (w-1),-\sqrt{2} \kappa,\sqrt{2} \lambda +2 \sqrt{3}\right\}$. The stability conditions are the same as for $K=+1$ (see \S  \ref{Subsect:IIB}). The deceleration parameter is $q=2$. Therefore, it represents a stiff fluid solution. It never inflates.

       \item $B: (-1, 1, 0, 0)$, with eigenvalues $\left\{\frac{4}{\sqrt{3}},-\sqrt{3} (w-1),\sqrt{2} \kappa, 2 \sqrt{3}-\sqrt{2} \lambda \right\}$. The stability conditions are the same as for $K=+1$ (see \S  \ref{Subsect:IIB}). The deceleration parameter is $q=2$. Therefore, it represents a stiff fluid solution. It never inflates.
       
       \item $C: (1, -1, 0, 0)$, with eigenvalues $\left\{-\frac{4}{\sqrt{3}},\sqrt{3} (w-1),-\sqrt{2} \kappa, \sqrt{2} \lambda -2 \sqrt{3}\right\}$. The stability conditions are the same as for $K=+1$ (see \S  \ref{Subsect:IIB}). The deceleration parameter is $q=2$. The deceleration parameter is $q=2$. Therefore, it represents a stiff fluid solution. It never inflates.
      
       \item  $D: (-1, -1, 0, 0)$, with eigenvalues $\left\{-\frac{4}{\sqrt{3}},\sqrt{3} (w-1),\sqrt{2} \kappa, -\sqrt{2} \lambda -2 \sqrt{3}\right\}$. The stability conditions are the same as for $K=+1$ (see \S  \ref{Subsect:IIB}). The deceleration parameter is $q=2$. Therefore, it represents a stiff fluid solution. It never inflates.

       \item $E: \left( -\frac{\lambda }{\sqrt{6}}, 1, 0, 0\right)$, with eigenvalues are $\left\{\frac{\lambda ^2-3 w-3}{\sqrt{3}},\frac{\lambda ^2-2}{\sqrt{3}},\frac{\lambda ^2-6}{2 \sqrt{3}},\frac{\lambda (\kappa +\lambda )-6}{\sqrt{3}}\right\}$. The stability conditions are the same as for $K=+1$ (see \S  \ref{Subsect:IIB}). The deceleration parameter is $q= \frac{1}{2} \left(\lambda ^2-2\right)$. 
   It is an inflationary solution for $\lambda ^2<2$.

       \item $F: \left( \frac{\lambda }{\sqrt{6}}, -1, 0, 0\right)$, with eigenvalues $\left\{\frac{-\lambda ^2+3 w+3}{\sqrt{3}},-\frac{\lambda ^2-2}{\sqrt{3}},-\frac{\lambda ^2-6}{2
   \sqrt{3}},\frac{6-\lambda  (\kappa +\lambda )}{\sqrt{3}}\right\}$. The stability conditions are the same as for $K=+1$ (see \S  \ref{Subsect:IIB}). The deceleration parameter is $q= \frac{1}{2} \left(\lambda ^2-2\right)$. 
   It is an inflationary solution for $\lambda ^2<2$.

        \item $G: \left( \frac{1}{\sqrt{\frac{1}{6} \kappa  (\kappa +\lambda )+1}}, -\frac{\kappa +\lambda }{\sqrt{\kappa  (\kappa +\lambda)+6}}, 0, 0\right)$, with eigenvalues \newline
\begin{footnotesize}
$\left\{\frac{2 (\kappa -2 \lambda )}{\sqrt{3} \sqrt{\kappa  (\kappa +\lambda )+6}},\frac{\kappa  (w+1)+\lambda 
   (w-1)}{\sqrt{\frac{1}{3} \kappa  (\kappa +\lambda )+2}},\frac{\sqrt{3} \kappa -\sqrt{-\kappa \left(4
   \left(\kappa ^2-6\right) \lambda +8 \kappa  \lambda ^2-27 \kappa +4 \lambda ^3\right)}}{2 \sqrt{\kappa  (\kappa +\lambda
   )+6}},\frac{\sqrt{3} \kappa +\sqrt{-\kappa \left(4 \left(\kappa ^2-6\right) \lambda +8 \kappa  \lambda ^2-27
   \kappa +4 \lambda ^3\right)}}{2 \sqrt{\kappa  (\kappa +\lambda )+6}}\right\}$
  \end{footnotesize}  
   The stability conditions are the same as for $K=+1$ (see \S  \ref{Subsect:IIB}). The deceleration parameter is $q= 2-\frac{3 \kappa }{\kappa +\lambda }$. It is an inflationary solution for $\frac{\kappa }{\kappa +\lambda }>\frac{2}{3}$. 

      \item $H: \left( -\frac{1}{\sqrt{\frac{1}{6} \kappa  (\kappa +\lambda )+1}}, \frac{\kappa +\lambda }{\sqrt{\kappa  (\kappa +\lambda)+6}}, 0, 0 \right)$, with eigenvalues \newline 
\begin{footnotesize}
$\left\{\frac{4 \lambda -2 \kappa }{\sqrt{3} \sqrt{\kappa  (\kappa +\lambda )+6}},-\frac{\kappa  (w+1)+\lambda 
   (w-1)}{\sqrt{\frac{1}{3} \kappa  (\kappa +\lambda )+2}},-\frac{\sqrt{3} \kappa +\sqrt{-\kappa \left(4
   \left(\kappa ^2-6\right) \lambda +8 \kappa  \lambda ^2-27 \kappa +4 \lambda ^3\right)}}{2 \sqrt{\kappa  (\kappa +\lambda
   )+6}},\frac{-\sqrt{3} \kappa +\sqrt{-\kappa \left(4 \left(\kappa ^2-6\right) \lambda +8 \kappa  \lambda ^2-27
   \kappa +4 \lambda ^3\right)}}{2 \sqrt{\kappa  (\kappa +\lambda )+6}}\right\}$.
\end{footnotesize} 
The stability conditions are the same as for $K=+1$ (see \S  \ref{Subsect:IIB}). The deceleration parameter is $q= 2-\frac{3 \kappa }{\kappa +\lambda }$. It is an inflationary solution for $\frac{\kappa }{\kappa +\lambda }>\frac{2}{3}$. 
  
        \item $O^{+}: (1, 0, 0, 0)$, with eigenvalues $\left\{\frac{\kappa }{\sqrt{2}},\sqrt{2} \kappa ,\sqrt{2} \kappa ,\sqrt{2} (\kappa +\lambda )\right\}$. The stability conditions are the same as for $K=+1$ (see \S  \ref{Subsect:IIB}). The deceleration parameter is $q=2$. Therefore, it represents a stiff fluid solution. It never inflates.

        \item $O^{-}: (-1, 0, 0, 0)$, with eigenvalues $\left\{-\sqrt{2} \kappa ,-\sqrt{2} \kappa ,-\frac{\kappa }{\sqrt{2}},-\sqrt{2} (\kappa +\lambda )\right\}$. The stability conditions are the same as for $K=+1$ (see \S  \ref{Subsect:IIB}). The deceleration parameter is $q=2$. Therefore, it represents a stiff fluid solution. It never inflates.

        \item $I: (0, -1, 0, 1)$, with eigenvalues $\left\{-\sqrt{3} (w-1),-\frac{1}{2} \sqrt{3} (w-1),-\sqrt{3} (w+1),-\frac{3 w+1}{\sqrt{3}}\right\}$. The stability conditions are the same as for $K=+1$ (see \S  \ref{Subsect:IIB}). The deceleration parameter is $q=\frac{1}{2} (3 w+1)$. It is an inflationary solution for $w<-\frac{1}{3}$. 
        
        \item $J: (0, 1, 0, 1)$, with eigenvalues $\left\{\frac{1}{2} \sqrt{3} (w-1),\sqrt{3} (w-1),\sqrt{3} (w+1),\frac{3 w+1}{\sqrt{3}}\right\}$. The stability conditions are the same as for $K=+1$ (see \S  \ref{Subsect:IIB}). The deceleration parameter is $q=\frac{1}{2} (3 w+1)$. It is an inflationary solution for $w<-\frac{1}{3}$.
        
        \item $K: \left( \frac{\sqrt{\frac{3}{2}} (w+1)}{\lambda }, -1, 0, 1-\frac{3 (w+1)}{\lambda ^2}\right)$, with eigenvalues \newline 
     \begin{footnotesize}
     $\left\{-\frac{3 w+1}{\sqrt{3}}, -\frac{\sqrt{3} (\kappa  (w+1)+\lambda  (w-1))}{\lambda },\frac{\sqrt{3} \left(\lambda(1-w) +\sqrt{w-1} \sqrt{\lambda ^2 (9 w+7)-24 (w+1)^2}\right)}{4 \lambda }, -\frac{\sqrt{3} \left(\sqrt{w-1}
   \sqrt{\lambda ^2 (9 w+7)-24 (w+1)^2}+\lambda  (w-1)\right)}{4 \lambda }\right\}$.
   \end{footnotesize} 
   The stability conditions are the same as for $K=+1$ (see \S  \ref{Subsect:IIB}). The deceleration parameter is $q=\frac{1}{2} (3 w+1)$. It is an inflationary solution for $w<-\frac{1}{3}$. 

       \item $L: \left( -\frac{\sqrt{\frac{3}{2}} (w+1)}{\lambda }, 1, 0, 1-\frac{3 (w+1)}{\lambda ^2}\right)$, with eigenvalues \newline 
          \begin{footnotesize} 
          $\left\{\frac{3 w+1}{\sqrt{3}},\frac{\sqrt{3} (\kappa  (w+1)+\lambda  (w-1))}{\lambda },-\frac{\sqrt{3} \left(\lambda (1-w)  +\sqrt{w-1} \sqrt{\lambda ^2 (9 w+7)-24 (w+1)^2}\right)}{4 \lambda },\frac{\sqrt{3} \left(\sqrt{w-1}    \sqrt{\lambda ^2 (9 w+7)-24 (w+1)^2}+\lambda  (w-1)\right)}{4 \lambda }\right\}$.
      \end{footnotesize} 
     The stability conditions are the same as for $K=+1$ (see \S  \ref{Subsect:IIB}). The deceleration parameter is $q= \frac{1}{2} (3 w+1)$. It is an inflationary solution for $w<-\frac{1}{3}$. 

       \item $M, N: \left(\pm\frac{w-1}{\sqrt{\frac{2 \kappa ^2}{3}+(w-1)^2}}, \pm \frac{\kappa }{\sqrt{\kappa ^2+\frac{3}{2} (w-1)^2}}, 0,\frac{2 \kappa ^2}{2 \kappa ^2+3 (w-1)^2}\right)$. The eigenvalues are \newline $\left\{\pm \frac{\kappa  (3 w+1)}{\sqrt{3 \kappa ^2+\frac{9}{2} (w-1)^2}}, \pm \frac{\mu_{11}}{6 \left(2 \kappa ^2+3
   (w-1)^2\right)^3}, \pm \frac{\mu_{12}}{6 \left(2 \kappa ^2+3 (w-1)^2\right)^3}, \pm \frac{\mu_{13}}{6
   \left(2 \kappa ^2+3 (w-1)^2\right)^3}\right\}$, where $\mu_{1 j}, j=1,2,3$ are the roots of 
       $P_1(\mu):= \mu^3-3 \sqrt{6} \mu^2 \left(2 \kappa ^2+3 (w-1)^2\right)^{5/2} (\kappa +3 \kappa  w+2 \lambda 
   (w-1))+108 \mu \kappa  (w-1) \left(2 \kappa ^2+3 (w-1)^2\right)^5 (2 \kappa +\lambda  (w-1))+648 \sqrt{6}
   \kappa ^2 (w-1)^2 \left(2 \kappa ^2+3 (w-1)^2\right)^{15/2} (\kappa  (w+1)+\lambda  (w-1))$. The deceleration parameter is  $q=\frac{1}{2} (3 w+1)$. They are inflationary solutions for $w<-\frac{1}{3}$. 
   The stability of the points $M$ and $N$ (which have opposite dynamical behaviours) is examined numerically. For dust matter ($w=0$), the signs of the real parts of the three complicated eigenvalues of $M$ are represented in Fig.  \ref{fig:MN-K-minus1}. Therefore, for dust, the points are saddles. 

   \begin{figure}
    \centering
    \includegraphics[scale=0.7]{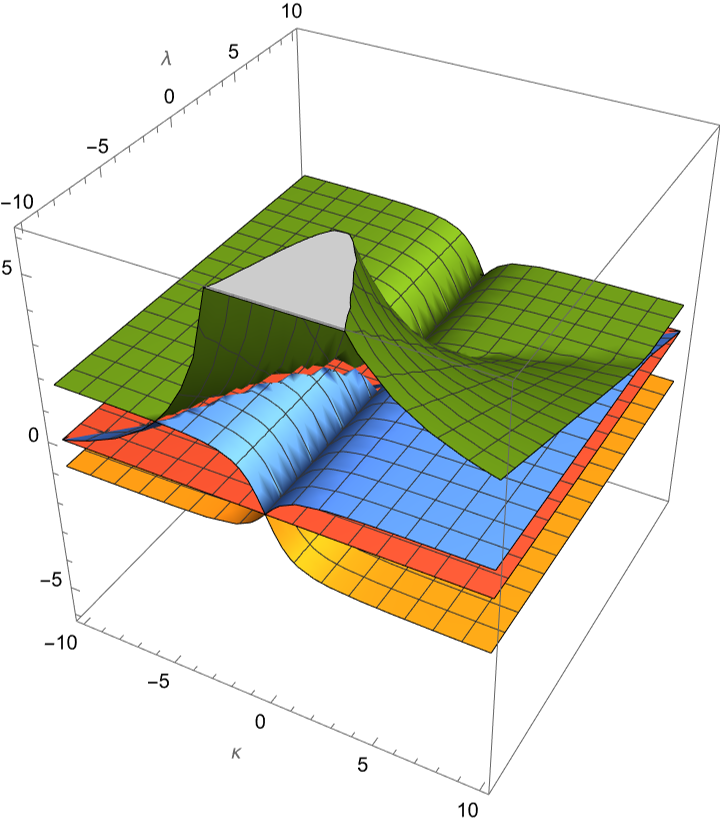}
    \caption{Real parts of the eigenvalues $\left\{\frac{\mu_{11}}{6 \left(2 \kappa ^2+3
   (w-1)^2\right)^3}, \frac{\mu_{12}}{6 \left(2 \kappa ^2+3 (w-1)^2\right)^3},  \frac{\mu_{13}}{6
   \left(2 \kappa ^2+3 (w-1)^2\right)^3}\right\}$, where $\mu_{1j}, j=1,2,3$ are the
roots of the  polynomial 
$P_1(\mu):= \mu^3-3 \sqrt{6} \mu^2 \left(2 \kappa ^2+3 (w-1)^2\right)^{5/2} (\kappa +3 \kappa  w+2 \lambda 
   (w-1))+108 \mu \kappa  (w-1) \left(2 \kappa ^2+3 (w-1)^2\right)^5 (2 \kappa +\lambda  (w-1))+648 \sqrt{6}
   \kappa ^2 (w-1)^2 \left(2 \kappa ^2+3 (w-1)^2\right)^{15/2} (\kappa  (w+1)+\lambda  (w-1))$ related to $M$, compared with zero for dust. This diagram shows its saddle nature.}
    \label{fig:MN-K-minus1}
\end{figure}

   \item $P: \left(-\frac{\sqrt{\frac{2}{3}}}{\lambda }, 1, 1-\frac{2}{\lambda ^2}, 0\right)$, with eigenvalues $\left\{-\frac{3 w+1}{\sqrt{3}},\frac{2 (\kappa -2 \lambda )}{\sqrt{3} \lambda }, -\frac{\lambda+ \sqrt{8 -3 \lambda ^2}}{\sqrt{3} \lambda}, -\frac{\lambda- \sqrt{8 -3 \lambda ^2}}{\sqrt{3} \lambda}\right\}$. $P$ is nonhyperbolic for $\{\lambda \neq 0, \kappa =2 \lambda \}\cup \left\{w=-\frac{1}{3} \right\} \cup\{\lambda =-\sqrt{2}\} \cup \{\lambda =\sqrt{2}\}$, is a sink for $\left\{-\frac{1}{3}<w<1, \lambda <-\sqrt{2}, \kappa >2 \lambda \right\}\cup\left\{-\frac{1}{3}<w<1, \lambda >\sqrt{2}, \kappa <2 \lambda \right\}$. It is a saddle otherwise. The deceleration parameter is $q=0$. It never inflates. 

     \item $Q: \left( \frac{\sqrt{\frac{2}{3}}}{\lambda }, -1, 1-\frac{2}{\lambda ^2}, 0\right)$, with eigenvalues $\left\{\frac{3 w+1}{\sqrt{3}}, -\frac{2 (\kappa -2 \lambda )}{\sqrt{3} \lambda }, \frac{\lambda+ \sqrt{8 -3 \lambda ^2}}{\sqrt{3} \lambda}, \frac{\lambda- \sqrt{8 -3 \lambda ^2}}{\sqrt{3} \lambda}\right\}$. $Q$ is nonhyperbolic for $\{\lambda \neq 0, \kappa =2 \lambda \}\cup \left\{w=-\frac{1}{3} \right\} \cup\{\lambda =-\sqrt{2}\} \cup \{\lambda =\sqrt{2}\}$, is a source for $\left\{-\frac{1}{3}<w<1, \lambda <-\sqrt{2}, \kappa >2 \lambda \right\}\cup\left\{-\frac{1}{3}<w<1, \lambda >\sqrt{2}, \kappa <2 \lambda \right\}$. It is a saddle otherwise. The deceleration parameter is $q=0$. It never inflates.

      \item $R, S: \left(\mp\frac{4}{\sqrt{6 \kappa ^2+16}}, \pm \frac{\kappa }{\sqrt{\kappa ^2+\frac{8}{3}}}, \frac{3 \kappa ^2}{3 \kappa ^2+8},  0\right)$, with eigenvalues \newline $\left\{\mp\frac{\kappa  (3 w+1)}{\sqrt{3 \kappa ^2+8}},\pm\frac{\mu_{2 1}}{6 \left(3 \kappa ^2+8\right)^3}, \pm\frac{\mu_{2 2}}{6 \left(3 \kappa ^2+8\right)^3}, \pm\frac{\mu_{2 3}}{6 \left(3 \kappa^2+8\right)^3}\right\}$, where $\mu_{2j}, j=1,2,3$ are the roots of 
$P_2(\mu)= \mu^3+24 \mu^2 \left(3 \kappa ^2+8\right)^{5/2} \lambda -144 \mu \kappa  \left(3 \kappa
   ^2+8\right)^5 (3 \kappa -2 \lambda )+3456 \kappa ^2 \left(3 \kappa ^2+8\right)^{15/2} (\kappa -2 \lambda )$. The deceleration parameter is $q=0$. They never inflate.
     The stability of the points $R$ and $S$ (which have opposite dynamical behaviours) is examined numerically. They are saddles because at least two eigenvalues have real parts of different signs, as presented in Fig.  \ref{fig:RS-K-minus1}.

   \begin{figure}
    \centering
    \includegraphics[scale=0.7]{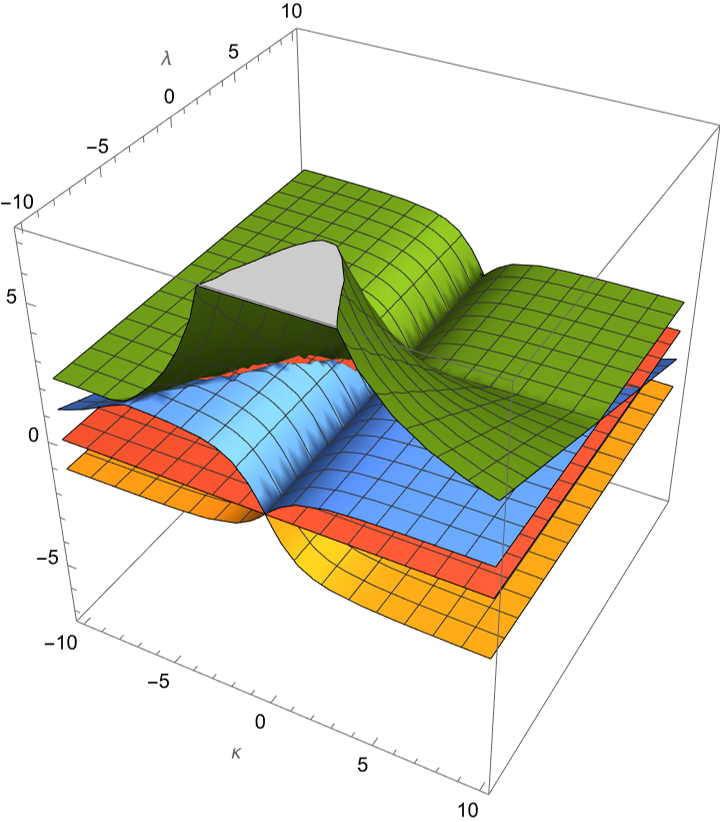}
    \caption{Real parts of the eigenvalues $\left\{\frac{\mu_{2 1}}{6 \left(3 \kappa ^2+8\right)^3}, \frac{\mu_{2 2}}{6 \left(3 \kappa ^2+8\right)^3}, \frac{\mu_{2 3}}{6 \left(3 \kappa ^2+8\right)^3}\right\}$, where $\mu_{2j}, j=1,2,3$ are the
roots of the  polynomial 
$P_2(\mu)= \mu^3+24 \mu^2 \left(3 \kappa ^2+8\right)^{5/2} \lambda -144 \mu \kappa  \left(3 \kappa
   ^2+8\right)^5 (3 \kappa -2 \lambda )+3456 \kappa ^2 \left(3 \kappa ^2+8\right)^{15/2} (\kappa -2 \lambda )$ related to $R$, compared with zero. This diagram shows its saddle nature.}
    \label{fig:RS-K-minus1}
\end{figure}
   
      \item $T: (0, 1, 1, 0)$, with eigenvalues $\left\{-\frac{4}{\sqrt{3}},-\frac{2}{\sqrt{3}},\frac{2}{\sqrt{3}},-\frac{3 w+1}{\sqrt{3}}\right\}$. It is always a saddle point.  The deceleration parameter is $q=0$.  It never inflates.
 
        \item $U: (0, -1, 1, 0)$, with eigenvalues $\left\{\frac{4}{\sqrt{3}},-\frac{2}{\sqrt{3}},\frac{2}{\sqrt{3}},\frac{3 w+1}{\sqrt{3}}\right\}$. It is always a saddle point. The deceleration parameter is $q=0$. It never inflates.

        \item $MC: \left(0, 0, \frac{3 (\kappa  (w+1)+\lambda  (w-1))}{\lambda +3 \lambda  w}, \frac{4 \lambda -2 \kappa }{\lambda +3\lambda  w}\right)$. The eigenvalues are $-\mu_1, \mu_1, -\mu_2, \mu_2$, where \newline 
  \begin{footnotesize} 
 $\mu_1= \frac{\sqrt{\lambda  \left(\kappa  \lambda ^2+\lambda  \left(\kappa ^2+2 w-2\right)-\sqrt{\kappa ^4 \lambda
   ^2+2 \kappa ^3 \lambda  \left(\lambda ^2+w+1\right)+\kappa ^2 \left(\lambda ^4-2 \lambda ^2 (w+9)+(w+1)^2\right)-4
   \kappa  \lambda  (w-1) \left(\lambda ^2+w+1\right)+4 \lambda ^2 (w-1)^2}-\kappa  (w+1)\right)}}{\sqrt{2} \lambda 
   \text{sgn}(3 w+1)}$, \\
   $\mu_2= \frac{\sqrt{\lambda  \left(\kappa  \lambda ^2+\lambda  \left(\kappa ^2+2 w-2\right)+\sqrt{\kappa ^4 \lambda
   ^2+2 \kappa ^3 \lambda  \left(\lambda ^2+w+1\right)+\kappa ^2 \left(\lambda ^4-2 \lambda ^2 (w+9)+(w+1)^2\right)-4
   \kappa  \lambda  (w-1) \left(\lambda ^2+w+1\right)+4 \lambda ^2 (w-1)^2}-\kappa  (w+1)\right)}}{\sqrt{2} \lambda 
   \text{sgn}(3 w+1)}$.
      \end{footnotesize} 
      The points are nonhyperbolic for $w= \frac{\lambda -\kappa }{\kappa +\lambda }$, $\kappa=0$, or $\lambda= \frac{\kappa
   }{2}$, or a saddle otherwise. 
      The deceleration parameter is $q=2$.  Therefore, it represents a stiff fluid solution. It never inflates.

\end{itemize}

\subsubsection{Vacuum Model ($\rho_m=0$) with negative curvature ($K=-1$)}

With $\rho_m=0$ and thus $\bar{\Omega}=0$, the system \eqref{k-1sys1}-\eqref{k-1sys4} becomes
\begin{align}
    {\bar{\Phi}}^\prime &=\kappa \bar{h}^2 (1-\bar{\Phi}^2)/\sqrt{2}+\sqrt{3}{\bar{h}}\bar{\Phi} \left({\bar{\Phi}}^2+ {\bar{\Omega}}_K/3-1\right) +\left[(\kappa +\lambda )({\bar{\Phi}} ^2-1)+\lambda {\bar{\Omega}}_K\right]/\sqrt{2},\label{k-1sys1b} \\ 
    {\bar{h}}^{\prime}  & =(\bar{h}^2-1)\left(-\kappa\bar{h}{\bar{\Phi}}/\sqrt{2} +\sqrt{3} \left[{\bar{\Phi}}^2 +{\bar{\Omega}}_K/3-1\right]\right),\\
    {\bar{\Omega}}_K^{\prime}  & = 2\bar{\Omega}_K\left( -\kappa(\bar{h}^2-1){\bar{\Phi}}/\sqrt{2}+\sqrt{3}{\bar{h}}\left[ {\bar{\Phi}}^2 +\bar{\Omega}_K/3-1/3\right]\right), \label{k-1sys4b}
\end{align}

\noindent where the bounds on the variables are now
\begin{equation}
    0 \leq \bar{h}^2, \bar{\Phi}^2, \bar{\eta}^2, \bar{\Psi}, \bar{\Omega}_K \leq 1,  
\end{equation}

\noindent and satisfy $\bar{h}^2 + \bar{\eta}^2= \bar{\Phi}^2+ \bar{\Psi} + \bar{\Omega}_K=1$. As such, the surface $\bar{\Omega}_K= 1 - \bar{\Phi}^2$ (which is $\bar{\Psi}=0$) is invariant given that 
\begin{equation}
 (\bar{\Omega}_K +\bar{\Phi}^2-1)^{\prime}=  (\bar{\Omega}_K +\bar{\Phi}^2-1)\cdot2\sqrt{3}\left[ \left(\kappa +\lambda-\kappa\,\bar{h}^2 \right)\Phi/\sqrt{6} + \bar{h} \left(\bar{\Phi}^2 +\bar{\Omega}_K/3\right)\right]. 
\end{equation}

The main focus of the analysis is to investigate whether different orbits inflate. The condition for a point in the phase space to be inflationary, using the deceleration parameter, $q$, is defined by \eqref{q.def}, which, in terms of the dimensionless variables, becomes
\begin{equation}
    q= 2+ \frac{3 \bar{\Phi}^2+\bar{\Omega}_{K}-3}{\bar{h}^2}<0. 
\end{equation}

The equilibrium points of the system \eqref{k-1sys1b} - \eqref{k-1sys4b} are $(\bar{\Phi} , \bar{h}, \bar{\Omega} _K)$:

\begin{itemize}
    \item $A:(1,1,0)$, with eigenvalues $\left\{ \frac{4}{\sqrt{3}},-\sqrt{2}\kappa ,\sqrt{2}\lambda +2\sqrt{3}\right\} $. It is a saddle for $\{\kappa >0,\lambda <-\sqrt{6}\}$, or $\{\kappa >0,\lambda >-\sqrt{6}\}$, or $ \{\kappa <0,\lambda <-\sqrt{6}\}$ or a source for $\kappa <0,\lambda >-\sqrt{6}$. It is a solution dominated by the kinetic term of quintessence field. The deceleration parameter is $q=2$. Therefore, it never inflates.

    \item $B: (-1,1,0)$, with eigenvalues $\left\{\frac{4}{\sqrt{3}},\sqrt{2} \kappa ,2 \sqrt{3}-\sqrt{2} \lambda \right\}$. It is is a saddle for $\{\kappa<0, \lambda >\sqrt{6}\} $, or $ \{\kappa<0, \lambda <\sqrt{6}\} $, or $ \{\kappa>0, \lambda >\sqrt{6}\}$ or a source for $\kappa>0, \lambda <\sqrt{6}$. It is a solution dominated by the kinetic term of quintessence field. The deceleration parameter is $q=2$. Therefore, it never inflates.

    \item $C:(1,-1,0)$, with eigenvalues $\left\{ -\frac{4}{\sqrt{3}},-\sqrt{2}\kappa ,\sqrt{2}\lambda -2\sqrt{3}\right\} $. It is a sink for $\kappa >0,\lambda <\sqrt{6}$, a saddle for $\{\kappa >0,\lambda >\sqrt{6}\}$, or $ \{\kappa <0,\lambda <\sqrt{6}\}$, or $ \{\kappa <0,\lambda >\sqrt{6}\}$. It is a solution dominated by the kinetic term of quintessence field. The deceleration parameter is $q=2$. Therefore, it never inflates.

    \item $D: (-1,-1,0)$, with eigenvalues $\left\{-\frac{4}{\sqrt{3}},\sqrt{2} \kappa ,-\sqrt{2}\lambda -2 \sqrt{3}\right\}$. It is a sink for $\kappa<0, \lambda >-\sqrt{6}$, a saddle for $\{\kappa<0, \lambda <-\sqrt{6}\} $, or $ \{\kappa>0, \lambda >-\sqrt{6}\}$, or $\{\kappa>0, \lambda <-\sqrt{6}\}$.  It is a solution dominated by the kinetic term of quintessence field. The deceleration parameter is $q=2$. Therefore, it never inflates.

    \item $E:\left( -\frac{\lambda }{\sqrt{6}},1,0\right)$, with eigenvalues $\left\{ \frac{\lambda ^{2}-2}{\sqrt{3}},\frac{\lambda^2-6}{2\sqrt{3}},\frac{\lambda (\kappa +\lambda)-6}{\sqrt{3}}\right\} $.  It is a sink for $\kappa \in \mathbb{R}$ and \newline $\left\{ -\sqrt{2}<\lambda<0,\kappa >\frac{6-\lambda ^{2}}{\lambda }\right\} $, or $ \{\lambda =0\}$, or $\left\{ 0<\lambda <\sqrt{2},\kappa <\frac{6-\lambda ^{2}}{\lambda }\right\} $. It is a saddle for $\left\{ -\sqrt{6}<\lambda <-\sqrt{2}\land \left(\kappa <\frac{6-\lambda ^{2}}{\lambda }\vee \kappa >\frac{6-\lambda ^{2}}{\lambda }\right) \right\} $, or $ \left\{ -\sqrt{2}<\lambda <0,\kappa <\frac{6-\lambda ^{2}}{\lambda }\right\} $ , or $ \left\{ 0<\lambda <\sqrt{2},\kappa >\frac{6-\lambda ^{2}}{\lambda }\right\}$, or $ \left\{ \sqrt{2}<\lambda <\sqrt{6}\land \left( \kappa <\frac{6-\lambda ^{2}}{\lambda }\vee \kappa >\frac{6-\lambda ^{2}}{\lambda }\right) \right\} $.

    \item $F:\left( \frac{\lambda }{\sqrt{6}},-1,0\right)$, with eigenvalues $\left\{ -\frac{\lambda ^{2}-2}{\sqrt{3}},-\frac{\lambda^{2}-6}{2\sqrt{3}},\frac{6-\lambda (\kappa +\lambda )}{\sqrt{3}}\right\} $. It is a source for $\kappa \in \mathbb{R}$ and $\left\{ -\sqrt{2}<\lambda<0,\kappa >\frac{6-\lambda ^{2}}{\lambda }\right\} $, or $ \{\lambda =0\}$, or $\left\{ 0<\lambda <\sqrt{2}, \kappa <\frac{6-\lambda ^{2}}{\lambda }\right\} $. It is a saddle for $\left\{ -\sqrt{6}<\lambda <-\sqrt{2}\land\left( \kappa <\frac{6-\lambda ^{2}}{\lambda }\vee \kappa >\frac{6-\lambda^{2}}{\lambda }\right) \right\} $, or $ \left\{ -\sqrt{2}<\lambda <0,\kappa <\frac{6-\lambda ^{2}}{\lambda }\right\}$ , or $ \left\{ 0<\lambda <\sqrt{2},\kappa >\frac{6-\lambda ^{2}}{\lambda }\right\} $, or $ \left\{ \sqrt{2}<\lambda <\sqrt{6}\land \left( \kappa <\frac{6-\lambda^{2}}{\lambda }\vee\kappa>\frac{6-\lambda^{2}}{\lambda }\right) \right\} $. 

    \item $G: \left(\frac{\sqrt{6}}{\sqrt{\kappa^2 + \kappa \lambda +6}}, -\frac{\kappa+ \lambda}{\sqrt{\kappa^2 + \kappa \lambda +6}},0\right)$, with  eigenvalues \newline $\left\{\frac{2 (\kappa -2 \lambda )}{\sqrt{3} \sqrt{\kappa(\kappa +\lambda )+6}},\frac{\sqrt{3} \kappa -\sqrt{-\kappa \left(4 \left(\kappa ^2-6\right) \lambda +8 \kappa \lambda ^2-27\kappa +4 \lambda ^3\right)}}{2 \sqrt{\kappa (\kappa +\lambda )+6}},\frac{\sqrt{3} \kappa +\sqrt{-\kappa \left(4 \left(\kappa ^2-6\right) \lambda +8 \kappa \lambda ^2-27\kappa +4 \lambda ^3\right)}}{2 \sqrt{\kappa (\kappa +\lambda )+6}}\right\}$. It is a saddle for $\left\{0<\kappa \leq 2 \sqrt{2}, -\kappa <\lambda <\frac{\kappa}{2}\right\}$, or $ \left\{\kappa >2 \sqrt{2}, -\kappa <\lambda <\frac{\sqrt{\kappa ^2+24}}{2}-\frac{\kappa }{2}\right\}$, or $ \left\{-2 \sqrt{2}<\kappa<0, -\frac{\sqrt{\kappa ^2+24}}{2}-\frac{\kappa }{2}<\lambda <\frac{\kappa }{2}\right\}$, or $ \left\{0<\kappa <2 \sqrt{2}, \frac{\kappa }{2}<\lambda <\frac{\sqrt{\kappa ^2+24}}{2}-\frac{\kappa }{2}\right\}$, or $ \left\{\kappa \leq -2\sqrt{2}, -\frac{\sqrt{\kappa ^2+24}}{2}-\frac{\kappa }{2}<\lambda <-\kappa\right\}$, or $ \left\{-2 \sqrt{2}<\kappa <0, \frac{\kappa }{2}<\lambda<-\kappa \right\}$, \newline or $ \left\{0<\kappa \leq 2 \sqrt{2}, -\kappa <\lambda <\frac{\kappa }{2}\right)$,  or $ \left\{\kappa >2 \sqrt{2}, -\kappa <\lambda <\frac{\sqrt{\kappa ^2+24}}{2}-\frac{\kappa }{2}\right\}$, \newline or $ \left\{-2 \sqrt{2}<\kappa <0, -\frac{\sqrt{\kappa ^2+24}}{2}-\frac{\kappa }{2}<\lambda <\frac{\kappa }{2}\right\}$.  The deceleration parameter is $q= 2-\frac{3 \kappa }{\kappa +\lambda }$. It is an inflationary solution for $\frac{\kappa }{\kappa +\lambda }>\frac{2}{3}$. 

    \item $H: \left(-\frac{\sqrt{6}}{\sqrt{\kappa^2 + \kappa \lambda +6}}, \frac{\kappa+ \lambda}{\sqrt{\kappa^2 + \kappa \lambda +6}},0\right)$, with eigenvalues \newline $\left\{\frac{4 \lambda -2 \kappa }{\sqrt{3} \sqrt{\kappa (\kappa +\lambda)+6}},\frac{-\sqrt{3} \kappa -\sqrt{-\kappa \left(4 \left(\kappa ^2-6\right) \lambda +8 \kappa \lambda ^2-27\kappa +4 \lambda ^3\right)}}{2 \sqrt{\kappa (\kappa +\lambda )+6}},\frac{-\sqrt{3} \kappa +\sqrt{-\kappa \left(4 \left(\kappa ^2-6\right) \lambda +8 \kappa \lambda ^2-27\kappa +4 \lambda ^3\right)}}{2 \sqrt{\kappa (\kappa +\lambda )+6}}\right\}$. It is a saddle for $\left\{0<\kappa <2 \sqrt{2}, \frac{\kappa }{2}<\lambda <\frac{\sqrt{\kappa^2+24}}{2}-\frac{\kappa }{2}\right\}$, or $ \left\{\kappa \leq -2 \sqrt{2}, -\frac{\sqrt{\kappa ^2+24}}{2}-\frac{\kappa }{2}<\lambda <-\kappa\right\}$, or $ \left\{-2 \sqrt{2}<\kappa <0, \frac{\kappa }{2}<\lambda<-\kappa \right)$, or $ \left\{0<\kappa \leq 2 \sqrt{2}, -\kappa <\lambda <\frac{\kappa }{2}\right\}$, \newline or $ \left\{\kappa >2 \sqrt{2}, -\kappa <\lambda <\frac{\sqrt{\kappa ^2+24}}{2}-\frac{\kappa }{2}\right\}$ , or $ \left(-2 \sqrt{2}<\kappa <0, -\frac{\sqrt{\kappa ^2+24}}{2}-\frac{\kappa }{2}<\lambda <\frac{\kappa }{2}\right\}$, or $ \left\{0<\kappa <2 \sqrt{2}, \frac{\kappa }{2}<\lambda <\frac{\sqrt{\kappa ^2+24}}{2}-\frac{\kappa }{2}\right\}$, or $\left\{\kappa \leq -2 \sqrt{2}, -\frac{\sqrt{\kappa ^2+24}}{2}-\frac{\kappa }{2}<\lambda <-\kappa \right\}$, or $ \left\{-2 \sqrt{2}<\kappa <0,\frac{\kappa }{2}<\lambda <-\kappa \right\}$. The deceleration parameter is $q= 2-\frac{3 \kappa }{\kappa +\lambda }$. It is an inflationary solution for $\frac{\kappa }{\kappa +\lambda }>\frac{2}{3}$. 

    \item $O^{\pm }:(\pm 1,0,0)$, with eigenvalues $\left\{ \pm \frac{\kappa }{\sqrt{2}},\pm \sqrt{2}\kappa,\pm \sqrt{2}(\kappa +\lambda )\right\} $. $O^{+}$ is a sink for $\{\kappa<0,\lambda <-\kappa\}$, a saddle for $\{\kappa <0,\lambda >-\kappa \}$, or $\{\kappa >0,\lambda <-\kappa \}$ , a source for $\{\kappa >0,\lambda >-\kappa\}$. $O^{-}$ is a sink for $\{\kappa >0,\lambda >-\kappa\}$, a saddle for $\{\kappa <0,\lambda >-\kappa \}$, or $ \{\kappa >0,\lambda <-\kappa \}$, a source for $\{\kappa <0,\lambda <-\kappa\}$. The deceleration parameter is $q=2$. Therefore, it represents a stiff fluid solution. It never inflates.

    \item $P=\left( -\frac{\sqrt{6}}{3\lambda },1,1-\frac{2}{\lambda ^{2}}\right),$ exists for $\lambda^2>2$. The eigenvalues of the linearization are $\left\{ \frac{2\left( \kappa -2\lambda \right) }{\sqrt{3}\lambda },-\frac{\lambda +\sqrt{8-3\lambda ^{2}}}{\sqrt{3}\lambda },-\frac{\lambda -\sqrt{8-3\lambda ^{2}}}{\sqrt{3}\lambda }\right\} $. $P$ is a sink for $\left\{ \lambda >\sqrt{2},\kappa <2\lambda \right\} $ or $\left\{\lambda <-\sqrt{2},\kappa <2\lambda \right\} $, otherwise is a saddle point. The deceleration parameter is $q=0$. It never inflates. 

    \item $Q=\left(\frac{\sqrt{6}}{3\lambda} , -1,1-\frac{2}{\lambda ^{2}}\right),$ exists for $\lambda^2>2$. The eigenvalues of the linearization are  $\left\{ -\frac{2\left( \kappa -2\lambda \right) }{\sqrt{3} \lambda },\frac{\lambda +\sqrt{8-3\lambda ^{2}}}{\sqrt{3}\lambda },\frac{\lambda -\sqrt{8-3\lambda ^{2}}}{\sqrt{3}\lambda }\right\} $. $Q$ is a saddle point for $\left\{ \lambda >\sqrt{2},\kappa >2\lambda \right\} $ or $\left\{ \lambda <-\sqrt{2},\kappa <2\lambda \right\} $, otherwise is a source. The deceleration parameter is $q=0$. It never inflates. 

    \item $R=\left( -\frac{4}{\sqrt{16+6\kappa ^{2}}},\frac{\kappa }{\sqrt{\frac{8}{3}+\kappa ^{2}}},\frac{3\kappa ^{2}}{8+3\kappa ^{2}}\right)$, with  eigenvalues $\left\{ -\frac{4\kappa }{\sqrt{8+3\kappa ^{2}}},\frac{2\kappa }{\sqrt{8+3\kappa ^{2}}},\frac{2\left( \kappa -2\lambda \right) }{\sqrt{8+3\kappa ^{2}}}\right\} $. $R$ is always a saddle point. The deceleration parameter is $q=0$. It never inflates.  

    \item $S=\left( \frac{4}{\sqrt{16+6\kappa ^{2}}},-\frac{\kappa }{\sqrt{\frac{8}{3}+\kappa ^{2}}},\frac{3\kappa ^{2}}{8+3\kappa ^{2}}\right)$, with  eigenvalues $\left\{ \frac{4\kappa }{\sqrt{8+3\kappa ^{2}}},-\frac{2\kappa }{\sqrt{8+3\kappa ^{2}}},-\frac{2\left( \kappa -2\lambda \right) }{\sqrt{8+3\kappa ^{2}}}\right\} $. $S$ is always a saddle point. The deceleration parameter is $q=0$. It never inflates. 
    
    \item $T= \left(0,1,0\right)$,  with eigenvalues $\left\{-\frac{4}{\sqrt{3}},-\frac{2}{\sqrt{3}},-\frac{2}{\sqrt{3}}\right\}$. $T$ is always a saddle point. The deceleration parameter is $q=0$. It never inflates. 
    
    \item $U=  \left(0,-1,0\right)$,  with eigenvalues $\left\{\frac{4}{\sqrt{3}},-\frac{2}{\sqrt{3}},\frac{2}{\sqrt{3}}\right\}$. $U$ is always a saddle point. The deceleration parameter is $q=0$. It never inflates. 
\end{itemize}

\subsubsection{Discussion}

\begin{figure}[h!]
	\centering
	\includegraphics[scale=0.4]{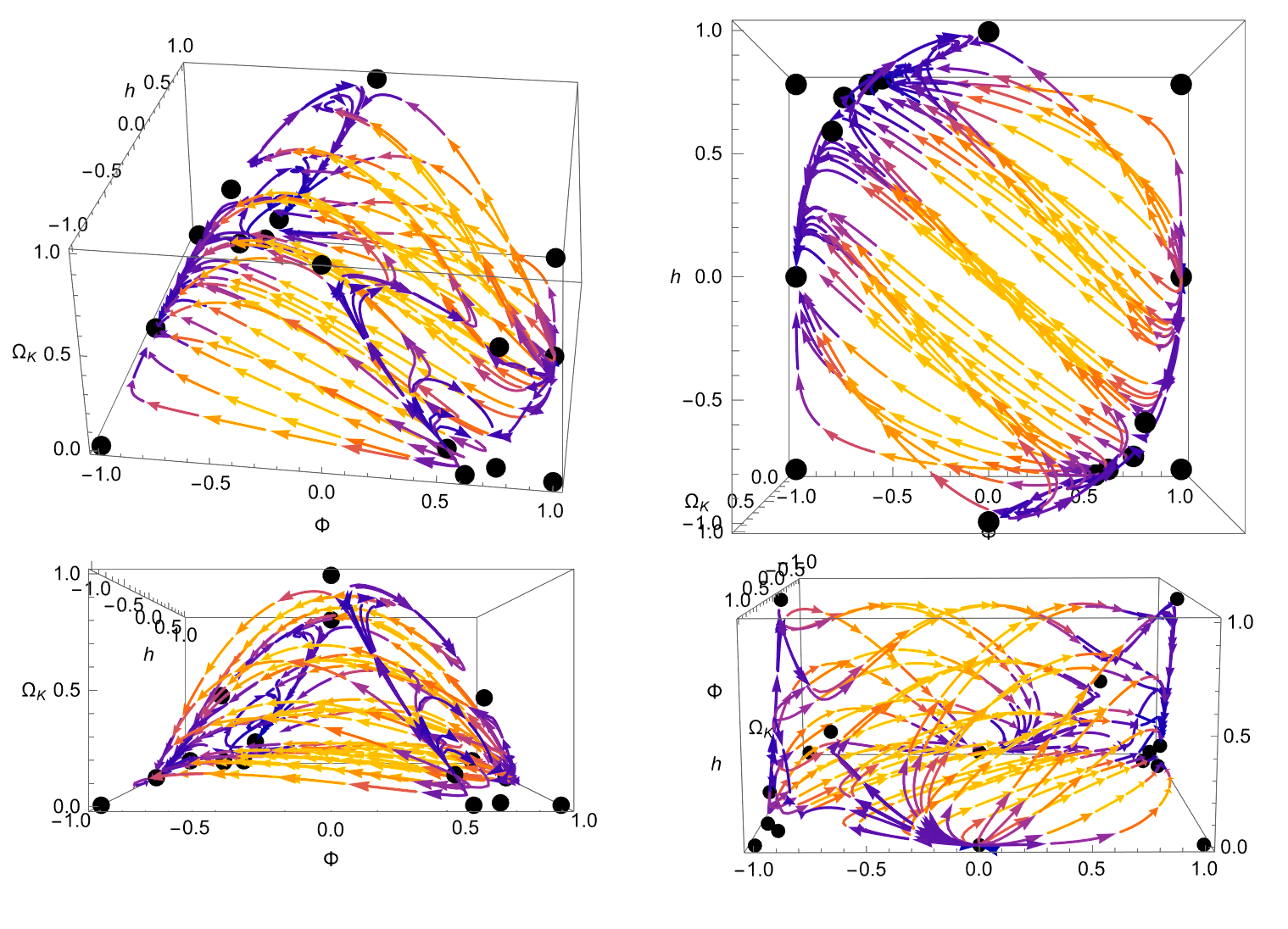}
	\caption{Different views of the flow \eqref{k-1sys1b}-\eqref{k-1sys4b} for $\lambda=\kappa=1.5$.}
	\label{k-1M1flow}
\end{figure}

We follow a similar analysis for the positive curvature model in \S \ref{m1k+1}. Fig. \ref{k-1M1flow} shows the equilibrium points and flow of the system \eqref{k-1sys1b}-\eqref{k-1sys4b} for parameters $\lambda=\kappa=1.5$.

\begin{figure}[h!]
	\centering
	\subfigure[\label{k-1series1}]{\includegraphics[scale=0.5]{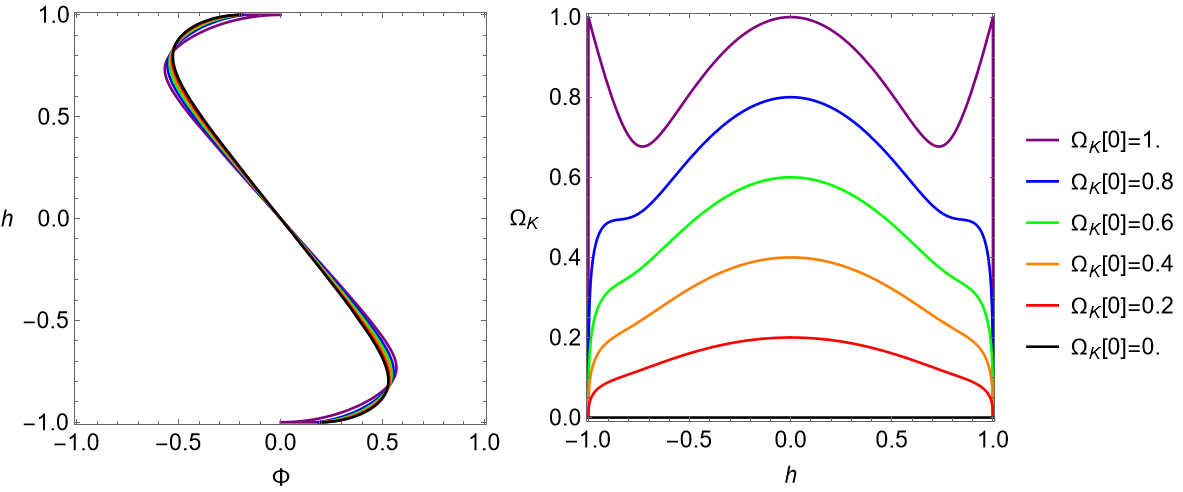}}
	\subfigure[\label{k-1series2}]{\includegraphics[scale=0.5]{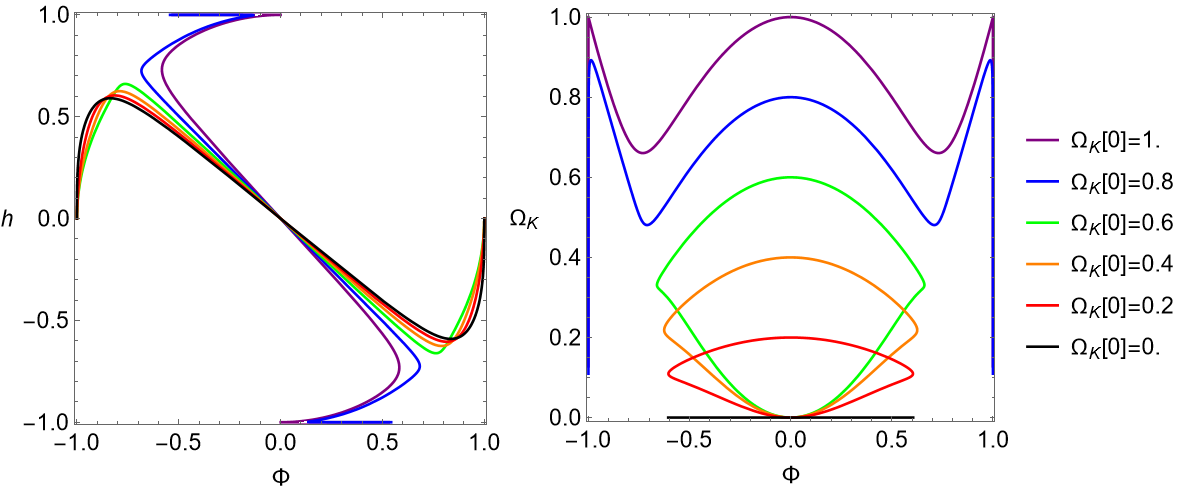}}
	\caption{Time-symmetric solutions of \eqref{k-1sys1b}-\eqref{k-1sys4b}, with varying initial value $\Omega_K(0)$ and a) $\lambda=0.5,\,\kappa=1.5$ b) $\lambda=\kappa=1.5$.}
	\label{k-1series}
\end{figure}

Fig. \ref{k-1series1} highlights a handful of solutions with initial conditions $\Phi=h=0$ and varying $\Omega_K$, for parameters $\lambda=0.5,\,\kappa=1.5$.  These time-symmetric solutions correspond to bouncing cosmologies with varying curvature values at the bounce. Fig. \ref{k-1series2} shows a similar series of solutions for parameter values $\lambda=\kappa=1.5$, where we see a new behaviour not seen in the $K=+1$ case (\S\ref{m1k+1}): for low curvature values these solutions go from early-time $O^+$ to late-time $O^-$, while for curvature values above a certain critical value, the solutions run from $Q$ to $P$, which are respectively a source and sink for these parameter values. As before, these transitions are due to interactions with saddle points (in this case, the points $R, S$), which are in the invariant surface $\lbrace\bar{\Phi}^2+\bar{\Omega}_K=1\rbrace$.

There is a bifurcation at $\abs{\lambda}=\sqrt{2}$, associated with introducing $P, Q$, and stability is transferred from $E, F$ to $P, Q$. However, for $\lambda^2<2$, $E$ and $F$ have $q<0$, so these early- and late-time epochs are inflationary. The same cannot be said of $P$ and $Q$, where $q=0$.

\begin{figure}[h!]
	\centering
	\subfigure[\label{k-1bouncing}]{\includegraphics[scale=0.42]{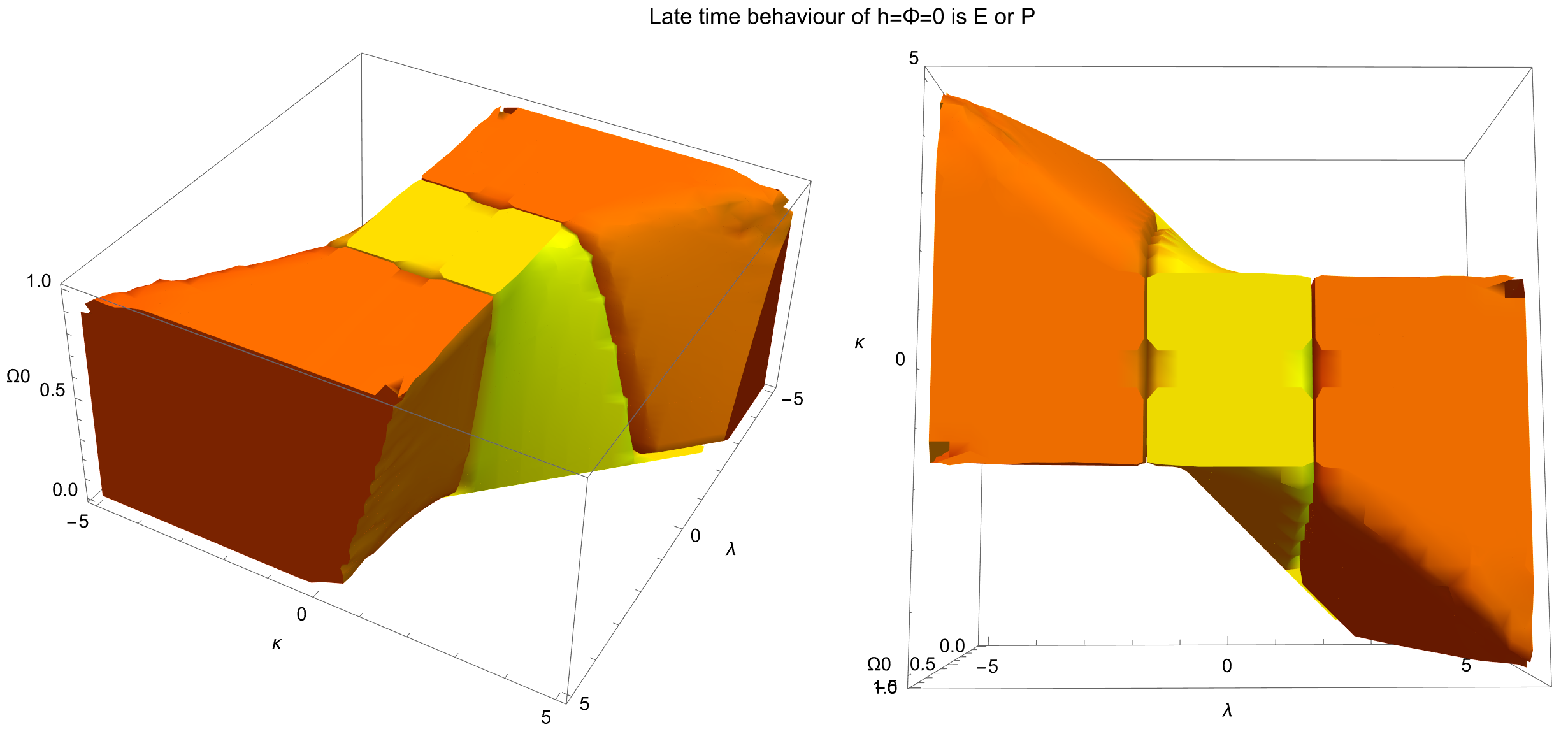}}
	\subfigure[\label{k-1crunching}]{\includegraphics[scale=0.42]{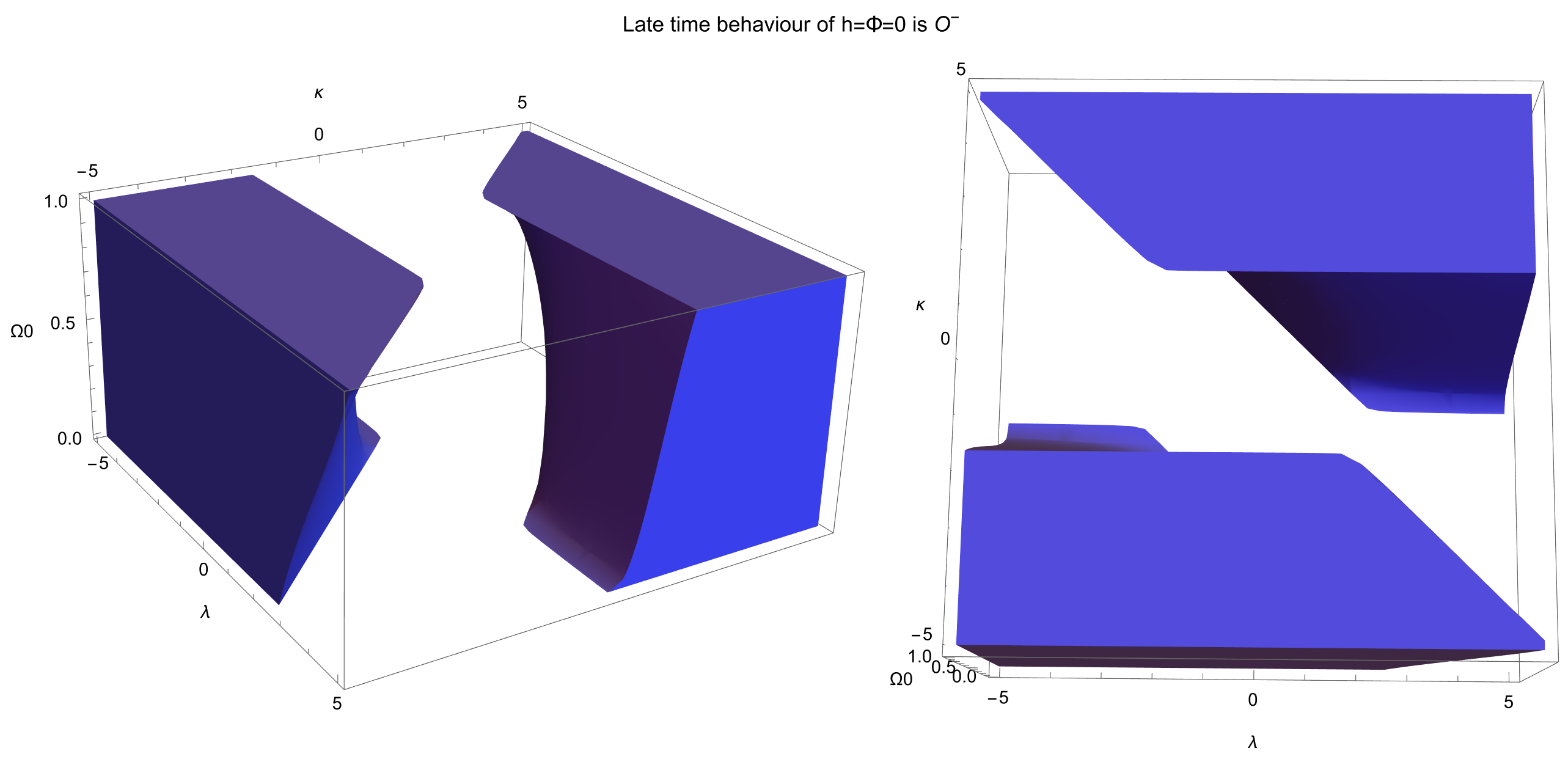}}
	\caption{Parameter values for which the late time cosmology of $h=\Phi=0$ is either (a) $E$ (yellow) or $P$ (orange), (b) $O^-$ (blue), depending on varying initial value of $\Omega_K=\Omega0$.}
	\label{k-1regions}
\end{figure}

The projected system in the invariant surface is
\begin{align}
	\bar{\Phi}^\prime&=(\bar{\Phi}^2-1)[2\bar{\Phi}\bar{h}/3+\kappa(1-h^2)/\sqrt{6}] \label{inv-1.1} \\
	\bar{h}^\prime&=-(\bar{h}^2-1)[2(1-\bar{\Phi}^2)/3+\kappa\,\bar{\Phi}\bar{h}/\sqrt{6}]. \label{inv-1.2} 
\end{align}
Interestingly, the unit circle is invariant to this flow
\begin{align}
	\left(\bar{\Phi}^2+\bar{h}^2\right)^\prime_{\eqref{inv-1.1},\eqref{inv-1.2}}&=-(\bar{\Phi}^2+\bar{h}^2-1)\left[2\bar{h}(1-\bar{\Phi}^2)/3-\kappa\bar{\Phi}(1-\bar{h}^2)/\sqrt{6}\right].
\end{align}
and the points $R,S$ lie on this circle.  The flow \eqref{inv-1.1},\eqref{inv-1.2} at the point (0,0,1) is orthogonal to $\tfrac{2}{3}\bar{\Phi}+\tfrac{\kappa}{\sqrt{6}}\bar{h}=0$.  As before, we seek the saddle-connection bifurcation by a polynomial division calculation:
\begin{align}
	\left(\tfrac{2}{3}\bar{\Phi}+\tfrac{\kappa}{\sqrt{6}}\bar{h}\right)^\prime_{\eqref{inv-1.1},\eqref{inv-1.2}}=&-\left(\tfrac{2}{3}\bar{\Phi}+\tfrac{\kappa}{\sqrt{6}}\bar{h}\right)\left[2\bar{h}(1-\bar{\Phi}^2)/3+\kappa\,\bar{\Phi}\bar{h}^2/\sqrt{6}-\kappa^2\bar{h}/4\right] \nonumber \\
	&+\frac{\bar{h}^2}{12\sqrt{6}}\kappa(8-3\kappa^2)
\end{align}

This gives the horizontal boundaries seen in Fig. \ref{k-1regions} for $\Omega0=1$, which are $\kappa=\pm\sqrt{8/3}$.

\subsection{The effect of curvature on the dynamics}

Points $A$, $B$, $C$, $D$, $E$, $F$, $G$, $H$ and $O^\pm$ have the same existence conditions and physical interpretation as the points with the same labels for the flat, positively and negatively curved case. For the negatively curved models, stability conditions are precisely the same as for the positively curved case, and they slightly differ from the flat case. 
However, the effect of curvature on the dynamics is due to the new equilibrium points. 
For positively curved models, new saddle points appear where curvature has a role, say $P$ exists for $-\sqrt{2}\leq \lambda \leq \sqrt{2}$. It is nonhyperbolic for $w=-\frac{1}{3}$, or $\kappa=2\lambda$, or  $\lambda^2=2$. It is a saddle otherwise. $Q$ exists for $-\sqrt{2}\leq \lambda \leq \sqrt{2}$. It is nonhyperbolic for $w=-\frac{1}{3}$, or $\kappa=2\lambda$, or  $\lambda^2=2$. It is a saddle otherwise. Finally, point $MC$ exists for $3 \kappa  (w+1)-4 \lambda \neq 0$. It is nonhyperbolic for $\kappa\in\left\{2 \lambda, -\lambda(w-1)/(w+1)\right\}$ or a saddle, otherwise. The effect on the dynamics is the change in stability of $E$ and $F$. In the flat case, $E$ is a sink and $F$ is a source for $\kappa \in \mathbb{R}$ and $\left\{-\sqrt{6}<\lambda <0, \kappa >\frac{6-\lambda^2}{\lambda }\right\}$, or $ \{\lambda =0\} $, or $ \left\{0<\lambda <\sqrt{6}, \kappa   <\frac{6-\lambda ^2}{\lambda }\right\}$, or a saddle for $\left\{-\sqrt{6}<\lambda <0, \kappa <\frac{6-\lambda ^2}{\lambda }\right\}$,  or $ \left\{0<\lambda <\sqrt{6}, \kappa >\frac{6-\lambda ^2}{\lambda }\right\}$. However, when $\Omega_K\neq 0$, $E$ is a sink and $F$ is a source for $\kappa \in \mathbb{R}$ and $\left\{-\sqrt{2}<\lambda <0,  \kappa >\frac{6-\lambda   ^2}{\lambda }\right\}$, or $ \{\lambda =0\} $, or $ \left\{0<\lambda <\sqrt{2},  \kappa
   <\frac{6-\lambda ^2}{\lambda }\right\}$. They are saddles otherwise. 

For negatively curved models, the new points are $P,~Q$, $R$ and $S$, for which we calculate$~\left( \Psi \left(P\right),\eta \left(P\right) \right) =\left( \frac{4}{3}\lambda^2,0\right) $, $~\left( \Psi \left(Q\right),\eta \left( Q\right) \right)=\left( \frac{4}{3}\lambda ^{2},0\right) ~$, $\left( \Psi \left(R\right),\eta \left(R\right) \right) =\left( 0,2\sqrt{\frac{2}{8+3\kappa^2}}\right) $, $\left( \Psi \left(S\right),\eta \left(S\right) \right)=\left( 0,2\sqrt{\frac{2}{8+3\kappa^2}}\right) $ while the deceleration parameter in all cases is zero, which means that the points describe Milne universes.
Furthermore, $T$ and $U$ corresponds to the Curvature dominated Milne solution $( \bar{\eta}^2=   \bar{\Phi}^2 =  \bar{\Psi}= \Omega  = 0, \bar{h}^2=\bar{\Omega}_K = 1, k =-1$) [Ref. \cite{wainwright_ellis_1997}, Sect. 9.1.6, Eq. (9.8)]).  The new early and late-time attractors are $P$ that exists for $\lambda^2>2$ and is a sink for $\left\{ \lambda >\sqrt{2},\kappa <2\lambda \right\} $ or $\left\{\lambda <-\sqrt{2},\kappa <2\lambda \right\} $, otherwise is a saddle point. Moreover, $Q$ which exists for $\lambda^2>2$, is a saddle point for $\left\{ \lambda >\sqrt{2},\kappa >2\lambda \right\} $ or $\left\{ \lambda <-\sqrt{2},\kappa <2\lambda \right\} $, otherwise is a source. The points $R, S, T$ and $U$ are always saddles. 

For $\Omega_K\neq 0$, the saddle points $G$, and $H$ remain saddle, but the stability conditions slightly change due to the extra eigenvalue. The stability conditions of the points $O^\pm$ remains the same in presence of curvature, say,  $O^{+}$ is a sink for $\{\kappa<0,\lambda <-\kappa\}$, a saddle for $\{\kappa <0,\lambda >-\kappa \}$, or $\{\kappa >0,\lambda <-\kappa \}$ , a source for $\{\kappa >0,\lambda >-\kappa\}$. $O^{-}$ is a sink for $\{\kappa >0,\lambda >-\kappa\}$, a saddle for $\{\kappa <0,\lambda >-\kappa \}$, or $ \{\kappa >0,\lambda <-\kappa \}$, a source for $\{\kappa <0,\lambda <-\kappa\}$. Finally, for $\Omega_K=0$,   $K$ is a source  and $L$ is a sink for $\left\{-1<w<1, \lambda <-\sqrt{3 w+3}, \kappa
   >\frac{\lambda(1-w)}{w+1}\right\}$,  or 
   $\left\{-1<w<1, \lambda >\sqrt{3 w+3}, \kappa
   <\frac{\lambda(1-w)}{w+1}\right\}$. When $\Omega_K>0$, the stability conditions changes to: $K$ is a source  and $L$ is a sink for $\left\{-1<w<-\frac{1}{3},  \lambda <-\sqrt{3 w+3},  \kappa
   >\frac{\lambda(1-w)}{w+1}\right\}$, or $\left\{-1<w<-\frac{1}{3}, \lambda >\sqrt{3 w+3}, \kappa <\frac{\lambda(1-w)}{w+1}\right\}$.

\section{Second model}
\label{sec3}

We consider the two-scalar field model with action integral
\begin{equation}
	S=\int\sqrt{-g}dx^{4}\left(  R+\frac{1}{2}g^{\mu\nu}\nabla_{\mu}\phi
	\nabla_{\nu}\phi - \frac{1}{2}g^{\mu\nu}e^{\kappa\; \phi}\nabla_{\mu}\psi
	\nabla_{\nu}\psi-V\left(  \phi\right)  \right) + S_{\text{matter}}, \label{Secondsp.01}
\end{equation}
where the two scalar fields are $\phi\left(  x^{\mu}\right)$, which is a phantom-like field, and $\psi\left(x^{\nu}\right)$, which is a quintessence-like field have kinetic terms which lie on a two-dimensional manifold. For the background space, we again assume the FLRW universe
\begin{equation}
	ds^2 = -dt^2 +a^2(t)\left( \frac{ dr^2}{1-Kr^2} +r^2\left( d\theta^2 +\sin^2\theta\,d\varphi^2 \right)\right),\label{Secondsp.02}
\end{equation}
where $K$ is the spatial curvature for the three-dimensional hypersurface  (for $K=0$, we have a spatially flat universe; for $K=1$ we have a closed universe; and for $K=-1$, the line element (\ref{Secondsp.02}) describes an open universe). The quantity $S_{\text{matter}}$ corresponds to the matter action considered as a perfect fluid, with energy density $\rho_m$, pressure $p_m$ with the constant equation of state parameter $p_m= w \rho_m$. 

Now we study the cosmological model with action integral (\ref{Secondsp.01}) for the line element (\ref{Secondsp.02}). The gravitational field equations are
\begin{align}
& -3H^2 -\frac{1}{2}\dot{\phi}^2 +\frac{1}{2}e^{\kappa \phi}\dot{\psi}^2
	+V\!\left( \phi \right) + \rho_m   -\frac{3K}{a^2}=0, \label{M20sp.04} \\
& 2\dot{H} +3H^2 -\frac{1}{2}\dot{\phi}^2 +\frac{1}{2}e^{\kappa  \phi}\dot{\psi}^2	-V\!\left( \phi \right) + w \rho_m +  \frac{K}{a^2}=0, \label{M20sp.05} 	\\
& \ddot{\phi}+ 3H\dot{\phi} +\frac{\kappa}{2}e^{\kappa\, \phi}\dot{\psi}^2 - 	V'\left( \phi \right) =0, \label{M20sp.06} 	\\
& \ddot{\psi} +3H\dot{\psi} +\kappa \dot{\phi}  \dot{\psi} =0, \label{M20sp.07} \\
&\dot{\rho_m} + 3 H(1+w) \rho_m=0. \label{M2matter}
\end{align}
where $H=\frac{\dot{a}}{a}$ is the Hubble function.

\subsection{Dynamical systems formulation ($K=0$)}
\label{subsect:ModelII-IIIA}

Using the  variables 
\begin{equation}
\label{chi_lambdaM20}
    \chi^2 = \frac{1}{2}\dot{\phi}^2
	+3H^2, \quad V(\phi)= V_0 e^{\lambda \phi},
\end{equation}
 and
\begin{align}
	&h^2= \frac{3 H^2}{\chi^2}, \; \eta^2= \frac{e^{\kappa \phi}\dot{\psi}^2}{2 \chi^2}, \;  \Phi^2 = \frac{\dot{\phi}^2}{2 \chi^2}, \; \Psi= \frac{V(\phi)}{\chi^2}, \; \Omega= \frac{\rho_m}{\chi^2},\label{ZerovarsM20}
\end{align}
 which satisfy
\begin{equation}
	h^2 +\Phi^2 = \eta^2 + \Psi + \Omega=1,  \label{eq102M2}
\end{equation}
we have then bounded variables
\begin{equation}
	0 \leq h^2, \Phi^2, \eta^2,\Psi,\Omega \leq 1.  
\end{equation}
The field equations are 
\begin{align}
& \Phi^{\prime} = \frac{\left(\Phi ^2-1\right) \left(\eta ^2 (\kappa +\lambda )+\lambda  (\Omega -1)\right)}{\sqrt{2}}+\frac{1}{2} \sqrt{3} h \Phi  \left(2 \eta ^2+w \Omega +\Omega
   -2\right),\\
& h^{\prime}= \frac{1}{6} \Big(\sqrt{3} h^2 \left(6 \eta ^2+3 (w+1) \Omega -2\right)+3 \sqrt{2} h \Phi  \left(\eta ^2 (\kappa +\lambda )+\lambda  (\Omega -1)\right) \nonumber \\
& +\sqrt{3} \left(-6 \eta ^2+4 \Phi ^2-3 (w+1) \Omega +2\right)\Big),\\
& \eta^{\prime}=\frac{1}{2} \eta  \left(\sqrt{2} \Phi  \left(\left(\eta ^2-1\right) (\kappa +\lambda )+\lambda  \Omega \right)+\sqrt{3} h \left(2 \eta ^2+(1+w) \Omega -2\right)\right),\\
& \Omega^{\prime}= \Omega  \left(\sqrt{2} \Phi  \left(\eta ^2 (\kappa +\lambda )+\lambda  (\Omega -1)\right)+\sqrt{3} h \left(2 \eta ^2+w (\Omega -1)+\Omega
   -1\right)\right). 
\end{align}
The deceleration parameter reads
\begin{equation}
 q= \frac{6 \eta ^2-4 \Phi ^2+3 (w+1) \Omega -2}{2 h^2}. 
\end{equation}

Given that 
\begin{align}
{\mathcal{C}}^{\prime}=  \mathcal{C} \left[\sqrt{2} \Phi  \left(\eta ^2 (\kappa +\lambda )+\lambda  (\Omega -1)\right)+\frac{h \left(6 \eta ^2+3 (w+1) \Omega -2\right)}{\sqrt{3}}\right],  \quad \mathcal{C} := 1-h^2-\Phi^2,
\end{align}
we have that $ \mathcal{C} =0$ is an invariant set. 

We use the polar coordinates \begin{align}
    h= \cos(\varphi), \quad \Phi=\sin(\varphi) \implies\vartheta= \arctan(\Phi/h).
\end{align}
We then have 
\begin{align}
\vartheta^{\prime} & =   \frac{h \Phi^{\prime}}{h^2+\Phi ^2}-\frac{\Phi  h^{\prime}}{h^2+\Phi ^2}= h \Phi^{\prime} - \Phi  h^{\prime}\nonumber \\
 &=\frac{1}{6} \left(-4 \sqrt{3} h^2 \Phi -3 \sqrt{2} h \left(\eta ^2 (\kappa +\lambda )+\lambda  (\Omega -1)\right)+\sqrt{3} \Phi  \left(6 \eta ^2-4 \Phi ^2+3 (w+1) \Omega -2\right)\right).
\end{align}
Hence, replacing $ h\mapsto  \cos(\varphi), \;  \Phi\mapsto \sin(\varphi) $, we have the system 
\begin{align}
& \eta^{\prime}=  \frac{1}{2} \eta  \left(\sqrt{2} \sin (\vartheta ) \left(\left(\eta ^2-1\right) (\kappa +\lambda )+\lambda  \Omega \right)+\sqrt{3} \cos (\vartheta ) \left(2 \eta ^2+w \Omega +\Omega
   -2\right)\right),\\
&  \vartheta^{\prime}= \frac{1}{2} \left(\sqrt{3} \sin (\vartheta ) \left(2 \eta ^2+(w+1) \Omega -2\right)-\sqrt{2} \cos (\theta ) \left(\eta ^2 (\kappa +\lambda )+\lambda  (\Omega -1)\right)\right), \\
& \Omega^{\prime}= \Omega  \left(\sqrt{2} \sin (\vartheta ) \left(\eta ^2 (\kappa +\lambda )+\lambda  (\Omega -1)\right)+\sqrt{3} \cos (\vartheta ) \left(2 \eta ^2+w (\Omega -1)+\Omega
   -1\right)\right).
\end{align}
We can find the inflationary condition with these variables where $q$, the deceleration parameter, is negative:
\begin{align}
    q =\frac{3}{2 \cos^2(\vartheta)} \left(2 \eta ^2+(w+1) \Omega -2\right)+2<0.
\end{align}

\subsubsection{ Vacuum model ($\rho_m=0$) with zero curvature ($K =0$)}
\label{SectIIIA1}

In this example, we have 
\begin{align}
\vartheta^{\prime} & =   \frac{h \Phi^{\prime}}{h^2+\Phi ^2}-\frac{\Phi  h^{\prime}}{h^2+\Phi ^2}= h \Phi^{\prime} - \Phi  h^{\prime}\nonumber \\
 &= \frac{\Phi  \left(3 \eta ^2-2 \Phi ^2-1\right)}{\sqrt{3}}-\frac{2 h^2 \Phi }{\sqrt{3}}+\frac{h \left(\lambda -\eta ^2
   (\kappa +\lambda )\right)}{\sqrt{2}}.
\end{align}
Hence, replacing $ h\mapsto  \cos(\varphi), \;  \Phi\mapsto \sin(\varphi) $, we have the system 
\begin{align}
\eta^{\prime} & =    \frac{1}{2} \eta  \left(\eta ^2-1\right) \left(\sqrt{2} \sin (\vartheta ) (\kappa +\lambda )+2 \sqrt{3} \cos (\vartheta
   )\right), \label{EQ.(120)}\\
   \vartheta^{\prime} & = 
\frac{\cos (\vartheta) \left(\lambda -\eta ^2 (\kappa +\lambda )\right)}{\sqrt{2}}+\sqrt{3} \left(\eta ^2-1\right) \sin
   (\vartheta). \label{EQ.(121)}
\end{align}

We can also find the inflationary condition with these variables when $q$, the deceleration parameter, is negative: 
\begin{align}
    q =\frac{3 \left(\eta ^2-1\right)}{\cos^2(\vartheta )}+2<0.
\end{align}

We have the equilibrium points $(\eta,\vartheta)$ (we use different labels as for \S \ref{sec2}): 
\begin{itemize}
    \item $A,B: \left(\pm1,\frac{\pi}{2}\right)$, which  exists for all parameter values. The eigenvalues of the linearization are $\left\{ \sqrt{2}(\kappa+\lambda),\frac{1}{\sqrt{2}}\kappa \right\}$. Therefore $A$ and $B$ are sinks for $\{\kappa<0,\lambda<-\kappa\}$, saddle for $\left\{\kappa<0,\lambda>-\kappa \right\} $, or $ \left\{\kappa>0,\lambda<-\kappa\right\}$, and a source for $\{\kappa>0,\lambda>-\kappa\}$.  It never inflates. Compared with the results of \S \ref{Subsect:II1}, these points represent $O^+$. 
    
    \item $C,D: \left(\pm1,\frac{3\pi}{2}\right)$, which  exists for all parameter values. The eigenvalues of the linearization are $\left\{ -\sqrt{2}(\kappa+\lambda),-\frac{1}{\sqrt{2}}\kappa \right\}$. Therefore $C$ and $D$ are sinks for $\{\kappa>0,\lambda>-\kappa\}$, saddle for $\left\{\kappa<0,\lambda>-\kappa \right\} $, or $ \left\{\kappa>0,\lambda< -\kappa\right\}$, and a source for $\{\kappa<0,\lambda< - \kappa\}$. It never inflates.  Compared with the results of \S \ref{Subsect:II1}, these points represent $O^-$.

   The following points are unrelated to those examined in the \S \ref{Subsect:II1} with the same labels. 

    \item $E: \left(0,\arctan\left(\frac{\lambda}{\sqrt{6}}\right)\right).$  The eigenvalues of the linearization are $\left\{ -\frac{1}{\sqrt{2}}\sqrt{\lambda^2+6},-\frac{1}{\sqrt{2}}\frac{\kappa\lambda+\lambda^2+6}{\sqrt{\lambda^2+6}} \right\}$. Therefore $E$ is a sink for $\{\kappa\lambda+\lambda^2+6>0\}$, saddle for $\{\kappa\lambda+\lambda^2+6<0\}$. The deceleration parameter is $q=-1-\frac{\lambda ^2}{2}$. Inflates for all choice of parameters.  Compared with the results of \S \ref{Subsect:II1}, this point is new and corresponds to a phantom-dominated solution. 
    
    \item $F: \left(0,\arctan\left(\frac{\lambda}{\sqrt{6}}\right)+\pi\right)$, which  exists for all choice of parameters. The eigenvalues of the linearization are $\left\{ \frac{1}{\sqrt{2}}\sqrt{\lambda^2+6},\frac{1}{\sqrt{2}}\frac{\kappa\lambda+\lambda^2+6}{\sqrt{\lambda^2+6}} \right\}$. Therefore $F$ is a source for $\{\kappa\lambda+\lambda^2+6>0\}$, saddle for $\{\kappa\lambda+\lambda^2+6<0\}$. The deceleration parameter is $q=-1-\frac{\lambda ^2}{2}$. Inflates for all choices of parameters. Compared with the results of \S \ref{Subsect:II1}, this point is new and corresponds to a phantom-dominated solution.

    \item $I_{+},I_{-}: \left(\pm \sqrt{\frac{\kappa\lambda+\lambda^2+6}{(\kappa+\lambda)^2+6}}, -\arctan\left(\frac{\sqrt{6}}{\kappa+\lambda}\right)+\pi\right)$, which  satisfies $-1\leq \eta\leq 1$  for $\{\kappa \in \mathbb{R}, \lambda =0\}\cup \{\lambda >0, \kappa \geq 0\} \cup\left\{\lambda <0, -\lambda \leq \kappa \leq \frac{-\lambda ^2-6}{\lambda }\right\}\cup \{\lambda <0, \kappa \leq 0\}\cup \left\{\lambda >0,  \frac{-\lambda ^2-6}{\lambda }\leq \kappa \leq -\lambda \right\}$. The eigenvalues of the linearization are \newline  $\left\{ \frac{\sqrt{3}\kappa}{2\sqrt{(\kappa+\lambda)^2+6}} \pm \frac{1}{2}\sqrt{\frac{\kappa(4\kappa^2\lambda+8\kappa\lambda^2+4\lambda^3+27\kappa+24\lambda)}{(\kappa+\lambda)^2+6}} \right\}$. Therefore $I_{+}$ and $I_{-}$ are saddles whenever they exist. The deceleration parameter is $q= 2-\frac{3 \kappa }{\kappa +\lambda }$. Inflation occurs for  $\frac{\kappa }{\kappa +\lambda }>\frac{2}{3}$. 
    
    \item $J_{+},J_{-}: \left(\pm \sqrt{\frac{\kappa\lambda+\lambda^2+6}{(\kappa+\lambda)^2+6}}, -\arctan\left(\frac{\sqrt{6}}{\kappa+\lambda}\right)\right)$, which  satisfies $-1\leq \eta\leq 1$  for $\{\kappa \in \mathbb{R}, \lambda =0\}\cup \{\lambda >0, \kappa \geq 0\} \cup\left\{\lambda <0, -\lambda \leq \kappa \leq \frac{-\lambda ^2-6}{\lambda }\right\}\cup \{\lambda <0, \kappa \leq 0\}\cup \left\{\lambda >0,  \frac{-\lambda ^2-6}{\lambda }\leq \kappa \leq -\lambda \right\}$. The eigenvalues of the linearization are \newline $\left\{  -\frac{\sqrt{3}\kappa}{2\sqrt{(\kappa+\lambda)^2+6}} \pm \frac{1}{2}\sqrt{\frac{\kappa(4\kappa^2\lambda+8\kappa\lambda^2+4\lambda^3+27\kappa+24\lambda)}{(\kappa+\lambda)^2+6}} \right\}$. Therefore $J_{+}$ and $J_{-}$ are saddles whenever they exist. The deceleration parameter is $q= 2-\frac{3 \kappa }{\kappa +\lambda }$. Inflation occurs for  $\frac{\kappa }{\kappa +\lambda }>\frac{2}{3}$. 
    
\end{itemize}

\subsubsection{Discussion}
Perhaps, the most interesting system is when we have two periods of inflation, when $E$ is a sink, $F$ is a source, and $I_{\pm}, J_{\pm}$ are saddles, which are all inflationary.
Hence, orbits with two periods of inflation occur for the parameters that agree with the following conditions: $\left\{ \kappa\lambda+\lambda^2+6>0, \frac{\kappa }{\kappa +\lambda }>\frac{2}{3} \right\}$.

\begin{figure}[h!]
    \centering
    \includegraphics[scale=0.15]{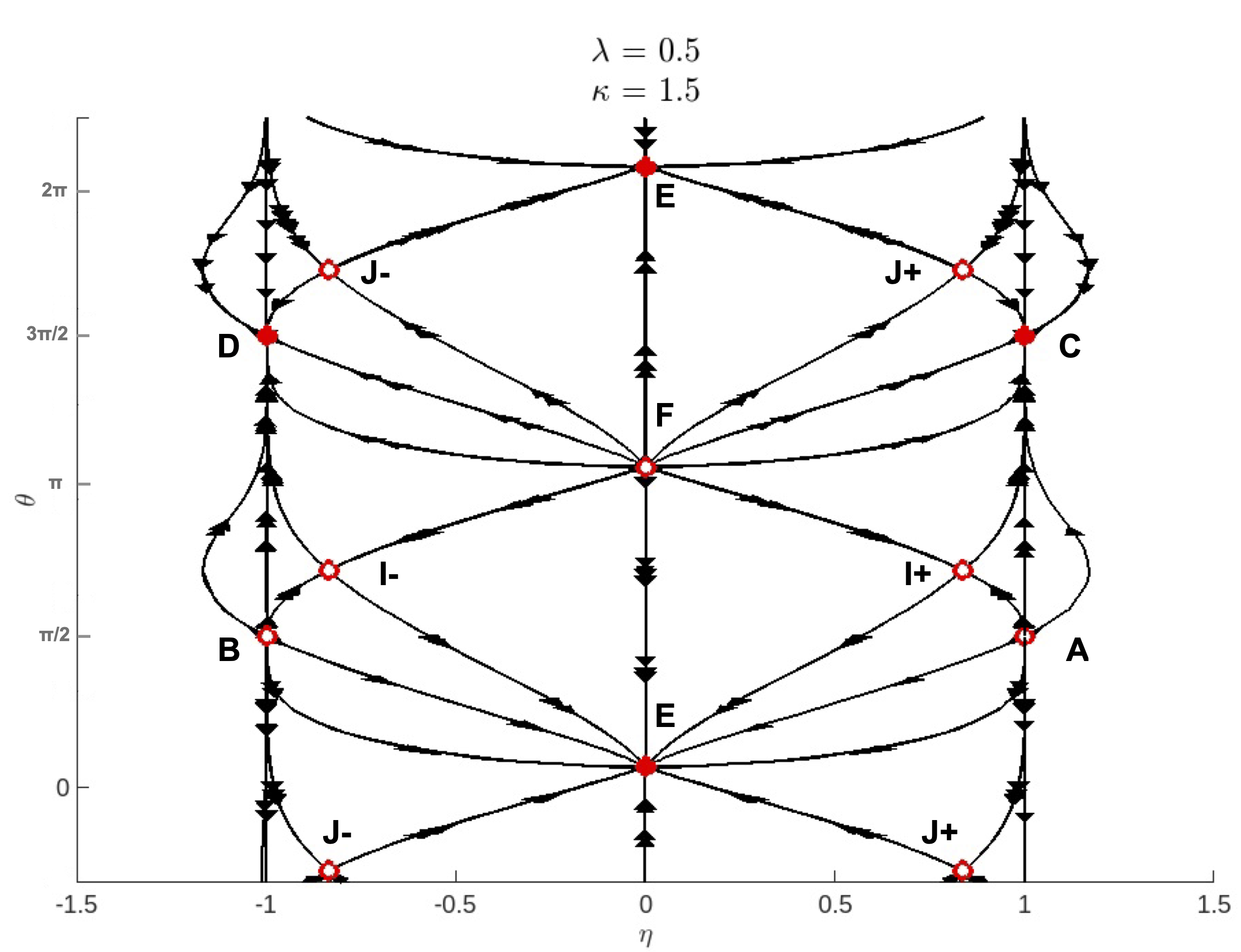}
    \caption{This figure shows a skeleton of the dynamical system \eqref{EQ.(120)}-\eqref{EQ.(121)}  within the phase-space $\eta\in[-1,1],\vartheta\in[0,2\pi]$, for parameter values $\lambda=0.5,\kappa=1.5$.  The stable and unstable manifolds of the saddles $I_{\pm}$ and $J_{\pm}$ are shown. Orbits emerging near $F$, moving close to $I_{\pm}$ and finally approaching the sink $E$, experience two periods of inflation.}
    \label{skeleton_2}
\end{figure}

 Fig.  \ref{skeleton_2} shows a skeleton of the dynamical system \eqref{EQ.(120)}-\eqref{EQ.(121)} within the phase-space $\eta\in[-1,1],\vartheta\in[0,2\pi]$, for parameter values $\lambda=0.5,\kappa=1.5$.  The stable and unstable manifolds of the saddles $I_{\pm}$ and $J_{\pm}$ are shown. Orbits are emerging near $F$, moving close to $I_{\pm}$ and finally approaching the sink $E$, experiencing two periods of inflation.

\subsection{Dynamical systems formulation ($K=+1$)}
\label{SectIIIB}
We define
\begin{equation}
\label{chi_lambdaM2}
    \chi^2 = \frac{1}{2}\dot{\phi}^2
	+3H^2 + \frac{3}{a^2} , \quad V(\phi)= V_0 e^{\lambda \phi},
\end{equation}
 and
\begin{align}
	&h^2= \frac{3 H^2}{\chi^2}, \; \eta^2= \frac{e^{\kappa \phi}\dot{\psi}^2}{2 \chi^2}, \;  \Phi^2 = \frac{\dot{\phi}^2}{2 \chi^2}, \; \Psi= \frac{V(\phi)}{\chi^2}, \; \Omega= \frac{\rho_m}{\chi^2}, \; 	\Omega_K=\frac{3}{a^2 \chi^2},\label{PositivevarsM2}
\end{align}
 which satisfy
\begin{equation}
	h^2+ \Phi^2  + \Omega_K= \eta^2 + \Psi + \Omega=1.  \label{PositiveeqM2}
\end{equation}
 Using \eqref{PositiveeqM2}, we have then bounded variables
\begin{equation}
	0 \leq h^2, \Phi^2, \eta^2,\Psi, \Omega_K,\Omega \leq 1.  
\end{equation}
 Then, equations \eqref{M20sp.04}, \eqref{M20sp.05}, \eqref{M20sp.06}, \eqref{M20sp.07} and \eqref{M2matter} become 
\begin{align}
 \frac{\dot{H}}{\chi^2} & =   \frac{1}{6} \left(-3 \left(\eta ^2+h^2-\Phi ^2+w \Omega -\Psi \right)-\Omega_K\right),\\
 \frac{\ddot{\phi}}{\chi^2} &= - \kappa \eta ^2-\sqrt{6} h \Phi +\lambda  \Psi ,\\
 \frac{\ddot{\psi}}{\dot{\psi}\chi}  &= -\sqrt{3} h-\sqrt{2} \kappa  \Phi ,\\
  \frac{\dot{\rho_m}}{\chi^2} &= -\sqrt{3} h (1+w) \chi  \Omega
\end{align} 

Introducing the time derivative $d\tau=\chi d t$ and taking the derivatives of the variables with respect to the new time variable, the field equations are transformed to 
\begin{align}
 \Phi^{\prime} & =   -\frac{\eta ^2 \kappa }{\sqrt{2}}-\sqrt{3} h \Phi +\frac{\lambda  \Psi }{\sqrt{2}}-\frac{\Phi  \dot{\chi}}{\chi
   ^2},\\
 h^{\prime} & =  -\frac{3 \left(\eta ^2+h^2-\Phi ^2+w \Omega -\Psi \right)+\Omega_K}{2 \sqrt{3}}-\frac{h \dot{\chi}}{\chi ^2},\\
   \eta^{\prime} &= -\frac{\eta  \kappa  \Phi }{\sqrt{2}}-\sqrt{3} \eta  h-\frac{\eta  \dot{\chi}}{\chi ^2},\\
   \Psi^{\prime} &= 
   \sqrt{2}
   \lambda  \Phi  \Psi -\frac{2 \Psi  \dot{\chi}}{\chi ^2},\\
  \Omega^{\prime} &= -\Omega\left[ \sqrt{3} h (1+w)   +\frac{2 \dot{\chi}}{\chi
   ^2}\right], \\
     \Omega_K^{\prime} &= -2 \Omega_K\left[\frac{h}{\sqrt{3}}-\frac{\dot{\chi}}{\chi ^2}\right].
\end{align}
Next, substituting 
\begin{align} 
\frac{\dot{\chi}}{\chi ^2 }& = -\frac{1}{2} \left(\sqrt{2} \Phi  \left(\eta ^2 \kappa -\lambda  \Psi \right)+\sqrt{3} h^3+\sqrt{3} h
   \left(\eta ^2+\Phi ^2+w \Omega -\Psi +\Omega_K\right)\right),\\
\Psi & = 1 -\eta ^2-\Omega,\\
 \Omega_{K}& = 1 -h^2-\Phi ^2, 
\end{align}
we obtain 
\begin{align}
 \Phi^{\prime} & =  \frac{\left(\Phi ^2-1\right) \left(\eta ^2 (\kappa +\lambda )+\lambda  (\Omega
   -1)\right)}{\sqrt{2}}+\frac{1}{2} \sqrt{3} h \Phi  \left(2 \eta ^2+(w+1)\Omega -2\right), \label{eq:128}\\
 h^{\prime} & =   \frac{1}{6}
   \Big(\sqrt{3} h^2 \left(6 \eta ^2+3 (1+w) \Omega -2\right)+3 \sqrt{2} h \Phi  \left(\eta ^2 (\kappa +\lambda
   )+\lambda  (\Omega -1)\right) \nonumber \\
   & +\sqrt{3} \left(-6 \eta ^2+4 \Phi ^2-3 (1+w) \Omega +2\right)\Big), \label{eq:129}\\
    \eta^{\prime} & = \frac{1}{2} \eta
    \left(\sqrt{2} \Phi  \left(\left(\eta ^2-1\right) (\kappa +\lambda )+\lambda  \Omega \right)+\sqrt{3} h \left(2
   \eta ^2+(w+1)\Omega -2\right)\right), \label{eq:130}\\
  \Omega^{\prime} & =  \Omega  \left(\sqrt{2} \Phi  \left(\eta ^2 (\kappa +\lambda )+\lambda 
   (\Omega -1)\right)+\sqrt{3} h \left(2 \eta ^2+w (\Omega -1)+\Omega -1\right)\right),  \label{eq:131}
\end{align}
and the auxiliary equations
\begin{align}
   \Psi^{\prime}&= \left(1-\eta ^2-\Omega  \right) \left(\sqrt{2} \Phi  \left(\eta ^2 (\kappa +\lambda )+\lambda  \Omega
   \right)+\sqrt{3} h \left(2 \eta ^2+(w+1)\Omega \right)\right), \label{eq:132}\\
   \Omega_{K}^{\prime}& = \frac{1}{3} \left(1-h^2-\Phi ^2\right) \left(3
   \sqrt{2} \Phi  \left(\eta ^2 (\kappa +\lambda )+\lambda  (\Omega -1)\right)+\sqrt{3} h \left(6 \eta ^2+3 (1+w)
   \Omega -2\right)\right).
\end{align}

\begin{table}[]
    \centering
\resizebox{\textwidth}{!}{  
\begin{tabular}{|ccccccccc|}\hline
Label & $\Phi$ & $h$ & $\eta$ & $\Omega$ & $k_1$ & $k_2$ &$k_3$ & $k_4$ \\\hline
$A, B$ &$ 1$ &$ 0$ & $\pm 1$ & $0 $& $\frac{\kappa }{\sqrt{2}}$ & $\sqrt{2} \kappa$  &$ \sqrt{2} \kappa $ & $\sqrt{2} (\kappa +\lambda ) $\\\hline
$C, D$ &$ -1 $& $0$ & $\pm 1$ & $0$ & $-\sqrt{2} \kappa$  & $-\sqrt{2} \kappa$  & $-\frac{\kappa }{\sqrt{2}}$ &$ -\sqrt{2} (\kappa +\lambda )$ \\\hline
$E, F$ &$ 1$& $-\frac{1}{2} \sqrt{\frac{3}{2}} \kappa $ & $\pm 1$ & $0$ &$ -\frac{\kappa }{\sqrt{2}}$ &$ \sqrt{2} \kappa$  &$ \frac{\kappa  (3 w+1)}{2 \sqrt{2}}$ &$ -\frac{\kappa -2 \lambda }{\sqrt{2}}$ \\\hline
$G, H$ &$ -1$ &$ \frac{1}{2} \sqrt{\frac{3}{2}} \kappa $ & $\pm 1$ & $0$ & $-\sqrt{2} \kappa$ & $\frac{\kappa }{\sqrt{2}} $&$ -\frac{\kappa  (3 w+1)}{2 \sqrt{2}} $& $\frac{\kappa -2 \lambda }{\sqrt{2}}$ \\\hline
$I_+$ &$ \frac{\lambda }{\sqrt{\lambda ^2+6}}$ & $\frac{\sqrt{6}}{\sqrt{\lambda ^2+6}}$ & $0$ & $0$ & $-\frac{\lambda  (\kappa +\lambda )+6}{\sqrt{2} \sqrt{\lambda ^2+6}}$ & $-\frac{\sqrt{2} \left(\lambda
   ^2+3(w+1)\right)}{\sqrt{\lambda ^2+6}}$ &$ -\frac{\sqrt{2} \left(\lambda ^2+2\right)}{\sqrt{\lambda ^2+6}}$ & $-\frac{\sqrt{\lambda ^2+6}}{\sqrt{2}} $\\\hline
$I_-$ & $-\frac{\lambda }{\sqrt{\lambda ^2+6}}$ &$ -\frac{\sqrt{6}}{\sqrt{\lambda ^2+6}}$ & $0$ & $0$ &$ \frac{\lambda  (\kappa +\lambda )+6}{\sqrt{2} \sqrt{\lambda ^2+6}}$ &$ \frac{\sqrt{2} \left(\lambda
   ^2+3(w+1)\right)}{\sqrt{\lambda ^2+6}}$ &$ \frac{\sqrt{2} \left(\lambda ^2+2\right)}{\sqrt{\lambda ^2+6}}$ & $\frac{\sqrt{\lambda ^2+6}}{\sqrt{2}} $\\\hline
$J_\pm$ &  $\frac{\sqrt{6}}{\sqrt{(\kappa +\lambda )^2+6}}$ &$ -\frac{\kappa +\lambda }{\sqrt{(\kappa +\lambda )^2+6}} $& $\pm \frac{\sqrt{\lambda  (\kappa +\lambda )+6}}{\sqrt{(\kappa +\lambda )^2+6}}$ &$0$ &
  $ \frac{\sqrt{3} (\kappa  (w+1)+\lambda  (w-1))}{\sqrt{(\kappa +\lambda )^2+6}}$ & $\frac{\mu_{11}}{6 \left((\kappa +\lambda )^2+6\right)^{5/2}}$ &$ \frac{\mu_{12}}{6 \left((\kappa +\lambda )^2+6\right)^{5/2}} $& $\frac{\mu_{13}}{6 \left((\kappa +\lambda )^2+6\right)^{5/2}}$ \\\hline
$K_\pm$ & $ -\frac{\sqrt{6}}{\sqrt{(\kappa +\lambda )^2+6}}$ &$ \frac{\kappa +\lambda }{\sqrt{(\kappa +\lambda )^2+6}}$ &$ \pm \frac{\sqrt{\lambda  (\kappa +\lambda )+6}}{\sqrt{(\kappa +\lambda )^2+6}}$ &$ 0$ &
  $ -\frac{\sqrt{3} (\kappa  (w+1)+\lambda  (w-1))}{\sqrt{(\kappa +\lambda )^2+6}}$ & $\frac{-\mu_{11}}{6 \left((\kappa +\lambda )^2+6\right)^{5/2}}$ & $ \frac{-\mu_{12}}{6 \left((\kappa +\lambda )^2+6\right)^{5/2}}$ &$ \frac{-\mu_{13}}{6 \left((\kappa +\lambda )^2+6\right)^{5/2}}$ \\\hline
$L_+$ &$ -\frac{w+1}{\sqrt{\frac{2 \lambda ^2}{3}+(w+1)^2}}$ & $\frac{\lambda }{\sqrt{\lambda ^2+\frac{3}{2} (w+1)^2}}$ & $0$ & $\frac{2 \left(\lambda ^2+3(w+1)\right)}{2 \lambda ^2+3 (w+1)^2}$ &
 $  \frac{\kappa  (w+1)+\lambda  (w-1)}{\sqrt{\frac{4 \lambda ^2}{3}+2 (w+1)^2}} $& $\frac{\mu_{21}}{6 \left(2 \lambda ^2+3 (w+1)^2\right)^3}$ & $ \frac{\mu_{22}}{6 \left(2 \lambda ^2+3 (w+1)^2\right)^3}$ & $\frac{\mu_{23}}{6 \left(2 \lambda ^2+3 (w+1)^2\right)^3}$ \\\hline
$L_-$ & $\frac{w+1}{\sqrt{\frac{2 \lambda ^2}{3}+(w+1)^2}}$ &$ -\frac{\lambda }{\sqrt{\lambda ^2+\frac{3}{2} (w+1)^2}}$ & $0 $&$ \frac{2 \left(\lambda ^2+3(w+1)\right)}{2 \lambda ^2+3 (w+1)^2} $&
  $ -\frac{\kappa  (w+1)+\lambda  (w-1)}{\sqrt{\frac{4 \lambda ^2}{3}+2 (w+1)^2}}$ & $\frac{-\mu_{21}}{6 \left(2 \lambda ^2+3 (w+1)^2\right)^3}$ & $ \frac{-\mu_{22}}{6 \left(2 \lambda ^2+3 (w+1)^2\right)^3} $& $\frac{-\mu_{23}}{6 \left(2 \lambda ^2+3 (w+1)^2\right)^3}$ \\\hline
${}_+M_\pm$ & $\frac{w-1}{\sqrt{\frac{2 \kappa ^2}{3}+(w-1)^2}} $& $\frac{\kappa }{\sqrt{\kappa ^2+\frac{3}{2} (w-1)^2}} $& $\pm \frac{w-1}{\sqrt{\frac{2 \kappa ^2}{3}+(w-1)^2}}$ & $\frac{2 \kappa ^2}{2 \kappa ^2+3
   (w-1)^2}$ & $\frac{\mu_{31}}{6 \left(2 \kappa ^2+3 (w-1)^2\right)^{7/2}}$ & $\frac{\mu_{32}}{6 \left(2 \kappa ^2+3 (w-1)^2\right)^{7/2}}$ & $ \frac{\mu_{33}}{6 \left(2 \kappa ^2+3 (w-1)^2\right)^{7/2}}$ & $\frac{\mu_{34}}{6 \left(2 \kappa ^2+3 (w-1)^2\right)^{7/2}} $\\\hline
${}_-M_\pm$ &  $-\frac{w-1}{\sqrt{\frac{2 \kappa ^2}{3}+(w-1)^2}}$ & $-\frac{\kappa }{\sqrt{\kappa ^2+\frac{3}{2} (w-1)^2}} $& $\pm \frac{w-1}{\sqrt{\frac{2 \kappa ^2}{3}+(w-1)^2}}$ &$ \frac{2 \kappa ^2}{2 \kappa
   ^2+3 (w-1)^2}$ & $\frac{-\mu_{31}}{6 \left(2 \kappa ^2+3 (w-1)^2\right)^{7/2}}$ & $\frac{-\mu_{32}}{6 \left(2 \kappa ^2+3 (w-1)^2\right)^{7/2}}$ & $  \frac{-\mu_{33}}{6 \left(2 \kappa ^2+3 (w-1)^2\right)^{7/2}}$ &$ \frac{-\mu_{34}}{6 \left(2 \kappa ^2+3 (w-1)^2\right)^{7/2}} $\\\hline
$N_\pm$ & $ 0 $& $0$ & $\pm \frac{\sqrt{\lambda +3 \lambda  w}}{\sqrt{3} \sqrt{\kappa  (w+1)+\lambda  (w-1)}} $& $\frac{2 (\kappa -2 \lambda )}{3 (\kappa  (w+1)+\lambda  (w-1))} $& $\mu_1$ & $-\mu_1 $& $\mu_2$& $-\mu_2 $\\\hline
$I$&$0$ & $-1$ & $0$ &$1$ &$ -\frac{1}{2} \sqrt{3} (w-1) $&$ -\frac{1}{2} \sqrt{3} (w-1)$ & $-\sqrt{3} (w+1)$ & $-\frac{w+1}{\sqrt{3}} $\\\hline
$J$& $0$ &$1$ & $0$ & $1 $& $\frac{1}{2} \sqrt{3} (w-1)$ & $\frac{1}{2} \sqrt{3} (w-1)$ & $\sqrt{3} (w+1)$ & $\frac{3 w+1}{\sqrt{3}} $\\\hline
$P_\pm$& $\pm\frac{1}{2} \sqrt{3 w+1}$ &$ 0$ & $0$ & $1$ & $\frac{\sqrt{(w-1)(3 w+1)}}{\sqrt{2}}$ & $-\frac{\sqrt{(w-1)(3 w+1)}}{\sqrt{2}} $& $-\frac{\kappa  \sqrt{3 w+1}}{2 \sqrt{2}}$ &$ \frac{\lambda 
   \sqrt{3 w+1}}{\sqrt{2}}$ \\\hline
\end{tabular}
}
    \caption{Equilibrium points of the system \eqref{eq:128}, \eqref{eq:129}, \eqref{eq:130} and \eqref{eq:131} and their eigenvalues. }
    \label{tab:2A}
\end{table}
The equilibrium points of the system \eqref{eq:128}, \eqref{eq:129}, \eqref{eq:130} and \eqref{eq:131} and their eigenvalues are presented in Tab.  \ref{tab:2A}, 
where  $\mu_{i j}, i\in\{1,2,3\}, j\in\{1,2,3,4\}$ is the $j$-th root of the  $i$-th  polynomial defined below (some polynomials have three roots and other have four roots) \newline 
 $P_1(\mu):=-36 \kappa  \mu  \left(\left(\kappa ^2+10\right) \lambda +2 \kappa  \lambda ^2+4 \kappa +\lambda ^3\right) \left((\kappa +\lambda )^2+6\right)^4+2 \sqrt{3} \mu ^2 (5 \kappa -4 \lambda )
   \left((\kappa +\lambda )^2+6\right)^2-144 \sqrt{3} \kappa  (\kappa -2 \lambda ) (\kappa +\lambda ) (\lambda  (\kappa +\lambda )+6) \left((\kappa +\lambda )^2+6\right)^6+\mu ^3$,  \newline
$P_2(\mu):= \mu ^3-\sqrt{6} \lambda  \mu ^2 (9 w-1) \left(2 \lambda ^2+3 (w+1)^2\right)^{5/2}-36 \mu  (w-1) \left(2 \lambda ^2+9 (w+1)^2\right) \left(2 \lambda ^2+3 (w+1)^2\right)^5+216 \sqrt{6} \lambda
    (w-1) (w+1) (3 w+1) \left(\lambda ^2+3(w+1)\right) \left(2 \lambda ^2+3 (w+1)^2\right)^{15/2}$,
    \newline
    $P_3(\mu):= \mu ^4-\sqrt{6} \mu ^3 \left(2 \kappa ^2+3 (w-1)^2\right)^3 (5 \kappa  (3 w+1)+6 \lambda  (w-1))+36 \kappa  \mu ^2 \left(2 \kappa ^2+3 (w-1)^2\right)^6 (\lambda +\kappa  (3 w-1) (3
   w+5)+\lambda  w (9 w-10))+216 \sqrt{6} \kappa ^2 \mu  (w-1) \left(2 \kappa ^2+3 (w-1)^2\right)^9 (\kappa  (3 (w-2) w-5)-4 \lambda  (w-1))-7776 \kappa ^3 (w-1)^2 (3 w+1) \left(2 \kappa
   ^2+3 (w-1)^2\right)^{12} (\kappa  (w+1)+\lambda  (w-1))$,
and 
\newline
\begin{footnotesize}
$\mu_1=\frac{\sqrt{3 w+1} \sqrt{\kappa ^2 \lambda -\sqrt{\kappa ^2 \lambda ^4+2 \kappa  \lambda ^3 \left(\kappa ^2+2 w-2\right)-2 \kappa  \lambda  (w+1) \left(\kappa ^2+2 w-2\right)+\kappa
   ^2 (w+1)^2+\lambda ^2 \left(\kappa ^4+2 \kappa ^2 (w+9)+4 (w-1)^2\right)}+\kappa  \left(\lambda ^2+w+1\right)-2 \lambda  (w-1)}}{\sqrt{6} \sqrt{\kappa  (w+1)+\lambda 
   (w-1)}}$, \newline 
   $\mu_2= \frac{\sqrt{3 w+1} \sqrt{\kappa ^2 \lambda +\sqrt{\kappa ^2 \lambda ^4+2 \kappa  \lambda ^3 \left(\kappa ^2+2 w-2\right)-2 \kappa  \lambda  (w+1) \left(\kappa ^2+2
   w-2\right)+\kappa ^2 (w+1)^2+\lambda ^2 \left(\kappa ^4+2 \kappa ^2 (w+9)+4 (w-1)^2\right)}+\kappa  \left(\lambda ^2+w+1\right)-2 \lambda  (w-1)}}{\sqrt{6} \sqrt{\kappa  (w+1)+\lambda 
   (w-1)}}$. 
\end{footnotesize}

We can find the inflationary condition with these variables where $q$, the deceleration parameter, is negative:
\begin{align}
    q = \frac{6 \eta ^2-4 \Phi ^2+3 (w+1) \Omega -2}{2 h^2}<0.
\end{align}

The equilibrium points $\left(\Phi, h, \eta, \Omega\right)$ of the system \eqref{eq:128}, \eqref{eq:129}, \eqref{eq:130} and \eqref{eq:131} are the following. 

\begin{itemize}
\item $A, B: \left(1, 0, \pm 1, 0 \right)$ has eigenvalues $\left\{\frac{\kappa }{\sqrt{2}},\sqrt{2} \kappa, \sqrt{2} \kappa, \sqrt{2} (\kappa +\lambda ) \right\}$.  Therefore $A$ and $B$ are sinks for $\{\kappa<0,\lambda<-\kappa\}$, saddle for $\left\{\kappa<0,\lambda>-\kappa \right\} $, or $ \left\{\kappa>0,\lambda<-\kappa\right\}$, and a source for $\{\kappa>0,\lambda>-\kappa\}$.  Nonhyperbolic for $\kappa=0$ or $\kappa=-\lambda$. The deceleration parameter is $q=0$. Inflation does not occur.  Compared with the results of \S \ref{Subsect:II1}, these points represent $O^+$. 

\item $C, D: \left(-1 , 0, \pm 1, 0\right)$, with eigenvalues $\left\{-\sqrt{2} \kappa, -\sqrt{2} \kappa, -\frac{\kappa }{\sqrt{2}}, -\sqrt{2} (\kappa +\lambda )\right\}$. Therefore, $C$ and $D$ are a sink for $\{\kappa>0, \lambda >-\kappa\}$, a saddle for $\{\kappa<0, \lambda >-\kappa\} $, or $ \{\kappa>0, \lambda <-\kappa\}$ or a source for  $\{\kappa<0, \lambda <-\kappa\}$.  Nonhyperbolic for $\kappa=0$ or $\kappa=-\lambda$. The deceleration parameter is $q=0$. Inflation does not occur.  Compared with the results of \S \ref{Subsect:II1}, these points represent $O^-$.

\item $E, F: \left(1, -\frac{1}{2} \sqrt{\frac{3}{2}} \kappa, \pm 1, 0\right)$, with eigenvalues $\left\{-\frac{\kappa }{\sqrt{2}}, \sqrt{2} \kappa, \frac{\kappa  (3 w+1)}{2 \sqrt{2}}, -\frac{\kappa -2 \lambda }{\sqrt{2}}\right\}$. Exists for $\kappa^2\le\frac{8}{3}$. Saddle for all values of the parameters as long as points exist, except the values $\kappa=0$, $\kappa=2 \lambda$ or $w=-\frac{1}{3}$ when it is nonhyperbolic. The deceleration parameter is $q=0$. Inflation does not occur.

\item $G, H: \left(-1, \frac{1}{2} \sqrt{\frac{3}{2}} \kappa, \pm 1, 0\right)$, with eigenvalues $\left\{\frac{\kappa }{\sqrt{2}}, -\sqrt{2} \kappa,  -\frac{\kappa  (3 w+1)}{2 \sqrt{2}}, \frac{\kappa -2 \lambda }{\sqrt{2}}\right\}$. Exists for $\kappa^2\le\frac{8}{3}$. Saddle for all values of the parameters as long as points exist, except for the values $\kappa=0$, $\kappa=2 \lambda$ or $w=-\frac{1}{3}$ when it is nonhyperbolic. The deceleration parameter is $q=0$. Inflation does not occur.

\item $I_+: \left(\frac{\lambda }{\sqrt{\lambda ^2+6}}, \frac{\sqrt{6}}{\sqrt{\lambda ^2+6}}, 0, 0\right)$, with eigenvalues  $\left\{-\frac{\lambda  (\kappa +\lambda )+6}{\sqrt{2} \sqrt{\lambda ^2+6}}, -\frac{\sqrt{2} \left(\lambda
   ^2+3(w+1)\right)}{\sqrt{\lambda ^2+6}}, -\frac{\sqrt{2} \left(\lambda ^2+2\right)}{\sqrt{\lambda ^2+6}}, -\frac{\sqrt{\lambda ^2+6}}{\sqrt{2}}\right\}$. Exists for all $\lambda, \kappa$. Assuming $-1 \le w \le 1$, it is a sink for $\{-\lambda (\lambda + \kappa) < 6\}$; a source for $\{-\lambda (\lambda + \kappa) > 6\}$. The deceleration parameter reads $q= -1 - \lambda^2/2$. Inflation always occurs. 
   
\item $I_-: \left(-\frac{\lambda }{\sqrt{\lambda ^2+6}}, -\frac{\sqrt{6}}{\sqrt{\lambda ^2+6}}, 0, 0\right)$, with eigenvalues $\left\{\frac{\lambda  (\kappa +\lambda )+6}{\sqrt{2} \sqrt{\lambda ^2+6}}, \frac{\sqrt{2} \left(\lambda ^2+3(w+1)\right)}{\sqrt{\lambda ^2+6}}, \frac{\sqrt{2} \left(\lambda ^2+2\right)}{\sqrt{\lambda ^2+6}}, \frac{\sqrt{\lambda ^2+6}}{\sqrt{2}}\right\}$. Exists for all $\lambda, \kappa$. Assuming $-1 \leq w \leq 1$, it is a sink for $\{-\lambda (\lambda + \kappa) > 6\}$; a source for $\{-\lambda (\lambda + \kappa) < 6\}$. The deceleration parameter reads $q= -1 - \lambda^2/2$. Inflation always occurs.
   
\item $J_\pm : \left(\frac{\sqrt{6}}{\sqrt{(\kappa +\lambda )^2+6}}, -\frac{\kappa +\lambda }{\sqrt{(\kappa +\lambda )^2+6}}, \pm \frac{\sqrt{\lambda  (\kappa +\lambda )+6}}{\sqrt{(\kappa +\lambda )^2+6}}, 0\right)$, which exists for $0\le\kappa(\kappa+\lambda)$
with eigenvalues 
  $\left\{\frac{\sqrt{3} (\kappa  (w+1)+\lambda  (w-1))}{\sqrt{(\kappa +\lambda )^2+6}}, \frac{\mu_{11}}{6 \left((\kappa +\lambda )^2+6\right)^{5/2}}, \frac{\mu_{12}}{6 \left((\kappa +\lambda )^2+6\right)^{5/2}}, \frac{\mu_{13}}{6 \left((\kappa +\lambda )^2+6\right)^{5/2}}\right\}$. The deceleration parameter reads $q= 2-\frac{3 \kappa }{\kappa +\lambda }$. Then, it is inflationary for $\frac{\kappa }{\kappa +\lambda }>\frac{2}{3}$. 
  
\item $K_\pm : \left(-\frac{\sqrt{6}}{\sqrt{(\kappa +\lambda )^2+6}}, \frac{\kappa +\lambda }{\sqrt{(\kappa +\lambda )^2+6}},  \pm \frac{\sqrt{\lambda  (\kappa +\lambda )+6}}{\sqrt{(\kappa +\lambda )^2+6}}, 0\right)$, with eigenvalues \newline
  $\left\{-\frac{\sqrt{3} (\kappa  (w+1)+\lambda  (w-1))}{\sqrt{(\kappa +\lambda )^2+6}}, -\frac{\mu_{11}}{6 \left((\kappa +\lambda )^2+6\right)^{5/2}}, -\frac{\mu_{12}}{6 \left((\kappa +\lambda )^2+6\right)^{5/2}}, -\frac{\mu_{13}}{6 \left((\kappa +\lambda )^2+6\right)^{5/2}}\right\}$.
For simplicity, we impose the conditions $-1 \leq  w \leq 1, \kappa> 0, \lambda < -\kappa$. The deceleration parameter reads $q= 2-\frac{3 \kappa }{\kappa +\lambda }$. Then, it is inflationary for $\frac{\kappa }{\kappa +\lambda }>\frac{2}{3}$.

\item $L_+: \left(-\frac{w+1}{\sqrt{\frac{2 \lambda ^2}{3}+(w+1)^2}}, \frac{\lambda }{\sqrt{\lambda ^2+\frac{3}{2} (w+1)^2}}, 0, \frac{2 \left(\lambda ^2+3(w+1)\right)}{2 \lambda ^2+3 (w+1)^2}\right)$, with eigenvalues \newline
 $\left\{\frac{\kappa  (w+1)+\lambda  (w-1)}{\sqrt{\frac{4 \lambda ^2}{3}+2 (w+1)^2}}, \frac{\mu_{21}}{6 \left(2 \lambda ^2+3 (w+1)^2\right)^3},  \frac{\mu_{22}}{6 \left(2 \lambda ^2+3 (w+1)^2\right)^3}, \frac{\mu_{23}}{6 \left(2 \lambda ^2+3 (w+1)^2\right)^3}\right\}$. The deceleration parameter is $q=\frac{1}{2} (3 w+1)$. Inflation occurs for $w<-\frac{1}{3}$.
 
\item $L_-: \left(\frac{w+1}{\sqrt{\frac{2 \lambda ^2}{3}+(w+1)^2}}, -\frac{\lambda }{\sqrt{\lambda ^2+\frac{3}{2} (w+1)^2}}, 0, \frac{2 \left(\lambda ^2+3(w+1)\right)}{2 \lambda ^2+3 (w+1)^2}\right)$, with eigenvalues \newline 
  $\left\{-\frac{\kappa  (w+1)+\lambda  (w-1)}{\sqrt{\frac{4 \lambda ^2}{3}+2 (w+1)^2}},\frac{-\mu_{21}}{6 \left(2 \lambda ^2+3 (w+1)^2\right)^3}, \frac{-\mu_{22}}{6 \left(2 \lambda ^2+3 (w+1)^2\right)^3}, \frac{-\mu_{23}}{6 \left(2 \lambda ^2+3 (w+1)^2\right)^3}\right\}$. The deceleration parameter is $q=\frac{1}{2} (3 w+1)$. Inflation occurs for $w<-\frac{1}{3}$.
  
\item ${}_+M_\pm: \left(\frac{w-1}{\sqrt{\frac{2 \kappa ^2}{3}+(w-1)^2}}, \frac{\kappa }{\sqrt{\kappa ^2+\frac{3}{2} (w-1)^2}}, \pm \frac{w-1}{\sqrt{\frac{2 \kappa ^2}{3}+(w-1)^2}},\frac{2 \kappa ^2}{2 \kappa ^2+3
   (w-1)^2}\right)$, with eigenvalues \newline $\left\{\frac{\mu_{31}}{6 \left(2 \kappa ^2+3 (w-1)^2\right)^{7/2}},\frac{\mu_{32}}{6 \left(2 \kappa ^2+3 (w-1)^2\right)^{7/2}}, \frac{\mu_{33}}{6 \left(2 \kappa ^2+3 (w-1)^2\right)^{7/2}}, \frac{\mu_{34}}{6 \left(2 \kappa ^2+3 (w-1)^2\right)^{7/2}} \right\}$. The deceleration parameter is $q=\frac{1}{2} (3 w+1)$. Inflation occurs for $w<-\frac{1}{3}$.
   
\item ${}_-M_\pm: \left(-\frac{w-1}{\sqrt{\frac{2 \kappa ^2}{3}+(w-1)^2}}, -\frac{\kappa }{\sqrt{\kappa ^2+\frac{3}{2} (w-1)^2}}, \pm \frac{w-1}{\sqrt{\frac{2 \kappa ^2}{3}+(w-1)^2}}, \frac{2 \kappa ^2}{2 \kappa
   ^2+3 (w-1)^2}\right)$, with eigenvalues $\left\{\frac{-\mu_{31}}{6 \left(2 \kappa ^2+3 (w-1)^2\right)^{7/2}}, \frac{-\mu_{32}}{6 \left(2 \kappa ^2+3 (w-1)^2\right)^{7/2}},  \frac{-\mu_{33}}{6 \left(2 \kappa ^2+3 (w-1)^2\right)^{7/2}}, \frac{-\mu_{34}}{6 \left(2 \kappa ^2+3 (w-1)^2\right)^{7/2}}\right\}$. The deceleration parameter is $q=\frac{1}{2} (3 w+1)$. Inflation occurs for $w<-\frac{1}{3}$.

\item $N_\pm : \left(0, 0, \pm\frac{\sqrt{\lambda +3 \lambda  w}}{\sqrt{3} \sqrt{\kappa  (w+1)+\lambda  (w-1)}} , \frac{2 (\kappa -2 \lambda )}{3 (\kappa  (w+1)+\lambda  (w-1))}\right)$, with eigenvalues $\left\{\mu_1, -\mu_1, \mu_2, -\mu_2 \right\}$. It is a saddle. The deceleration parameter is $q=0$. Inflation does not occur.

\item $I: \left(0, -1, 0, 1\right)$, with eigenvalues $\left\{-\frac{1}{2} \sqrt{3} (w-1), -\frac{1}{2} \sqrt{3} (w-1), -\sqrt{3} (w+1),-\frac{w+1}{\sqrt{3}}\right\}$. It always exists. It is nonhyperbolic for $w\in\{-1,1\}$. The point is a saddle for $-1<w<1$.  The deceleration parameter is $q=\frac{1}{2} (3 w+1)$. Inflation occurs for $w<-\frac{1}{3}$.

\item $J: \left(0, 1, 0, 1 \right)$, with eigenvalues $\left\{\frac{1}{2} \sqrt{3} (w-1), \frac{1}{2} \sqrt{3} (w-1), \sqrt{3} (w+1), \frac{3 w+1}{\sqrt{3}}\right\}$. It always exists. It is nonhyperbolic for $w\in\{-1,1\}$. The point is a saddle for $-1<w<1$. The deceleration parameter is $q=\frac{1}{2} (3 w+1)$. Inflation occurs for $w<-\frac{1}{3}$.

\item $P_\pm : \left(\pm\frac{1}{2} \sqrt{3 w+1}, 0, 0, 1\right)$, with eigenvalues  $\left\{\frac{\sqrt{(w-1)(3 w+1)}}{\sqrt{2}}, -\frac{\sqrt{(w-1)(3 w+1)}}{\sqrt{2}}, -\frac{\kappa  \sqrt{3 w+1}}{2 \sqrt{2}}, \frac{\lambda   \sqrt{3 w+1}}{\sqrt{2}}\right\}$. Exists for $w>-\frac{1}{3}$. Nonhyperbolic with two reals and two imaginary eigenvalues.  Always a saddle as long as they exist.
   
\end{itemize}

\subsubsection{ Vacuum model ($\rho_m=0$) with positive curvature ($K = +1$)}
\label{SectIIIB1}
In this case, the restrictions become 
\begin{align}
  \Psi & = 1 -  \eta ^2,\\
 \Omega_{K}& = 1 -h^2-\Phi ^2, 
\end{align}
and we have the reduced system 
\begin{align}
  \Phi^{\prime} & =  \frac{\left(\Phi ^2-1\right) \left(\eta ^2 (\kappa +\lambda )-\lambda \right)}{\sqrt{2}}+\sqrt{3} \left(\eta
   ^2-1\right) h \Phi, \label{EQ.(148)}\\
\eta^{\prime} & =  \frac{1}{2} \eta 
   \left(\eta ^2-1\right) \left(2 \sqrt{3} h+\sqrt{2} \Phi  (\kappa +\lambda )\right), \label{EQ.(149)}\\
 h^{\prime} & =   \frac{1}{6} \left(\sqrt{3} \left(-6 \eta ^2+4 \Phi ^2+2\right)+\sqrt{3} \left(6 \eta
   ^2-2\right) h^2+3 \sqrt{2} h \Phi  \left(\eta ^2 (\kappa +\lambda )-\lambda \right)\right). \label{EQ.(150)}
\end{align}

We can find the inflationary condition with these variables where $q$, the deceleration parameter, is negative:
\begin{align}
    q = -\frac{1}{h^2}(2\Phi^2-3\eta^2+1)<0
\end{align}

The equilibrium points with coordinates $(\Phi,\eta,h)$ are as follows: 

\begin{itemize}
    \item $A,B: (1,\pm 1,0 )$, which exists for all parameter values. The eigenvalues of the linearization are $\left\{\sqrt{2}\left( \kappa + \lambda \right),\frac{1}{\sqrt{2}} \kappa, \sqrt{2} \kappa \right\}$. Therefore, $A$ and $B$ are a sink for $\{\kappa<0, \lambda <-\kappa\}$, a saddle for $\{\kappa<0, \lambda >-\kappa\} $, or $ \{\kappa>0, \lambda <-\kappa\}$ or a source for  $\{\kappa>0, \lambda >-\kappa\}$. Nonhyperbolic for $\kappa=0$ or $\kappa=-\lambda$. Inflation does not occur.  Compared with the results of \S \ref{Subsect:II1}, these points represent $O^+$. 

   \item $C,D: (-1,\pm 1,0 )$, which exists for all parameter values. The eigenvalues of the linearization are $\left\{-\sqrt{2}\left( \kappa + \lambda \right),-\frac{1}{\sqrt{2}} \kappa, -\sqrt{2} \kappa \right\}$. Therefore, $C$ and $D$ are a sink for $\{\kappa>0, \lambda >-\kappa\}$, a saddle for $\{\kappa<0, \lambda >-\kappa\} $, or $ \{\kappa>0, \lambda <-\kappa\}$ or a source for  $\kappa<0, \lambda <-\kappa$. Nonhyperbolic for $\kappa=0$ or $\kappa=-\lambda$. Inflation does not occur.  Compared with the results of \S \ref{Subsect:II1}, these points represent $O^-$. 
   
   \item $E,F: \left(1,\pm 1,-\frac{\sqrt{6}}{4}\kappa \right)$. Exists  for $\kappa^2\le\frac{8}{3}$. The eigenvalues of the linearization are $\left\{ -\frac{1}{\sqrt{2}}\left( \kappa-2\lambda \right), -\frac{1}{\sqrt{2}}\kappa, \sqrt{2}\kappa \right\}$. Saddle for all values of the parameters as long as points exist, except the values $\kappa=0$, $\kappa=2 \lambda$ or $w=-\frac{1}{3}$ when it is nonhyperbolic. The deceleration parameter is $q=0$. Inflation does not occur.
   
   \item $G,H: \left(-1,\pm 1,\frac{\sqrt{6}}{4}\kappa \right)$. Exists  for $\kappa^2\le\frac{8}{3}$. The eigenvalues of the linearization are $\left\{ \frac{1}{\sqrt{2}}\left( \kappa-2\lambda \right), \frac{1}{\sqrt{2}}\kappa, -\sqrt{2}\kappa \right\}$. Saddle for all values of the parameters as long as points exist, except the values $\kappa=0$, $\kappa=2 \lambda$ or $w=-\frac{1}{3}$ when it is nonhyperbolic. The deceleration parameter is $q=0$. Inflation does not occur.

   \item $I_{+}: \left(\frac{\lambda}{\sqrt{\lambda^2+6}},0,\sqrt{\frac{6}{\lambda^2+6}} \right)$. Exists  for $\lambda ^2<6$. The eigenvalues of the linearization are \\ $\left\{ -\frac{1}{\sqrt{2}}\frac{\kappa\lambda+\lambda^2+6}{\sqrt{\lambda^2+6}},-\frac{1}{\sqrt{2}}\sqrt{\lambda^2+6}, -\sqrt{2}\frac{\lambda^2+2}{\sqrt{\lambda^2+6}} \right\}$. Therefore $I_{+}$ is a sink for $\{\kappa\lambda+\lambda^2+6>0\}$ and a saddle for $\{\kappa\lambda+\lambda^2+6<0\}$, nonhyperbolic for $\kappa\lambda+\lambda^2+6=0$. The deceleration parameter is $q=-1-\frac{\lambda ^2}{2}$. This is a phantom-dominated solution that is always accelerating. 
   
   \item $I_{-}: \left(-\frac{\lambda}{\sqrt{\lambda^2+6}},0,-\sqrt{\frac{6}{\lambda^2+6}} \right)$. Exists  for $\lambda ^2<6$. The eigenvalues of the linearization are \\ $\left\{ \frac{1}{\sqrt{2}}\frac{\kappa\lambda+\lambda^2+6}{\sqrt{\lambda^2+6}},\frac{1}{\sqrt{2}}\sqrt{\lambda^2+6}, \sqrt{2}\frac{\lambda^2+2}{\sqrt{\lambda^2+6}} \right\}$. Therefore $I_{-}$ is a source for $\{\kappa\lambda+\lambda^2+6>0\}$ and a saddle for $\{\kappa\lambda+\lambda^2+6<0\}$, nonhyperbolic for $\kappa\lambda+\lambda^2+6=0$. The deceleration parameter is $q=-1-\frac{\lambda ^2}{2}$. This is a phantom-dominated solution that is always accelerating. 
   
   \item $J_{+},J_{-}: \left(\sqrt{\frac{6}{(\kappa+\lambda)^2+6}},\pm \sqrt{\frac{\kappa\lambda+\lambda^2+6}{(\kappa+\lambda)^2+6}},-\sqrt{\frac{(\kappa+\lambda)^2}{(\kappa+\lambda)^2+6}} \right)$. Exists  for $0\le\kappa\left( \kappa+\lambda\right)$. The eigenvalues of the linearization are \\ $\left\{ -\frac{\sqrt{3}\kappa}{2\sqrt{(\kappa+\lambda)^2+6}} \pm \frac{1}{2}\sqrt{\frac{\kappa(4\kappa^2\lambda+8\kappa\lambda^2+4\lambda^3+27\kappa+24\lambda)}{(\kappa+\lambda)^2+6}},\frac{2}{\sqrt{3}}\frac{\kappa-2\lambda}{\sqrt{(\kappa+\lambda)^2+6}} \right\}$. Always a saddle as long as they exist. The deceleration parameter reads $q= 2-\frac{3 \kappa }{\kappa +\lambda }$. Then, it is inflationary for $\frac{\kappa }{\kappa +\lambda }>\frac{2}{3}$. 
   
   \item $K_{+},K_{-}: \left(-\sqrt{\frac{6}{(\kappa+\lambda)^2+6}},\pm \sqrt{\frac{\kappa\lambda+\lambda^2+6}{(\kappa+\lambda)^2+6}},\sqrt{\frac{(\kappa+\lambda)^2}{(\kappa+\lambda)^2+6}} \right)$. Exists  for $0\le\kappa\left( \kappa+\lambda\right)$. The eigenvalues of the linearization are \\ $\left\{ \frac{\sqrt{3}\kappa}{2\sqrt{(\kappa+\lambda)^2+6}} \pm \frac{1}{2}\sqrt{\frac{\kappa(4\kappa^2\lambda+8\kappa\lambda^2+4\lambda^3+27\kappa+24\lambda)}{(\kappa+\lambda)^2+6}},-\frac{2}{\sqrt{3}}\frac{\kappa-2\lambda}{\sqrt{(\kappa+\lambda)^2+6}} \right\}$. Always a saddle as long as they exist. The deceleration parameter reads $q= 2-\frac{3 \kappa }{\kappa +\lambda }$. Then, it is inflationary for $\frac{\kappa }{\kappa +\lambda }>\frac{2}{3}$.

\end{itemize}

\subsubsection{Discussion}
We divide the results according to whether we have matter ($\rho_m>0$, \S \ref{SectIIIB}) and we have a vacuum ($\rho_m=0$, \S \ref{SectIIIB1}).   For $\rho_m>0$, $I_+$ exists for all $\lambda, \kappa$. Assuming $-1 \le w \le 1$, it is a sink for $\{-\lambda (\lambda + \kappa) < 6\}$ or a source for $\{-\lambda (\lambda + \kappa) > 6\}$. Inflation always occurs.  $I_-$ exists for all $\lambda, \kappa$. Assuming $-1 \leq w \leq 1$, it is a source for $\{-\lambda (\lambda + \kappa) < 6\}$ and sink for $\{-\lambda (\lambda + \kappa) > 6\}$.  Inflation always occurs. $J_\pm$ and    $K_\pm$, which exist for $\{0\le\kappa(\kappa+\lambda)\}$, are inflationary for $\{\frac{\kappa }{\kappa +\lambda }>\frac{2}{3}\}$. For the matter-scaling solutions $L_+$, $L_-$, ${}_+M_\pm$, 
${}_-M_\pm$, and the matter dominated solutions, $I$, and $J$, the deceleration parameter is $q=\frac{1}{2} (3 w+1)$, thus, inflation occurs for $w<-\frac{1}{3}$.
For a vacuum,  we have one early inflationary solution  $I_{-}$ for $\left\{\kappa\lambda+\lambda^2+6>0\right\}$, four intermediate inflationary solutions, $J_{+}, J_{-}$, $K_{+}, K_{-}$ for $\left\{0\le\kappa\left( \kappa+\lambda\right), \frac{\kappa }{\kappa +\lambda }>\frac{2}{3}\right\}$, and the late time solution $I_{+}$  that is a sink and inflates for $\left\{\kappa\lambda+\lambda^2+6>0\right\}$. 

\subsection{Dynamical systems formulation ($K=-1$)}
\label{SectIIIC}

Let
\begin{equation}
    \xi^2 =\frac{1}{2}\dot{\phi}^2 	+3H^2, \quad V(\phi)= V_0 e^{\lambda \phi}, \label{xi-lambda(1)}
\end{equation}
and
\begin{align}
    & \bar{h}^2= \frac{3 H^2}{\xi^2}, \; \bar{\eta}^2= \frac{e^{\kappa \phi}\dot{\psi}^2}{2 \xi^2}, \;  \bar{\Phi}^2 = \frac{\dot{\phi}^2}{2 \xi^2}, \; \bar{\Psi}= \frac{V(\phi)}{\xi^2}, \; \bar{\Omega}_K=\frac{3}{a^2 \xi^2}, \;  \bar{\Omega}= \frac{\rho_m}{\xi^2}\label{Negativem2(1)}
\end{align}
which satisfy
\begin{equation}
    \bar{h}^2 +\bar{\Phi}^2=  \bar{\eta}^2+ \bar{\Psi} + \bar{\Omega}_K +  \bar{\Omega}=1. \label{Negativem2}
\end{equation}

Using  \eqref{Negativem2}, we have bounded variables
\begin{equation}
    0 \leq \bar{h}^2, \bar{\Phi}^2, \bar{\eta}^2, \bar{\Psi}, \bar{\Omega}_K, \bar{\Omega} \leq 1.  
\end{equation}

 Then, equations \eqref{M20sp.04}, \eqref{M20sp.05}, \eqref{M20sp.06}, \eqref{M20sp.07} and \eqref{M2matter} become
 \begin{align}
 \frac{\dot{H}}{\xi^2} & = \frac{1}{6} \left(\bar{\Omega}_{K}-3 \left(\bar{\eta}^2+\bar{h}^2-\bar{\Phi}^2+w \bar{\Omega} -\bar{\Psi} \right)\right),\\
 \frac{\ddot{\phi}}{\xi^2} &= -\kappa \bar{\eta}^2 -\sqrt{6} \bar{h} \bar{\Phi} +\lambda  \bar{\Psi},\\
 \frac{\ddot{\psi}}{\dot{\psi}\xi}  &= -\sqrt{3} \bar{h}-\sqrt{2} \kappa  \bar{\Phi},\\
  \frac{\dot{\rho_m}}{\chi^2} &=-\sqrt{3} \bar{h} (1+w) \xi \bar{\Omega}.
\end{align} 
Introducing the time derivative $d\tau=\xi d t$ and taking the derivatives of the variables with respect to the new time variable, the field equations are transformed to
\begin{align}
 \bar{\Phi}^{\prime} & =-\frac{\bar{\eta}^2 \kappa }{\sqrt{2}}-\sqrt{3} \bar{h}- \bar{\Phi} +\frac{\lambda  \bar{\Psi} }{\sqrt{2}}-\frac{\bar{\Phi}  \dot{\xi}}{\xi
   ^2},
\\
 \bar{h}^{\prime} & =\frac{\bar{ \Omega}_{K}-3 \left(\bar{\eta} ^2+\bar{h}^2-\bar{\Phi} ^2+w \bar{\Omega} -\bar{\Psi} \right)}{2 \sqrt{3}}-\frac{\bar{h} \dot{\xi}}{\xi
   ^2},
\\
   \bar{\eta}^{\prime} &=-\frac{\bar{\eta}  \kappa  \bar{\Phi} }{\sqrt{2}}-\sqrt{3} \bar{\eta}  \bar{h}-\frac{\bar{\eta}  \dot{\xi}}{\xi ^2},
\\
  \bar{\Psi}^{\prime} &= \bar{\Psi} \left[\sqrt{2} \lambda  \bar{\Phi}  -\frac{2   \dot{\xi}}{\xi ^2}\right],
\\
  \bar{\Omega}^{\prime} &=-\bar{\Omega} \left[\sqrt{3} \bar{h} (1+w)  -\frac{2 \dot{\xi}}{\xi ^2}\right], 
\\
    \bar{\Omega}_K^{\prime} &=-2 \bar{\Omega}_K\left[\frac{  \bar{h} }{\sqrt{3}}+\frac{ \dot{\xi}}{\xi ^2}\right].
\end{align}
Next, substituting 
 \begin{align}
    &\frac{\dot{\xi}}{ \xi^2 } =\frac{1}{6}\left[3 \sqrt{2} \bar{\Phi}  \left(\lambda  \bar{\Psi} -\bar{\eta}^2 \kappa \right)-3 \sqrt{3} \bar{h}^3+\sqrt{3} \bar{h}
   \left(\bar{\Omega}_{K}-3 \left(\bar{\eta} ^2+\bar{\Phi}^2+w \bar{\Omega} -\bar{\Psi} \right)\right)\right] \\
    &  \bar{h}^2+\bar{\Phi}^2=  1,\\
    &  \bar{\Psi}= 1 -\bar{\eta}^2-\bar{\Omega} - \bar{\Omega}_{K},
 \end{align}
we obtain 
\begin{align}
  \bar{\Phi}^{\prime} & =  \frac{1}{6} \Big[3 \sqrt{2} \left(\bar{\Phi} ^2-1\right) \left(\bar{\eta} ^2 (\kappa +\lambda )+\lambda  (\bar{\Omega}
   +\bar{\Omega}_K-1)\right)+3 \sqrt{3} \bar{h}^3 \bar{\Phi} \nonumber \\
   & +\sqrt{3} h \bar{\Phi}  \left(6 \bar{\eta} ^2+3 \left(\bar{\Phi} ^2+w \bar{\Omega} +\bar{\Omega}
   -3\right)+2 \bar{\Omega}_K\right)\Big],
\\
     \bar{h}^{\prime} & =  \frac{1}{6} \Big[3 \sqrt{3} \bar{h}^4+\sqrt{3} \bar{h}^2 \left(6 \bar{\eta} ^2+3
   \left(\bar{\Phi} ^2+w \bar{\Omega} +\bar{\Omega} -2\right)+2 \bar{\Omega}_K\right) \nonumber\\
   & +3 \sqrt{2} \bar{h} \bar{\Phi}  \left(\bar{\eta} ^2 (\kappa
   +\lambda )+\lambda  (\bar{\Omega} +\bar{\Omega}_K-1)\right) \nonumber \\
   & +\sqrt{3} \left(-6 \bar{\eta} ^2+3 \bar{\Phi} ^2-3 (1+w) \bar{\Omega} -2
   \bar{\Omega}_K+3\right)\Big],
\\
     \bar{\eta}^{\prime} & =  \frac{1}{6} \bar{\eta}  \Big[3 \sqrt{2} \bar{\Phi}  \left(\left(\bar{\eta} ^2-1\right) \kappa
   +\lambda  \left(\bar{\eta} ^2+\bar{\Omega} +\bar{\Omega}_K-1\right)\right) \nonumber \\
   & +3 \sqrt{3} \bar{h}^3 +\sqrt{3} \bar{h} \left(6 \bar{\eta} ^2+3
   \left(\bar{\Phi} ^2+w \bar{\Omega} +\bar{\Omega} -3\right)+2 \bar{\Omega}_K\right)\Big],
\\
     \bar{\Omega}^{\prime} & = \sqrt{2} \bar{\Phi}  \bar{\Omega}  \left[\bar{\eta} ^2
   (\kappa +\lambda )+\lambda  (\bar{\Omega} +\bar{\Omega}_K-1)\right]+\sqrt{3} \bar{h}^3 \bar{\Omega} \nonumber \\
   & +\frac{\bar{h} \bar{\Omega}  \left[3
   \left(2 \bar{\eta} ^2+\bar{\Phi} ^2+w (\bar{\Omega} -1)+\bar{\Omega} -2\right)+2 \bar{\Omega}_K\right]}{\sqrt{3}}, 
\\
   \bar{\Omega}_K^{\prime} & =\sqrt{2} \bar{\Phi} 
   \bar{\Omega}_K \left(\bar{\eta} ^2 (\kappa +\lambda )+\lambda  (\bar{\Omega} +\bar{\Omega}_K-1)\right) \nonumber \\
   & +\sqrt{3} \bar{h}^3
   \bar{\Omega}_K+\frac{\bar{h} \bar{\Omega}_K \left[6 \bar{\eta} ^2+3 \bar{\Phi} ^2+3 (1+w) \bar{\Omega} +2 \bar{\Omega}_K-5\right]}{\sqrt{3}}. 
\end{align}
and the auxiliary equation 
\begin{align}
  & \bar{\Psi}^{\prime}=  -\frac{1}{3} \left(\bar{\eta} ^2+\bar{\Omega} +\bar{\Omega}_K-1\right)   \Big[3 \sqrt{2} \bar{\Phi}  \left(\bar{\eta} ^2 (\kappa
   +\lambda )+\lambda  (\bar{\Omega} +\bar{\Omega}_K)\right) \nonumber \\
   & +3 \sqrt{3} h^3+\sqrt{3} h \left(6 \bar{\eta} ^2+3 \left(\bar{\Phi} ^2+w
   \bar{\Omega} +\bar{\Omega} -1\right)+2 \bar{\Omega}_K\right)\Big].
\end{align}
As  before, we use polar coordinates 
\begin{align}
    \bar{h}= \cos(\vartheta), \quad \bar{\Phi}=\sin(\vartheta) \implies\vartheta= \arctan(\bar{\Phi}/\bar{h}). \label{polar}
\end{align}
Finally, 
\begin{align}
\bar{\eta}^{\prime}&= \frac{1}{6} \Big[3 \sqrt{2} \bar{\eta}  \sin (\vartheta ) \left(\left(\bar{\eta} ^2-1\right) \kappa +\lambda  \left(\bar{\eta}
   ^2+\bar{\Omega} +\bar{\Omega}_K-1\right)\right) \nonumber \\
   & +\sqrt{3} \bar{\eta}  \cos (\vartheta ) \left(6 \bar{\eta} ^2+3 w \bar{\Omega} +3 \bar{\Omega} +2
   \bar{\Omega}_K-6\right)\Big], \label{eq176}\\
\bar{\Omega}^{\prime}&=  \sqrt{2} \bar{\Omega}  \sin (\vartheta ) \left(\bar{\eta} ^2 (\kappa +\lambda )+\lambda  (\bar{\Omega}
   +\bar{\Omega}_K-1)\right) \nonumber \\
   & +\frac{\bar{\Omega}  \cos (\vartheta ) \left(6 \bar{\eta} ^2+3 w (\bar{\Omega} -1)+3 \bar{\Omega} +2 \bar{\Omega}_{K}-3\right)}{\sqrt{3}}, \label{eq177}
\\
\bar{\Omega}_K^{\prime}&= \sqrt{2} \bar{\Omega}_K \sin (\vartheta ) \left(\bar{\eta} ^2 (\kappa +\lambda )+\lambda  (\bar{\Omega}
   +\bar{\Omega}_K-1)\right) \nonumber \\
   & +\frac{\bar{\Omega}_K \cos (\vartheta ) \left(6 \bar{\eta} ^2+3 w \bar{\Omega} +3 \bar{\Omega} +2
   \bar{\Omega}_K-2\right)}{\sqrt{3}}, \label{eq178}
   \end{align}
   \begin{align}
 \vartheta^{\prime} & =   \frac{1}{6} \Big[\sqrt{3} \sin (\vartheta ) \left(6 \bar{\eta} ^2+3 w \bar{\Omega} +3 \bar{\Omega} +2 \bar{\Omega}_K-6\right) \nonumber \\
 & -3
   \sqrt{2} \cos (\vartheta ) \left(\bar{\eta} ^2 (\kappa +\lambda )+\lambda  (\bar{\Omega} +\bar{\Omega}_K-1)\right)\Big] \label{eq179}
\end{align}

\begin{table}[] 
    \centering
    \begin{tabular}{|c|c|c|c|c|}\hline
Label & $\bar{\eta}$ & $\bar{\Omega}$ & $\bar{\Omega}_K$ & $\vartheta$ \\\hline
$A, B$ & $\pm1$ & $0$ & $0$ & $\frac{\pi}{2}$ \\\hline
$C, D$ & $\pm 1$ &$ 0$ & $0$ & $\frac{3 \pi}{2} $\\\hline
$E$ & $0$ &$0$ & $0$ & $\arctan\left(\frac{\lambda }{\sqrt{6}}\right) $\\\hline
$F$ & $0$ & $0$ & $1$ & $0$ \\\hline
$G$ & $0$ & $0$ & $1$ & $\pi$ \\\hline
$H$ & $0$ & $0$ & $\frac{3 \left(\lambda ^2+2\right)}{3 \lambda ^2+2}$ & $-\arctan\left(\frac{\sqrt{\frac{2}{3}}}{\lambda
   }\right) $\\\hline
$I_{\pm}$ &$\pm\frac{2 \sqrt{2}}{\sqrt{3 \kappa ^2+8}} $& $0$ & $\frac{3 \kappa ^2}{3 \kappa ^2+8}$ & $-\arctan\left(\frac{2
   \sqrt{\frac{2}{3}}}{\kappa }\right) $\\\hline
$J_{\pm}$ & $\pm\frac{\sqrt{\kappa  \lambda +\lambda ^2+6}}{\sqrt{\kappa ^2+2 \kappa  \lambda +\lambda ^2+6}}$ & $0$ & $0 $&                        $-\arctan\left(\frac{\sqrt{6}}{\kappa +\lambda }\right)$ \\\hline
$K_{\pm}$ & $\pm \frac{\sqrt{3} (w-1)}{\sqrt{2 \kappa ^2+3 w^2-6 w+3}}$ & $\frac{2 \kappa ^2}{2 \kappa ^2+3 w^2-6 w+3}$ &$ 0$ & $-\arctan\left(\frac{\sqrt{\frac{3}{2}} (1-w)}{\kappa }\right)$ \\\hline
$L$ & $0 $& $\frac{2 \left(\lambda ^2+3(w+1)\right)}{2 \lambda ^2+3 w^2+6 w+3}$ & $0$ & $-\arctan\left(\frac{\sqrt{\frac{3}{2}}
   (w+1)}{\lambda }\right) $\\\hline
$M$& $0$ & $\frac{4}{3 w+1}$ & $\frac{3 (w-1)}{3 w+1}$ & $\frac{ \pi}{2}$ \\\hline
$N$& $0$ & $\frac{4}{3 w+1}$ & $\frac{3 (w-1)}{3 w+1}$ & $\frac{3\pi}{2}$ \\\hline
$O$& $0$ & $1$ & $0$ & $0$ \\\hline
$P$ &$0$ & $1$ & $0 $& $\pi$ \\\hline
    \end{tabular}
    \caption{The classes of equilibrium points  $\left(\bar{\eta}, \bar{\Omega}, \bar{\Omega}_K, \vartheta\right)$ of the system \eqref{eq176}, \eqref{eq177}, \eqref{eq178} and \eqref{eq179}.}
    \label{tab:V}
\end{table}

We can find the inflationary condition with these variables where $q$, the deceleration parameter, is negative: 
\begin{equation}
    q =\frac{1}{\cos ^2(\vartheta )} \left(3 \bar{\eta}^2+\frac{3}{2} (w+1) \bar{\Omega} +\bar{\Omega}_{K}-3\right)+2 < 0.
\end{equation}

In Tab.  \ref{tab:V} are presented the equilibrium points  $\left(\bar{\eta}, \bar{\Omega}, \bar{\Omega}_K, \vartheta\right)$ of the system \eqref{eq176}, \eqref{eq177}, \eqref{eq178} and \eqref{eq179}. They are as follows:

\begin{itemize}
\item $A,B:  \left(\pm1, 0, 0, \frac{\pi}{2}\right)$, which  exists for all parameter values. The eigenvalues of the linearization are $\left\{ \sqrt{2}(\kappa + \lambda), \frac{\sqrt{2}}{2} \kappa,\sqrt{2}\kappa, \sqrt{2}\kappa \right\} $. It is a source for $\{\lambda > -\kappa$, $\kappa > 0\}$; a saddle for $\{\lambda > -\kappa$, $\kappa < 0\}$ or $\{\lambda < -\kappa$, $\kappa > 0\}$; a sink for $\{\lambda < -\kappa$, $\kappa < 0\}$. Inflation does not occur.  Compared with the results of \S \ref{Subsect:II1}, these points represent $O^+$. 

\item $C,D:  \left(\pm1, 0, 0, \frac{3\pi}{2}\right)$, which  exists for all parameter values. The eigenvalues of the linearization are $\left\{ -\sqrt{2}(\kappa + \lambda), -\frac{\sqrt{2}}{2} \kappa,-\sqrt{2}\kappa, -\sqrt{2}\kappa \right\} $. It is a source for $\{\lambda < -\kappa$, $\kappa < 0\}$; a saddle for $\{\lambda > -\kappa, \kappa < 0\}$ or $\{\lambda < -\kappa, \kappa > 0\}$; a sink for $\{\lambda > -\kappa, \kappa > 0\}$. Inflation does not occur.  Compared with the results of \S \ref{Subsect:II1}, these points represent $O^-$.

\item $E: \left(0, 0, 0, \arctan\left(\frac{\lambda }{\sqrt{6}}\right) \right).$  The eigenvalues of the linearization are \newline $\left\{ -\frac{\lambda(\kappa+ \lambda)}{\sqrt{2}\sqrt{\lambda^2+6}},-\frac{\sqrt{2}(\lambda^2+3(w+1))}{\sqrt{\lambda^2+6}}, -\frac{\sqrt{2}(\lambda^2+2)}{\sqrt{\lambda^2+6}},-\frac{\sqrt{\lambda^2+6}}{\sqrt{2}} \right\}$. Assuming $-1 \le w \le 1$ we have a sink for $\{\lambda(\kappa+\lambda) > 0\}$. A saddle for $\{\lambda(\kappa+\lambda) < 0\}$. The deceleration parameter is $q=-1-\frac{\lambda ^2}{2}$. This solution is always inflationary, corresponding to a phantom-dominated solution. 

\item $E_{+\pi}: \left(0, 0, 0, \arctan\left(\frac{\lambda }{\sqrt{6}}\right) + \pi \right)$, which  exists for all parameter values. $ F$ denotes this point in \S \ref{subsect:ModelII-IIIA}. The eigenvalues of the linearization are \newline $\left\{ \frac{\lambda(\kappa+ \lambda)}{\sqrt{2}\sqrt{\lambda^2+6}},\frac{\sqrt{2}(\lambda^2+3(w+1))}{\sqrt{\lambda^2+6}}, \frac{\sqrt{2}(\lambda^2+2)}{\sqrt{\lambda^2+6}},\frac{\sqrt{\lambda^2+6}}{\sqrt{2}} \right\}$. Assuming $-1 \le w \le 1$ we have a source for $\{\lambda(\kappa+\lambda) > 0\}$. A saddle for $\{\lambda(\kappa+\lambda) < 0\}$. The deceleration parameter is $q=-1-\frac{\lambda ^2}{2}$. This solution is always inflationary, corresponding to a phantom-dominated solution.

\item $F: \left(0, 0, 1, 0\right)$. Exists for all parameter values. The eigenvalues of the linearization are $\left\{ -\frac{3w+1}{\sqrt{3}},\frac{2}{\sqrt{3}},-\frac{2}{\sqrt{3}},-\frac{2}{\sqrt{3}} \right\}$. Always a saddle. The deceleration parameter is $q=0$. Inflation does not occur. 

\item $G: \left(0, 0, 1, \pi \right)$. Exists for all parameter values. The eigenvalues of the linearization are $\left\{ \frac{3w+1}{\sqrt{3}},-\frac{2}{\sqrt{3}},\frac{2}{\sqrt{3}},\frac{2}{\sqrt{3}} \right\}$. Always a saddle. The deceleration parameter is $q=0$. Inflation does not occur. 

Points $F$ and $G$ in this section correspond to the points $T$ and $U$ studied in \S \ref{Sect:IIC}. They represent the Curvature dominated Milne solution $( \bar{\eta}^2=   \bar{\Phi}^2 =  \bar{\Psi}= \Omega  = 0, \bar{h}^2=\bar{\Omega}_K = 1, k =-1$) [Ref. \cite{wainwright_ellis_1997}, Sect. 9.1.6, Eq. (9.8)]). 

\item $H: \left(0, 0, \frac{3 \left(\lambda ^2+2\right)}{3 \lambda ^2+2}, -\arctan\left(\frac{\sqrt{\frac{2}{3}}}{\lambda}\right) \right).$ The deceleration parameter is $q=0$. Inflation does not occur. 

\item $H_{+\pi}: \left(0, 0, \frac{3 \left(\lambda ^2+2\right)}{3 \lambda ^2+2}, -\arctan\left(\frac{\sqrt{\frac{2}{3}}}{\lambda}\right)+\pi \right).$ 
Assuming $0\leq \bar{\Omega}_K\leq 1$, these points do not exist.  The deceleration parameter is $q=0$. Inflation does not occur. 
   
\item $I_{\pm}: \left(\pm \frac{2 \sqrt{2}}{\sqrt{3 \kappa ^2+8}}, 0, \frac{3 \kappa ^2}{3 \kappa ^2+8},  -\arctan\left(\frac{2 \sqrt{\frac{2}{3}}}{\kappa }\right) \right). $ 
The eigenvalues are \newline $\left\{-\frac{3 w+1}{\sqrt{\frac{8}{\kappa ^2}+3}},\frac{\mu_1}{2 \sqrt{3} \left(3 \kappa ^2+8\right)^3},\frac{\mu_2}{2 \sqrt{3}
   \left(3 \kappa ^2+8\right)^3},\frac{\mu_3}{2 \sqrt{3} \left(3 \kappa ^2+8\right)^3}\right\}$, where $\mu_i, i=1 \ldots 3$ are the roots of the polynomial
   $P(\mu)=\mu^3+8 \sqrt{3} \mu^2 \left(3 \kappa ^2+8\right)^{5/2} \lambda -48 \mu \kappa  \left(3 \kappa ^2+8\right)^5 (3 \kappa -2 \lambda )+384 \sqrt{3} \kappa ^2 \left(3 \kappa ^2+8\right)^{15/2} (\kappa -2 \lambda )$. The first eigenvalue is negative for $w > -\frac{1}{3}$ and positive for $w < -\frac{1}{3}$. From the other three eigenvalues, at least two have real parts of different signs, as presented in Fig.  \ref{fig:12}. Then, the equilibrium point is a saddle. The deceleration parameter is $q=0$. Inflation does not occur. 

\begin{figure}
    \centering
    \includegraphics[scale=0.7]{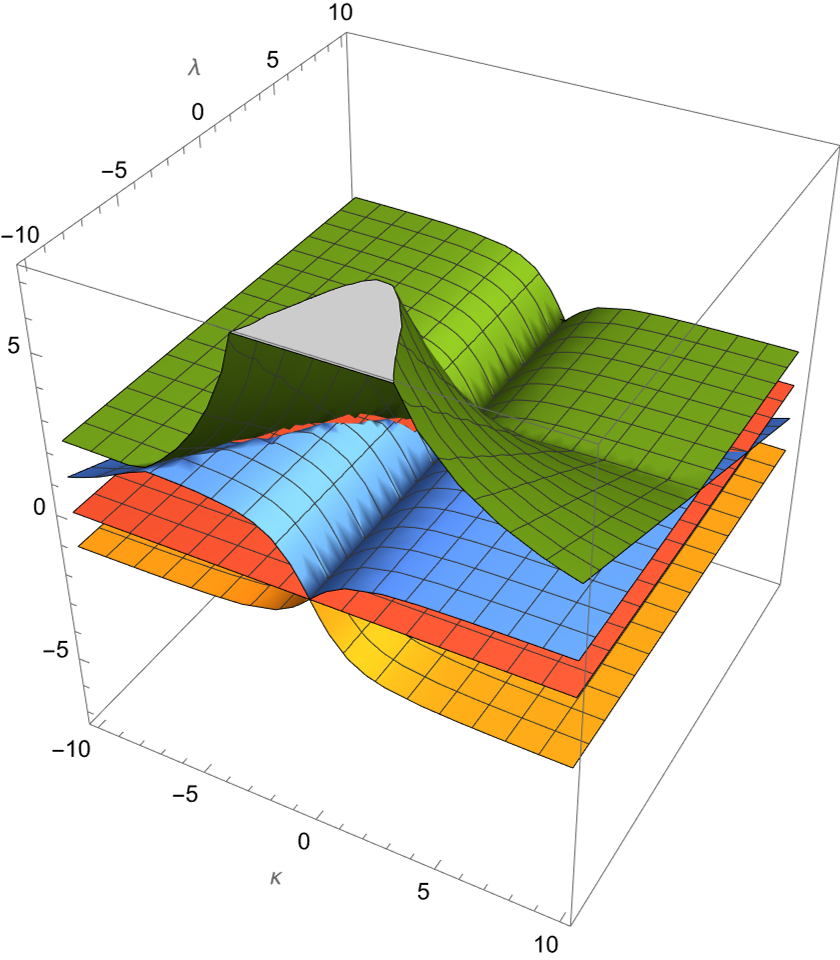}
    \caption{Reals parts of eigenvalues $\left\{\frac{\mu_1}{2 \sqrt{3} \left(3 \kappa ^2+8\right)^3},\frac{\mu_2}{2 \sqrt{3}
   \left(3 \kappa ^2+8\right)^3},\frac{\mu_3}{2 \sqrt{3} \left(3 \kappa ^2+8\right)^3}\right\}$ related to the roots of $P(\mu)=\mu^3+8 \sqrt{3} \mu^2 \left(3 \kappa ^2+8\right)^{5/2} \lambda -48 \mu \kappa  \left(3 \kappa ^2+8\right)^5 (3 \kappa -2 \lambda )+384 \sqrt{3} \kappa ^2 \left(3 \kappa ^2+8\right)^{15/2} (\kappa -2 \lambda )$, corresponding to the equilibrium point $I_{\pm}$ compared to zero. This diagram shows its saddle nature. }
    \label{fig:12}
\end{figure}

\item $I_{\pm}^{+\pi}: \left(\pm \frac{2 \sqrt{2}}{\sqrt{3 \kappa ^2+8}}, 0, \frac{3 \kappa ^2}{3 \kappa ^2+8},  -\arctan\left(\frac{2 \sqrt{\frac{2}{3}}}{\kappa }\right) + \pi\right). $ 
The eigenvalues are \newline  $\left\{\frac{3 w+1}{\sqrt{\frac{8}{\kappa ^2}+3}},\frac{\mu_1}{2 \sqrt{3} \left(3 \kappa ^2+8\right)^3},\frac{\mu_2}{2 \sqrt{3} \left(3 \kappa
   ^2+8\right)^3},\frac{\mu_3}{2 \sqrt{3} \left(3 \kappa ^2+8\right)^3}\right\}$ where $\mu_i, i=1 \ldots 3$ are the roots of the polynomial
   $P(\mu)=\mu^3-8 \sqrt{3} \mu^2 \left(3 \kappa ^2+8\right)^{5/2} \lambda -48 \mu \kappa  \left(3 \kappa ^2+8\right)^5 (3 \kappa -2 \lambda )-384 \sqrt{3} \kappa ^2 \left(3
   \kappa ^2+8\right)^{15/2} (\kappa -2 \lambda )$.  
   The first eigenvalue is negative for $w < -\frac{1}{3}$ and positive for $w > -\frac{1}{3}$. From the other three eigenvalues, at least two have real parts of different signs, as presented in Fig.  \ref{fig:13}. Then, the equilibrium point is a saddle. The deceleration parameter is $q=0$. Inflation does not occur. 

   \begin{figure}
    \centering
    \includegraphics[scale=0.7]{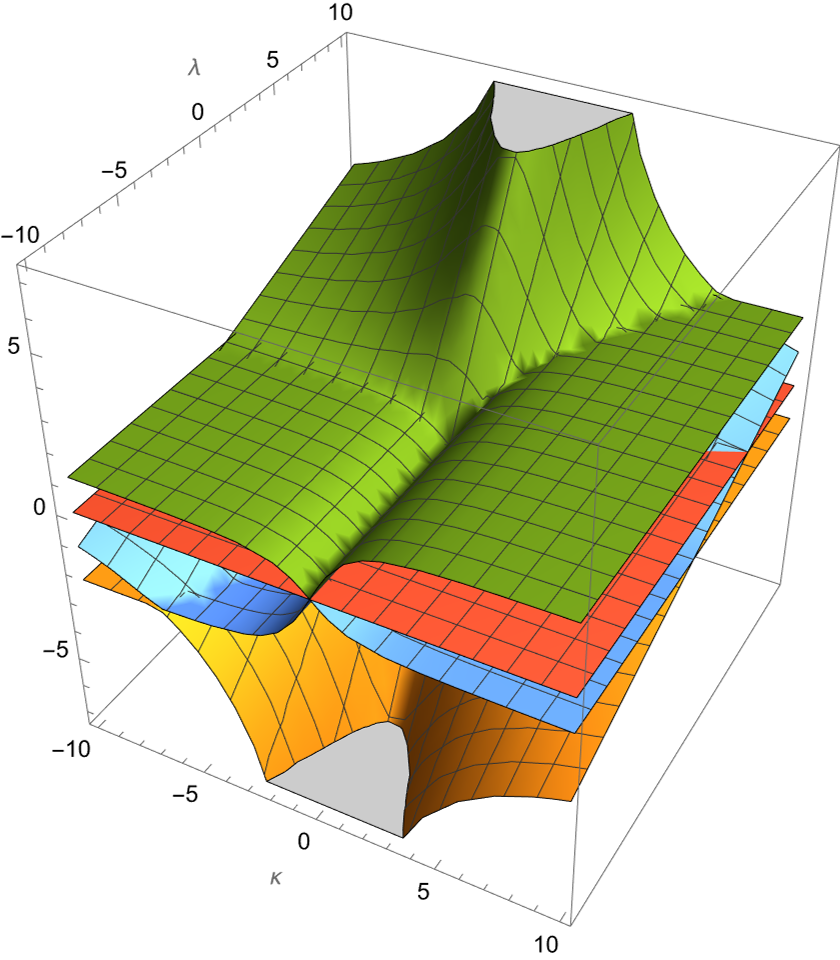}
    \caption{Reals parts of eigenvalues $\left\{\frac{\mu_1}{2 \sqrt{3} \left(3 \kappa ^2+8\right)^3},\frac{\mu_2}{2 \sqrt{3}
   \left(3 \kappa ^2+8\right)^3},\frac{\mu_3}{2 \sqrt{3} \left(3 \kappa ^2+8\right)^3}\right\}$ related to the roots of $P(\mu)=\mu^3-8 \sqrt{3} \mu^2 \left(3 \kappa ^2+8\right)^{5/2} \lambda -48 \mu \kappa  \left(3 \kappa ^2+8\right)^5 (3 \kappa -2 \lambda )-384 \sqrt{3} \kappa ^2 \left(3
   \kappa ^2+8\right)^{15/2} (\kappa -2 \lambda )$, corresponding to the equilibrium point $I_{\pm}^{+\pi}$ compared to zero. This diagram shows its saddle nature. }
    \label{fig:13}
\end{figure}

\item $J_{\pm}: \left(\pm\frac{\sqrt{\kappa  \lambda +\lambda ^2+6}}{\sqrt{\kappa ^2+2 \kappa  \lambda +\lambda ^2+6}}, 0, 0 , -\arctan\left(\frac{\sqrt{6}}{\kappa +\lambda }\right)\right). $
The eigenvalues are \\$\Big\{\frac{2 (\kappa -2 \lambda )}{\sqrt{3} \sqrt{(\kappa +\lambda )^2+6}},\frac{\sqrt{3} (\kappa  (w+1)+\lambda  (w-1))}{\sqrt{(\kappa +\lambda )^2+6}},\frac{\sqrt{\kappa  \left(4
   \left(\kappa ^2+6\right) \lambda +8 \kappa  \lambda ^2+27 \kappa +4 \lambda ^3\right) \left((\kappa +\lambda )^2+6\right)}+\sqrt{3} \kappa  \sqrt{(\kappa +\lambda )^2+6}}{2 \left((\kappa
   +\lambda )^2+6\right)},$ \newline $\frac{\sqrt{3} \kappa  \sqrt{(\kappa +\lambda )^2+6}-\sqrt{\kappa  \left(4 \left(\kappa ^2+6\right) \lambda +8 \kappa  \lambda ^2+27 \kappa +4 \lambda ^3\right)
   \left((\kappa +\lambda )^2+6\right)}}{2 \left((\kappa +\lambda )^2+6\right)}\Big\}$. 
   It is a sink for \newline $ \left\{-1<w<1, \lambda >0, \kappa <\frac{-\lambda ^2-6}{\lambda }\right\} \cup \left\{-1<w<1, \lambda >0, -\lambda <\kappa <0\right\}$. It is a source for $\left\{-1<w<1, \lambda <0, 0<\kappa <-\lambda\right\} \cup \left\{-1<w<1, \lambda <0, \kappa >\frac{-\lambda ^2-6}{\lambda }\right\}$. The deceleration parameter is $q= 2-\frac{3 \kappa }{\kappa +\lambda }$. Inflates for $\frac{\kappa }{\kappa +\lambda }>\frac{2}{3}$.

\item $J_{\pm}^{+\pi}: \left(\pm\frac{\sqrt{\kappa  \lambda +\lambda ^2+6}}{\sqrt{\kappa ^2+2 \kappa  \lambda +\lambda ^2+6}}, 0, 0 , -\arctan\left(\frac{\sqrt{6}}{\kappa +\lambda }\right) + \pi\right). $
The eigenvalues are \\
$\Big\{-\frac{2 (\kappa -2 \lambda )}{\sqrt{3} \sqrt{(\kappa +\lambda )^2+6}},-\frac{\sqrt{3} (\kappa  (w+1)+\lambda  (w-1))}{\sqrt{(\kappa +\lambda )^2+6}},\frac{\sqrt{\kappa  \left(4
   \left(\kappa ^2+6\right) \lambda +8 \kappa  \lambda ^2+27 \kappa +4 \lambda ^3\right) \left((\kappa +\lambda )^2+6\right)}-\sqrt{3} \kappa  \sqrt{(\kappa +\lambda )^2+6}}{2 \left((\kappa
   +\lambda )^2+6\right)},$ \newline $-\frac{\sqrt{\kappa  \left(4 \left(\kappa ^2+6\right) \lambda +8 \kappa  \lambda ^2+27 \kappa +4 \lambda ^3\right) \left((\kappa +\lambda )^2+6\right)}+\sqrt{3}
   \kappa  \sqrt{(\kappa +\lambda )^2+6}}{2 \left((\kappa +\lambda )^2+6\right)}\Big\}$. 
   It is a source for \newline $ \left\{-1<w<1, \lambda >0, \kappa <\frac{-\lambda ^2-6}{\lambda }\right\} \cup \left\{-1<w<1, \lambda >0, -\lambda <\kappa <0\right\}$. It is a sink for \newline $\left\{-1<w<1, \lambda <0, 0<\kappa <-\lambda\right\} \cup \left\{-1<w<1, \lambda <0,  \kappa >\frac{-\lambda ^2-6}{\lambda }\right\}$. The deceleration parameter is $q= 2-\frac{3 \kappa }{\kappa +\lambda }$. Inflates for $\frac{\kappa }{\kappa +\lambda }>\frac{2}{3}$.

\item  $K_{\pm}: \left(\pm \frac{\sqrt{3} (w-1)}{\sqrt{2 \kappa ^2+3 w^2-6 w+3}}, \frac{2 \kappa ^2}{2 \kappa ^2+3 w^2-6 w+3},  0, -\arctan\left(\frac{\sqrt{\frac{3}{2}} (1-w)}{\kappa }\right)\right). $  The deceleration parameter is $q=\frac{1}{2} (3 w+1)$. It is an inflationary solution for $w<-\frac{1}{3}$. 

The eigenvalues are complicated expressions of $(\lambda, \kappa, w)$. For simplicity, we choose the case $w=0$ corresponding to dust matter. In this case, the eigenvalues reduce in complexity. One real eigenvalue is always positive.  From the rest of the eigenvalues, at least two have real parts of different signs for all choices of $\lambda$ and $\kappa$, as presented in Fig.  \ref{fig:Fig13}. Hence, the points are saddles.

\begin{figure}
    \centering
    \includegraphics[scale=0.7]{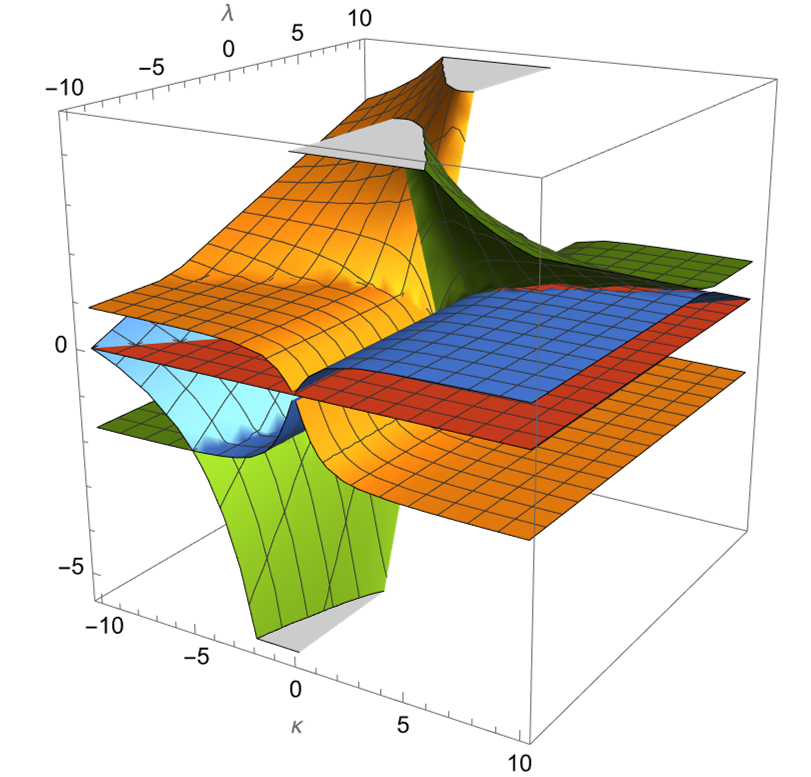}
    \caption{Real parts of the complicated eigenvalues related to $K_{\pm}$ compared with zero for dust.  This diagram shows its saddle nature.}
    \label{fig:Fig13}
\end{figure}
\item  $K_{\pm}^{+\pi}: \left(\pm \frac{\sqrt{3} (w-1)}{\sqrt{2 \kappa ^2+3 w^2-6 w+3}}, \frac{2 \kappa ^2}{2 \kappa ^2+3 w^2-6 w+3},  0, -\arctan\left(\frac{\sqrt{\frac{3}{2}} (1-w)}{\kappa }\right) + \pi\right). $  The deceleration parameter is $q=\frac{1}{2} (3 w+1)$. It is an inflationary solution for $w<-\frac{1}{3}$. 

The eigenvalues are complicated expressions of $(\lambda, \kappa, w)$. For simplicity, we choose the case $w=0$ corresponding to dust matter. In this case, the eigenvalues reduce in complexity. One real eigenvalue is always negative. From the rest of the eigenvalues, at least two have real parts of different signs for all choices of $\lambda$ and $\kappa$ as presented in Fig.  \ref{fig:Fig14}. Hence, the points are saddles. 

\begin{figure}
    \centering
    \includegraphics[scale=0.7]{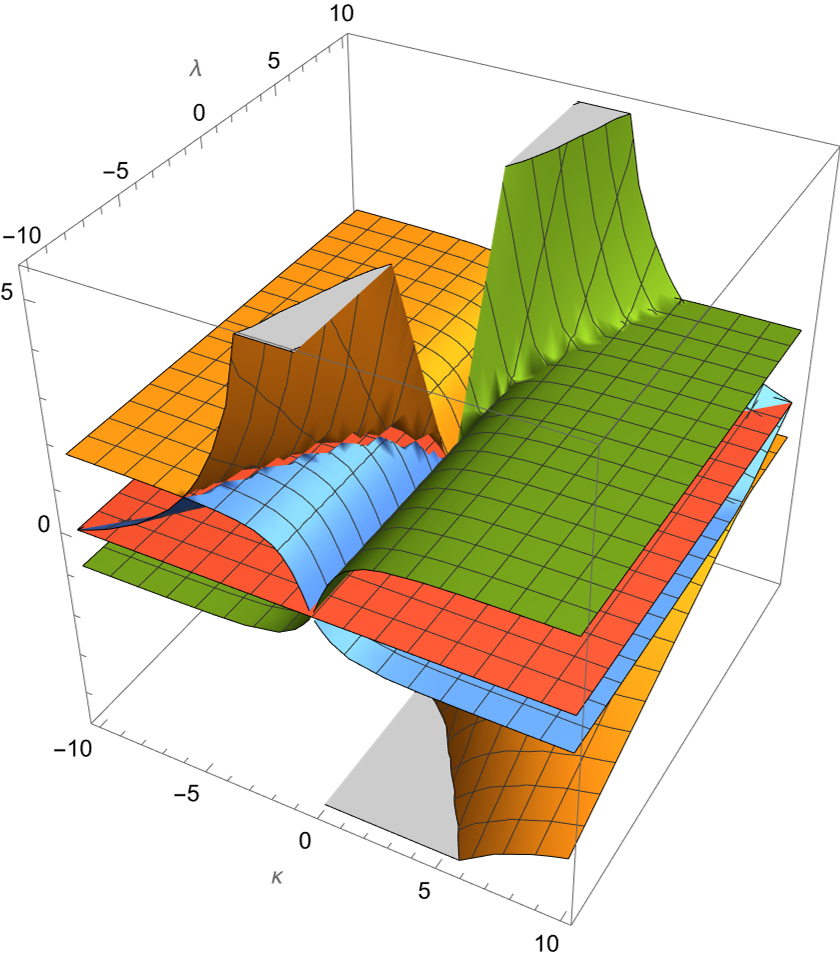}
    \caption{Real parts of the complicated eigenvalues related to $K_{\pm}^{+\pi}$compared with zero for dust. This diagram shows its saddle nature.}
    \label{fig:Fig14}
\end{figure}

 \item  $L : \left(0, \frac{2 \left(\lambda ^2+3(w+1)\right)}{2 \lambda ^2+3 w^2+6 w+3}, 0, -\arctan\left(\frac{\sqrt{\frac{3}{2}}(w+1)}{\lambda }\right) \right).$
The eigenvalues are \\ 
\begin{footnotesize}
$\left\{\frac{3 w+1}{\sqrt{\frac{9 (w+1)^2}{2 \lambda ^2}+3}},\frac{\kappa  (w+1)+\lambda  (w-1)}{\lambda  \sqrt{\frac{2 (w+1)^2}{\lambda ^2}+\frac{4}{3}}},-\frac{\lambda(1-w) +\sqrt{w-1} \sqrt{\lambda ^2 (9 w+7)+24 (w+1)^2}}{2 \lambda  \sqrt{\frac{2 (w+1)^2}{\lambda ^2}+\frac{4}{3}}},\frac{\sqrt{w-1} \sqrt{\lambda ^2 (9 w+7)+24 (w+1)^2}+\lambda  (w-1)}{2
   \lambda  \sqrt{\frac{2 (w+1)^2}{\lambda ^2}+\frac{4}{3}}}\right\}$.
   \end{footnotesize}
   It is a sink for $\left\{-1<w<-\frac{1}{3}, \lambda <0, \kappa >\frac{\lambda(1-w)}{w+1}\right\}\cup \left\{-1<w<-\frac{1}{3}, \lambda >0, \kappa <\frac{\lambda(1-w)}{w+1}\right\}$. It is a saddle otherwise.  The deceleration parameter is $q=\frac{1}{2} (3 w+1)$. It is an inflationary solution for $w<-\frac{1}{3}$. 
  
 \item  $L_{+\pi} : \left(0, \frac{2 \left(\lambda ^2+3(w+1)\right)}{2 \lambda ^2+3 w^2+6 w+3}, 0, -\arctan\left(\frac{\sqrt{\frac{3}{2}}(w+1)}{\lambda }\right) +\pi\right).$
The eigenvalues are \\ 
\begin{footnotesize}
$\left\{-\frac{3 w+1}{\sqrt{\frac{9 (w+1)^2}{2 \lambda ^2}+3}},-\frac{\kappa  (w+1)+\lambda  (w-1)}{\lambda  \sqrt{\frac{2 (w+1)^2}{\lambda ^2}+\frac{4}{3}}},\frac{\lambda(1-w) -\sqrt{w-1}\sqrt{\lambda ^2 (9 w+7)+24 (w+1)^2}}{2 \lambda  \sqrt{\frac{2 (w+1)^2}{\lambda ^2}+\frac{4}{3}}},\frac{\lambda(1-w) +\sqrt{w-1} \sqrt{\lambda ^2 (9 w+7)+24 (w+1)^2}}{2 \lambda  \sqrt{\frac{2 (w+1)^2}{\lambda ^2}+\frac{4}{3}}}\right\}$.
   \end{footnotesize}
   It is a source for $\left\{-1<w<-\frac{1}{3}, \lambda <0, \kappa >\frac{\lambda(1-w)}{w+1}\right\}\cup \left\{-1<w<-\frac{1}{3}, \lambda >0, \kappa <\frac{\lambda(1-w)}{w+1}\right\}$. It is a saddle otherwise.  The deceleration parameter is $q=\frac{1}{2} (3 w+1)$. It is an inflationary solution for $w<-\frac{1}{3}$.

 \item $M: \left(0, \frac{4}{3 w+1}, \frac{3 (w-1)}{3 w+1}, \frac{\pi}{2}\right). $ Exists for $w>1$, which does not exist under our assumption.  The deceleration parameter is $q=2$. Therefore, it represents a stiff fluid solution. It never inflates.
 
 \item $N: \left(0, \frac{4}{3 w+1}, \frac{3 (w-1)}{3 w+1}, \frac{3 \pi}{2}\right). $ Exists for $w>1$, which does not exist under our assumption.  The deceleration parameter is $q=2$. Therefore, it represents a stiff fluid solution. It never inflates.
 
 \item  $O: \left(0, 1, 0, 0\right). $ Exists for all parameter values. The eigenvalues of the linearization are \newline $\left\{ \frac{3w+1}{\sqrt{3}}, \sqrt{3}(w+1),\frac{\sqrt{3}}{2}(w-1),\frac{\sqrt{3}}{2}(w-1) \right\}$. Assuming $-1<w<1$, this is always a saddle. The deceleration parameter is $q=\frac{1}{2} (3 w+1)$. It is an inflationary solution for $w<-\frac{1}{3}$. 
 
 \item  $P: \left(0, 1, 0, \pi \right). $ Exists for all parameter values. The eigenvalues of the linearization are $\left\{ -\frac{3w+1}{\sqrt{3}}, -\sqrt{3}(w+1),\frac{\sqrt{3}}{2}(1-w),\frac{\sqrt{3}}{2}(1-w) \right\}$. Assuming $-1<w<1$, this is always a saddle. The deceleration parameter is $q=\frac{1}{2} (3 w+1)$. It is an inflationary solution for $w<-\frac{1}{3}$. 
\end{itemize}

\subsubsection{ Vacuum model ($\rho_m=0$) with negative curvature ($K = -1$)}
\label{SectIIIC1}

In this example, the restrictions become
\begin{align}
      &  \bar{h}^2+\bar{\Phi}^2=  1,\\
    &  \bar{\Psi}= 1 -\bar{\eta}^2- \bar{\Omega}_{K},
\end{align}
therefore, using polar coordinates, we have 
\begin{align}
  \bar{\eta}^{\prime} &=   \frac{\bar{\eta}  \sin (\vartheta ) \left(\left(\bar{\eta} ^2-1\right) (\kappa +\lambda )+\lambda \bar{\Omega}_{K}\right)}{\sqrt{2}}+\frac{\bar{\eta}  \left(3 \bar{\eta} ^2+\bar{\Omega}_K-3\right) \cos (\vartheta )}{\sqrt{3}}, \label{eq180}\\
   \bar{\Omega}_K^{\prime}&= \sqrt{2}
   \bar{\Omega}_K \sin (\vartheta ) \left(\bar{\eta} ^2 (\kappa +\lambda )+\lambda  (\bar{\Omega}_K-1)\right)+\frac{2
   \bar{\Omega}_K \left(3 \bar{\eta} ^2+\bar{\Omega}_K-1\right) \cos (\vartheta )}{\sqrt{3}}, \label{eq181}\\
    \vartheta^{\prime} & =  \frac{\left(3 \bar{\eta}
   ^2+\bar{\Omega}_K-3\right) \sin (\vartheta )}{\sqrt{3}}-\frac{\cos (\vartheta ) \left(\bar{\eta} ^2 (\kappa +\lambda
   )+\lambda  (\bar{\Omega}_K-1)\right)}{\sqrt{2}}. \label{eq182}
\end{align}

We can also find the inflationary condition with these variables when $q$, the deceleration parameter, is negative: 
\begin{equation}
    q =\frac{1}{\cos ^2(\vartheta )} \left(3 \bar{\eta}^2 +\bar{\Omega}_{K}-3\right)+2 < 0.
\end{equation}

\begin{table}[]
    \centering
    \resizebox{\textwidth}{!}{  
    \begin{tabular}{|c|c|c|c|c|c|c|}\hline
Label & $\bar{\eta}$  & $\bar{\Omega}_K$ & $\vartheta$ & $k_1$ & $k_2$ & $k_3$ \\\hline
$A,B$ & $\pm 1$ &$ 0$ & $\frac{\pi}{2}$ & $\frac{\kappa }{\sqrt{2}}$ & $\sqrt{2} \kappa $ & $\sqrt{2} (\kappa +\lambda )$ \\\hline
$C,D$ & $\pm 1$ & $0 $& $\frac{3 \pi}{2}$ & $-\frac{\kappa }{\sqrt{2}} $& $-\sqrt{2} \kappa$  & $-\sqrt{2} (\kappa +\lambda )$
\\\hline
$E$ & $ 0$ & $0$ & $\arctan\left(\frac{\lambda }{\sqrt{6}}\right)$ & $-\frac{\sqrt{\lambda ^2+6}}{\sqrt{2}}$ &
 $  -\frac{\lambda  (\kappa +\lambda )+6}{\sqrt{2} \sqrt{\lambda ^2+6}}$ & $-\frac{\sqrt{2} \left(\lambda^2+2\right)}{\sqrt{\lambda ^2+6}} $\\\hline
$E_{+ \pi}$ & $ 0$ & $0$ & $\arctan\left(\frac{\lambda }{\sqrt{6}} \right) +\pi$ & $\frac{\sqrt{\lambda ^2+6}}{\sqrt{2}}$ &
 $  \frac{\lambda  (\kappa +\lambda )+6}{\sqrt{2} \sqrt{\lambda ^2+6}}$ & $\frac{\sqrt{2} \left(\lambda^2+2\right)}{\sqrt{\lambda ^2+6}} $\\\hline

$F$ & $0$ & $1$ & $0$ & $-\frac{2}{\sqrt{3}}$ &$ -\frac{2}{\sqrt{3}}$ & $\frac{2}{\sqrt{3}}$\\\hline
$G$ & $0 $& $1$ & $\pi$& $\frac{2}{\sqrt{3}}$ & $\frac{2}{\sqrt{3}}$ & $-\frac{2}{\sqrt{3}}$\\\hline
$H$ & $0$ & $\frac{3 \left(\lambda ^2+2\right)}{3 \lambda ^2+2}$& $-\arctan\left(\frac{\sqrt{\frac{2}{3}}}{\lambda
   }\right)$ &$ \frac{2 \lambda -\kappa }{\sqrt{3 \lambda ^2+2}}$ & $\frac{\lambda -i \sqrt{3 \lambda
   ^2+8}}{\sqrt{3 \lambda ^2+2}}$ &$ \frac{\lambda +i \sqrt{3 \lambda ^2+8}}{\sqrt{3 \lambda ^2+2}}$\\\hline
   
$I_{\pm}$ & $\pm\frac{2}{\sqrt{\frac{3 \kappa ^2}{2}+4}}$ &$ \frac{3 \kappa ^2}{3 \kappa ^2+8}$ &$ -\arctan\left(\frac{2
   \sqrt{\frac{2}{3}}}{\kappa }\right)$ &$ -\frac{4 \kappa }{\sqrt{3 \kappa ^2+8}}$ &$ \frac{2 \kappa
   }{\sqrt{3 \kappa ^2+8}}$ & $\frac{2 (\kappa -2 \lambda )}{\sqrt{3 \kappa ^2+8}} $\\\hline

$I_{\pm}^{+\pi}$ & $\pm\frac{2}{\sqrt{\frac{3 \kappa ^2}{2}+4}}$ &$ \frac{3 \kappa ^2}{3 \kappa ^2+8}$ &$ -\arctan\left(\frac{2
   \sqrt{\frac{2}{3}}}{\kappa }\right)+\pi$ &$ \frac{4 \kappa }{\sqrt{3 \kappa ^2+8}}$ &$ -\frac{2 \kappa
   }{\sqrt{3 \kappa ^2+8}}$ & $-\frac{2 (\kappa -2 \lambda )}{\sqrt{3 \kappa ^2+8}} $\\\hline

$J_{\pm}$ & $\pm \frac{\sqrt{\kappa  \lambda +\lambda ^2+6}}{\sqrt{\kappa ^2+2 \kappa  \lambda +\lambda ^2+6}}$ &$ 0 $& $-\arctan\left(\frac{\sqrt{6}}{\kappa +\lambda }\right) $&$ \frac{2 (\kappa -2 \lambda )}{\sqrt{3}
   \sqrt{(\kappa +\lambda )^2+6}}$ & $\frac{\sqrt{3} \kappa -\sqrt{\kappa } \sqrt{4 \left(\kappa ^2+6\right) \lambda +8
   \kappa  \lambda ^2+27 \kappa +4 \lambda ^3}}{2 \sqrt{(\kappa +\lambda )^2+6}}$ & $\frac{\sqrt{\kappa } \sqrt{4
   \left(\kappa ^2+6\right) \lambda +8 \kappa  \lambda ^2+27 \kappa +4 \lambda ^3}+\sqrt{3} \kappa }{2 \sqrt{(\kappa
   +\lambda )^2+6}} $\\\hline

$J_{\pm}^{+\pi}$ & $\pm \frac{\sqrt{\kappa  \lambda +\lambda ^2+6}}{\sqrt{\kappa ^2+2 \kappa  \lambda +\lambda ^2+6}}$ &$ 0 $& $-\arctan\left(\frac{\sqrt{6}}{\kappa +\lambda }\right)+\pi $&$ -\frac{2 (\kappa -2 \lambda )}{\sqrt{3}
   \sqrt{(\kappa +\lambda )^2+6}}$ & $-\frac{\sqrt{3} \kappa -\sqrt{\kappa } \sqrt{4 \left(\kappa ^2+6\right) \lambda +8
   \kappa  \lambda ^2+27 \kappa +4 \lambda ^3}}{2 \sqrt{(\kappa +\lambda )^2+6}}$ & $-\frac{\sqrt{\kappa } \sqrt{4
   \left(\kappa ^2+6\right) \lambda +8 \kappa  \lambda ^2+27 \kappa +4 \lambda ^3}+\sqrt{3} \kappa }{2 \sqrt{(\kappa
   +\lambda )^2+6}} $\\\hline
    \end{tabular}}
    \caption{The equilibrium points  $\left(\bar{\eta}, \bar{\Omega}, \bar{\Omega}_K, \vartheta\right)$ of the system \eqref{eq180}, \eqref{eq181}, \eqref{eq182}. We assume $\kappa >0, \kappa +\lambda <0$. }
    \label{tab:VI}
\end{table}

In Tab.  \ref{tab:VI} are presented the equilibrium points  $\left(\bar{\eta}, \bar{\Omega}_K, \vartheta\right)$ of the system \eqref{eq180}, \eqref{eq181}, \eqref{eq182}. They are the following (for simplicity, we incorporated the conditions on the parameters $\kappa >0, \lambda <-\kappa $).

\begin{itemize}
    
\item $A,B:  \left(\pm1, 0, \frac{\pi}{2}\right)$. Exists for all parameter values. The eigenvalues are $\left\{\frac{\kappa }{\sqrt{2}}, \sqrt{2} \kappa, \sqrt{2} (\kappa +\lambda )\right\}$. It is a saddle under the conditions $\{\kappa >0, \lambda < -\kappa\}$. Inflation does not occur. Compared with the results of \S \ref{Subsect:II1}, these points represent $O^+$.

\item $C,D:  \left(\pm1, 0, \frac{3 \pi}{2}\right)$. Exist for all parameter values. The eigenvalues are \newline $\left\{-\frac{\kappa }{\sqrt{2}}, -\sqrt{2} \kappa, -\sqrt{2} (\kappa +\lambda )\right\}$. It is a saddle under the conditions $\{\kappa >0, \lambda < -\kappa\}$. Inflation does not occur. Compared with the results of \S \ref{Subsect:II1}, these points represent $O^-$.

\item $E:\left(0, 0, \arctan\left(\frac{\lambda }{\sqrt{6}}\right) \right).$  The eigenvalues are
$\left\{-\frac{\sqrt{\lambda ^2+6}}{\sqrt{2}},  -\frac{\lambda  (\kappa +\lambda )+6}{\sqrt{2} \sqrt{\lambda ^2+6}}, -\frac{\sqrt{2} \left(\lambda^2+2\right)}{\sqrt{\lambda ^2+6}}\right\}$. It is a sink because $\kappa >0, \lambda <-\kappa$ implies $\lambda  (\kappa +\lambda )+6>0$. The deceleration parameter is $q=-1-\frac{\lambda ^2}{2}$. This solution is always inflationary, corresponding to a phantom-dominated solution.

\item $E_{+ \pi}:\left(0, 0, \arctan\left(\frac{\lambda }{\sqrt{6}}\right)+\pi \right).$ The eigenvalues are 
$\left\{\frac{\sqrt{\lambda ^2+6}}{\sqrt{2}},  \frac{\lambda  (\kappa +\lambda )+6}{\sqrt{2} \sqrt{\lambda ^2+6}}, \frac{\sqrt{2} \left(\lambda^2+2\right)}{\sqrt{\lambda ^2+6}}\right\}$. It is a source because $\kappa >0, \lambda <-\kappa$ implies $\lambda  (\kappa +\lambda )+6>0$. $ F$ denotes this point in \S \ref{subsect:ModelII-IIIA}. 
The deceleration parameter is $q=-1-\frac{\lambda ^2}{2}$. This solution is always inflationary, corresponding to a phantom-dominated solution.

\item $ F:\left(0, 1, 0\right)$ The eigenvalues are $\left\{-\frac{2}{\sqrt{3}}, -\frac{2}{\sqrt{3}}, \frac{2}{\sqrt{3}}\right\}$. It is a saddle point. The deceleration parameter is $q=0$. Inflation does not occur. 

\item $G:\left(0, 1, \pi  \right)$ The eigenvalues are $\left\{\frac{2}{\sqrt{3}}, \frac{2}{\sqrt{3}}, \frac{2}{\sqrt{3}}\right\}$. It is a saddle point.
The deceleration parameter is $q=0$. Inflation does not occur. 

Points $F$ and $G$ in this section correspond to the points $T$ and $U$ studied in \S \ref{Sect:IIC}. They represent the Curvature-dominated Milne solutions.

\item $H:\left(0, \frac{3 \left(\lambda ^2+2\right)}{3 \lambda ^2+2}, -\arctan\left(\frac{\sqrt{\frac{2}{3}}}{\lambda}\right) \right)$ The eigenvalues are $\left\{\frac{2 \lambda -\kappa }{\sqrt{3 \lambda ^2+2}}, \frac{\lambda -i \sqrt{3 \lambda
   ^2+8}}{\sqrt{3 \lambda ^2+2}}, \frac{\lambda +i \sqrt{3 \lambda ^2+8}}{\sqrt{3 \lambda ^2+2}}´\right\}$. 
 The conditions  $\kappa >0, \lambda < -\kappa$ imply that the first eigenvalue is negative and the other two have negative real parts. Therefore, it is a sink. This point does not belong to the physical part of the phase space given by $0\leq \bar{\Omega}_K\leq 1$. The deceleration parameter is $q=0$. Inflation does not occur. 

 \item $H_{+\pi}: \left(0,  \frac{3 \left(\lambda ^2+2\right)}{3 \lambda ^2+2}, -\arctan\left(\frac{\sqrt{\frac{2}{3}}}{\lambda}\right)+\pi \right).$ The eigenvalues are \newline $\left\{-\frac{2 \lambda -\kappa }{\sqrt{3 \lambda ^2+2}}, -\frac{\lambda -i \sqrt{3 \lambda
   ^2+8}}{\sqrt{3 \lambda ^2+2}}, -\frac{\lambda +i \sqrt{3 \lambda ^2+8}}{\sqrt{3 \lambda ^2+2}}´\right\}$. 
 The conditions  $\kappa >0, \lambda < -\kappa$ imply that the first eigenvalue is positive and the other two have positive real parts. Therefore, it is a source.
Assuming $0\leq \bar{\Omega}_K\leq 1$, these points do not exist.  The deceleration parameter is $q=0$. Inflation does not occur. 

\item $I_{\pm}: \left(\pm \frac{2 \sqrt{2}}{\sqrt{3 \kappa ^2+8}},  \frac{3 \kappa ^2}{3 \kappa ^2+8},  -\arctan\left(\frac{2 \sqrt{\frac{2}{3}}}{\kappa }\right) \right) $ Exists for $\kappa<0$. The eigenvalues are
$\left\{ -\frac{4 \kappa }{\sqrt{3 \kappa ^2+8}}, \frac{2 \kappa}{\sqrt{3 \kappa ^2+8}}, \frac{2 (\kappa -2 \lambda )}{\sqrt{3 \kappa ^2+8}}\right\}$. It is a saddle. The deceleration parameter is $q=0$. Inflation does not occur. 

\item $I_{\pm}^{+\pi}: \left(\pm \frac{2 \sqrt{2}}{\sqrt{3 \kappa ^2+8}},  \frac{3 \kappa ^2}{3 \kappa ^2+8},  -\arctan\left(\frac{2 \sqrt{\frac{2}{3}}}{\kappa }\right)+\pi \right) $ Exists for all parameter values. The eigenvalues are
$\left\{ \frac{4 \kappa }{\sqrt{3 \kappa ^2+8}}, -\frac{2 \kappa}{\sqrt{3 \kappa ^2+8}}, -\frac{2 (\kappa -2 \lambda )}{\sqrt{3 \kappa ^2+8}}\right\}$. It is a saddle. The deceleration parameter is $q=0$. Inflation does not occur.

\item $J_{\pm}: \left(\pm \frac{\sqrt{\kappa  \lambda +\lambda ^2+6}}{\sqrt{\kappa ^2+2 \kappa  \lambda +\lambda ^2+6}}, 0, -\arctan\left(\frac{\sqrt{6}}{\kappa +\lambda }\right) \right) $ Exists for $0\le \kappa (\kappa+\lambda)$. The eigenvalues are \newline $\left\{\frac{2 (\kappa -2 \lambda )}{\sqrt{3} \sqrt{(\kappa +\lambda )^2+6}}, \frac{\sqrt{3} \kappa -\sqrt{\kappa } \sqrt{4 \left(\kappa ^2+6\right) \lambda +8
   \kappa  \lambda ^2+27 \kappa +4 \lambda ^3}}{2 \sqrt{(\kappa +\lambda )^2+6}},\frac{\sqrt{\kappa } \sqrt{4
   \left(\kappa ^2+6\right) \lambda +8 \kappa  \lambda ^2+27 \kappa +4 \lambda ^3}+\sqrt{3} \kappa }{2 \sqrt{(\kappa
   +\lambda )^2+6}}\right\}$. It is a source for $\lambda <0, 0<\kappa <-\lambda$. The deceleration parameter is $q= 2-\frac{3 \kappa }{\kappa +\lambda }$. Inflates for $\frac{\kappa }{\kappa +\lambda }>\frac{2}{3}$.

\item $J_{\pm}^{+\pi}: \left(\pm \frac{\sqrt{\kappa  \lambda +\lambda ^2+6}}{\sqrt{\kappa ^2+2 \kappa  \lambda +\lambda ^2+6}}, 0, -\arctan\left(\frac{\sqrt{6}}{\kappa +\lambda }\right)+\pi \right) $ Exists for $0\le \kappa (\kappa+\lambda)$. The eigenvalues are $\left\{-\frac{2 (\kappa -2 \lambda )}{\sqrt{3} \sqrt{(\kappa +\lambda )^2+6}}, -\frac{\sqrt{3} \kappa -\sqrt{\kappa } \sqrt{4 \left(\kappa ^2+6\right) \lambda +8
   \kappa  \lambda ^2+27 \kappa +4 \lambda ^3}}{2 \sqrt{(\kappa +\lambda )^2+6}},-\frac{\sqrt{\kappa } \sqrt{4
   \left(\kappa ^2+6\right) \lambda +8 \kappa  \lambda ^2+27 \kappa +4 \lambda ^3}+\sqrt{3} \kappa }{2 \sqrt{(\kappa
   +\lambda )^2+6}}\right\}$. It is a sink for $\lambda <0, 0<\kappa <-\lambda$.  The deceleration parameter is $q= 2-\frac{3 \kappa }{\kappa +\lambda }$. Inflates for $\frac{\kappa }{\kappa +\lambda }>\frac{2}{3}$.

\end{itemize}

\subsubsection{Discussion}
 We divide the results according to whether we have matter ($\rho_m>0$, \S \ref{SectIIIC}) and we have a vacuum ($\rho_m=0$, \S \ref{SectIIIC1}).   For $\rho_m>0$, The early and late-time attractors are as follows. $A, B
$ are sources for $\{\lambda > -\kappa$, $\kappa > 0\}$ and sinks for $\{\lambda < -\kappa$, $\kappa < 0\}$. $C, D$ are  sources for $\{\lambda < -\kappa$, $\kappa < 0\}$ and sinks for $\{\lambda > -\kappa, \kappa > 0\}$. Inflation does not occur for these points. $E$, assuming $-1 \le w \le 1$ we have a sink for $\{\lambda(\kappa+\lambda) > 0\}$. $E_{+\pi}$, assuming $-1 \le w \le 1$ we have a source for $\{\lambda(\kappa+\lambda) > 0\}$.
The deceleration parameter is $q=-1-\frac{\lambda ^2}{2}$. Therefore, these solutions are always inflationary, corresponding to phantom-dominated solutions. $J_{\pm}$ 
is a sink for  $ \left\{-1<w<1, \lambda >0, \kappa <\frac{-\lambda ^2-6}{\lambda }\right\} \cup \left\{-1<w<1, \lambda >0, -\lambda <\kappa <0\right\}$. It is a source for $\left\{-1<w<1, \lambda <0, 0<\kappa <-\lambda\right\} \cup \left\{-1<w<1, \lambda <0, \kappa >\frac{-\lambda ^2-6}{\lambda }\right\}$. $J_{\pm}^{+\pi}$ is a sink for $\left\{-1<w<1, \lambda <0, 0<\kappa <-\lambda\right\} \cup \left\{-1<w<1, \lambda <0,  \kappa >\frac{-\lambda ^2-6}{\lambda }\right\}$. For these solutions the deceleration parameter is $q= 2-\frac{3 \kappa }{\kappa +\lambda }$, therefore, they inflate for $\frac{\kappa }{\kappa +\lambda }>\frac{2}{3}$. $L$
is a sink for $\left\{-1<w<-\frac{1}{3}, \lambda <0, \kappa >\frac{\lambda(1-w)}{w+1}\right\}\cup \left\{-1<w<-\frac{1}{3}, \lambda >0, \kappa <\frac{\lambda(1-w)}{w+1}\right\}$ and 
$L_{+\pi}$ is a source for $\left\{-1<w<-\frac{1}{3}, \lambda <0, \kappa >\frac{\lambda(1-w)}{w+1}\right\}\cup \left\{-1<w<-\frac{1}{3}, \lambda >0, \kappa <\frac{\lambda(1-w)}{w+1}\right\}$.
The deceleration parameter for these solutions is $q=\frac{1}{2} (3 w+1)$, then, they are inflationary solutions for $w<-\frac{1}{3}$.

For $\rho_m=0$,  we incorporated the conditions on the parameters $\kappa >0, \lambda <-\kappa $. The early and late-time attractors are as follows. $E$ that is a sink because $\kappa >0, \lambda <-\kappa$ implies $\lambda  (\kappa +\lambda )+6>0$, $E_{+ \pi}$ that is a source because $\kappa >0, \lambda <-\kappa$ implies $\lambda  (\kappa +\lambda )+6>0$. They are noninflationary. $H$ under the conditions $\kappa >0, \lambda < -\kappa$ have the first eigenvalue negative, and the other two have negative real parts. Therefore, it is a sink. However, this point does not belong to the physical part of the phase space given by $0\leq \bar{\Omega}_K\leq 1$. The points $J_{\pm}$ exists for $0\le \kappa (\kappa+\lambda)$ and are sources for $\lambda <0, 0<\kappa <-\lambda$. $J_{\pm}^{+\pi}$ exists for $0\le \kappa (\kappa+\lambda)$ and are sinks for $\lambda <0, 0<\kappa <-\lambda$. They inflate for $\frac{\kappa }{\kappa +\lambda }>\frac{2}{3}$.

\subsection{The effect of curvature on the dynamics}

In section \ref{sec2}, we discussed the first model, and in section \ref{sec3}, we explored the second one. Like the previous model, it has similar existence conditions and physical interpretation for flat solutions, regardless of the curvature being positive or negative. The stability conditions for negatively curved models are the same as for positively curved models, with only slight differences from the flat case. The curvature introduces new equilibrium points that affect the dynamics, causing changes in the stability conditions of early and late-time attractors. However, the curvature has no impact on the conditions for inflation.

When comparing the results of sections \ref{sec2} and \ref{sec3}, the primary difference is the presence of phantom-dominated late-time solutions in the latter. These solutions are always accelerating  and inflationary, and always exist, replacing the usual quintessence-dominated solutions.

\section{Conclusions}
\label{sec4}

We have conducted a thorough mathematical analysis of FLRW cosmologies with a non-zero spatial curvature, with the energy-momentum tensor describing two scalar fields that follow the Chiral-quintom action integral. The two scalar fields, representing a quintessence and a phantom field, are minimally coupled to gravity, with their kinetic terms interacting. Additionally, we considered an ideal gas matter source. The kinetic energy of the scalar field Lagrangian defines a space of constant curvature with a Lorentzian signature. We have examined two Chiral-quintom models based on the Lorentz signature of the space of constant curvature.

We have derived the second-order field equations for these cosmological models. We have rewritten them through normalized variables, using dominant quantities for the normalization, which differs from the usual Hubble normalization. This approach has provided us with a set of algebraic equations necessary for reducing the dimension of the dynamical system. 

We have conducted a complete study of the asymptotic dynamics for the two Chiral-quintom models we considered. In particular, we have derived the equilibrium points that describe asymptotic solutions for the field equations. We have paid particular attention to the effect of spatial curvature on the local stability of the attractors in the corresponding flat models. We have discussed several physical properties of the models, including identifying those that undergo a bounce. We have paid particular attention to models that experience two different periods of inflation related to the possible Universe’s early and late-time acceleration phases.

\section*{Acknowledgments}

A.C. was supported by NSERC of Canada. Vicerrectoría de
Investigación y Desarrollo Tecnológico (Vridt) at Universidad Católica del Norte funded G. L.  through Resolución VRIDT No. 040/2022, Resolución VRIDT No. 054/2022,  Resolución VRIDT No. 026/2023 and Resolución VRIDT No. 027/2023.  He also thanks the support of Núcleo de Investigación Geometría Diferencial y Aplicaciones, Resolución Vridt N°096/2022 and the funding through Resolución VRIDT No. 026/2023 and Resolución VRIDT No. 027/2023.

    \bibliographystyle{unsrtnat}

\bibliography{refs.bib}

\end{document}